\documentclass[twoside,11pt]{article}

\usepackage{jmlr2e}

\usepackage{xcolor}

\newcommand{\revision}{\textcolor{black}}

\usepackage{caption}
\usepackage{subcaption}

\firstpageno{1}

\begin{document}

\title{Towards a robust out-of-the-box neural network model for genomic data}

\author{\name Zhaoyi Zhang\thanks{Joint first authors with equal contribution and randomly chosen order (reproducible script in Supplementary Material)}
\email zzhang825@wisc.edu \\
\name Songyang Cheng$^*$ 
\email scheng72@wisc.edu \\
       \addr Department of Computer Science\\
       University of Wisconsin-Madison\\
       Madison, WI 53715, USA
       \AND
       \name Claudia Sol\'is-Lemus\thanks{Corresponding author} \email solislemus@wisc.edu \\
       \addr Wisconsin Institute for Discovery \\
       and Department of Plant Pathology\\
       University of Wisconsin-Madison\\
       Madison, WI 53715, USA
       }


\maketitle

\begin{abstract}
  The accurate prediction of biological features from genomic data is
  paramount for precision medicine and sustainable agriculture. For decades, neural network models have been widely popular in fields like computer vision, astrophysics and targeted marketing given their prediction accuracy and their robust performance under big data settings. Yet neural network models have not made a successful transition into the medical and biological world due to the ubiquitous characteristics of biological data such as modest sample sizes, sparsity, and extreme heterogeneity. 

Here, we investigate the robustness, generalization potential and prediction accuracy of widely used convolutional neural network and natural language processing models with a variety of heterogeneous genomic datasets. 
Mainly, recurrent neural network models outperform convolutional
neural network models in terms of prediction accuracy, overfitting and
transferability across the datasets under study.
While the perspective of a robust out-of-the-box neural network model is out of reach, we identify certain model characteristics that translate well across datasets and could serve as a baseline model for translational researchers.
\end{abstract}


\section*{Background}

The ability to accurately predict phenotypes from genomic data is one of the most coveted goals of modern-day medicine and biology. Examples abound: from precision medicine where researchers want to predict a patient's disease susceptibility based on the genetic information \citep{ashley2015precision,rost2016protein,katuwal2016machine,krittanawong2017artificial,lee2018machine,ho2019machine} to prediction of antibiotic-resistant bacterial strains based on the genomes of pathogenic microbes \citep{fjell2009identification, coelho2013use, pesesky2016evaluation, kavvas2018machine, li2018tracking}. Examples extend beyond human health into soil and plant health such as the prediction of crops yield (or plant disease susceptibility) based on soil microbiome metagenomic data \citep{chang2017metagenome, bokulich2018q2, carrieri2019fast} and the prediction of pesticide-resistant microbial strains from plant bacterial pathogen genomes \citep{yang2017machine, ip2018big, maino2018climate, duarte2018evaluating}.
Our ability to anticipate outcomes from data is at our scientific core when we face human disease, environmental challenges, and climate change.

Naturally, biologists and medical researchers have turned to the machine-learning community for answers given the great success of machine-learning methods in a plethora of applications such as computer vision \citep{hjelmaas2001face,egmont2002image} and astrophysics \citep{kucuk2017large,jonas2018flare} to name a few.

However, the success of machine-learning methods on other fields has not been easily translated to the biological realm \citep{chen2018, ekins2019, teschendorff2019avoiding, dacrema2019}. Indeed, the complexity of biological omics data has hampered the adoption of machine-learning models, especially neural networks.
Among the main challenges of genomic data in neural network models are 1) smaller sample sizes compared to other fields, \revision{2) highly imbalanced datasets,} and 3) heterogeneity of training samples and testing samples.

First, despite the advance in high-throughput sequencing technologies, extracting whole genomes remains a time-consuming and expensive task when sample sizes must be in the order of thousands.
In addition, data privacy and restrictions on data sharing in medical research restrict scientists' ability to combine multiple smaller datasets into larger ones suitable for neural network modeling. 

Second, more important than sample size, the weak link of deep learning in biological applications is the assumption of homogeneity between training and testing samples. This assumption is violated, for example, in microbial datasets where laboratory samples (training data) and environmentally or clinically collected samples (testing data) can be intricately heterogeneous. This data heterogeneity can cause lack of robustness and generalization errors in neural network models. Robustness is the key ingredient that is needed for neural network models to translate into medical practice and into the phenotype prediction in the agricultural or environmental field.

In literature, there are multiple examples of successfully fit neural network models on biological or medical genomic data \citep{andrew2015weirauch, zhou2015predicting, kelley2016basset,Nguyen2016-jj,Zeng2016-nu, Agarwal2019-pr,Trabelsi2019-yt,Shadab2020-sb}. However, it remains uncertain whether the proposed models could be translated to other similar datasets with comparable performance. That is, we ask whether the neural network models proposed in literature are robust across heterogeneous (but similar in nature) datasets.

In addition, we approach the existing neural network models with the mindset of a biological or medical user. A biological researcher would see the neural network model in an existing publication and then would try to apply a similar model to their own dataset. First, we explore how easy it is to replicate the analysis on existing publications. Second, by making incremental changes to the model characteristics, we gauge the effect of each model component on the overall performance.

We learn mainly three things: 1) in multiple instances, we are not able to replicate the performance in existing publications either because data is not available, code is not available, or code is corrupted, incomplete or not well-documented; 2) most of the times the good performance of existing models does not translate to alternative datasets, yet we do encounter some model characteristics that are generally robust across datasets and that could serve as a potential baseline model -- albeit with modest performance -- to start the prediction process from a user perspective, and 3) we find that accurate prediction is a balancing game between underfitting and overfitting, and that small changes in the architecture can have unpredictable outcomes.

The quest for robust neural network models that could tackle the complexities of biological data (and its intrinsic heterogeneity) is imminent. Neural network models cannot be fully applicable in informed patient care, medical or agricultural framework if they cannot guarantee some level of generalization potential given that genomic data are not static but constantly evolving.
The difficulty of the prediction problem in biology or medicine is such that it would be naive to believe that there will exist an out-of-the-box model that will be fully transferable \revision{(recall the ``No Free Lunch" (NFL) theorem \citep{Wolpert1997}: improved performance over one class of problems is offset by performance over another class).}
Yet, from a user perspective, it is desirable to know if there are certain model characteristics that perform modestly under scrutiny from a variety of different datasets. 

While we advise biological or medical users against using out-of-the-box strategies, we conclude from our study that recurrent neural networks are relatively robust across genomics datasets and generally not affected by the size or type of the data. Overfitting is an issue on more complex CNN models (as expected), but it is relatively controlled via regularization schemes. We also found that a general LSTM layer for embedding performed relatively well across datasets and outperforms more intuitive data encoding schemes like doc2vec which performs poorly on all the scenarios we tested.
Finally, our work raises awareness to the importance of reproducibility and replicability. As machine-learning scientists, it is crucial to accompany our work with reproducible scripts that are relatively easy to follow by the scientific community so that our findings have an impact across fields, in particular, into the biological and medical community.


\section*{Methods}

We focus on convolutional neural networks (CNN) \citep{Zeng2016-nu, Nguyen2016-jj, Shadab2020-sb, Trabelsi2019-yt} and natural language processing (NLP) \citep{le2014distributed, Agarwal2019-pr, Dutta2018} on three datasets of increasing size from the available ones in the papers under study (Table \ref{tab:datasets}) and described below.

\subsubsection*{Splice data.}
In \citep{Nguyen2016-jj}, the authors included a splice dataset (also in the UCI machine learning repository \citep{Dua:2019}). Splice junctions are points on a DNA sequence at which superfluous DNA is removed during the process of protein creation in higher organisms. This dataset has 3190 sequences of length 60 bp and are classified into three classes: exon/intron boundaries (EI: 24\%), intron/exon boundaries (IE: 24\%), and non-splice (N: 52\%).

\subsubsection*{Histone data.}
In \citep{Nguyen2016-jj}, the authors included 10 datasets about DNA
sequences wrapping around histone proteins. We focus on the H3
occupancy from the histone dataset that has 14,965 sequences of length
500 bp. The H3 indicates the histone type, and the dataset has two
classes: the positive class includes DNA sequences that contain
regions wrapping around histone proteins (51\%) and the negative class
does not contain such regions (49\%).

\subsubsection*{Motif discovery data.}
In \citep{Zeng2016-nu}, the authors included two ChIP-seq datasets:
motif discovery and motif occupancy. These datasets contain the labels
of the binding affinity of transcription factors to DNA sequence in
690 different ChIP-seq experiments. We only focus on a subset of
269,100 sequences from the motif discovery data (out of 20,464,149) of
length 101 bp. The dataset contains two classes: positive class includes DNA sequences that are motif (50\%) and negative class includes DNA sequences that are not motif (50\%). 

\subsubsection*{Data splitting.}
For all CNN models, we use the following split of the data. The splice dataset is split into \revision{75\% for training}
and 25\% for testing \revision{with 15\% of training data used for validation}. The histone dataset is split into 70\% for training, 15\% for validation, and 15\% for testing. The motif discovery data is split into 48.7\% for training, 2.6\% for validation, and 48.7\% for testing. \revision{We note that the data partition for the motif discovery dataset deviates from the standard 70-15-15 or 75-25 data partitions. The rationale for this data partition is that the motif discovery data was stored in 690 different files each with a different number of sequences. Given that we do not know how these datasets were created, we wanted to have a uniform representation from all datasets in the training process. The smallest file had 190 sequences, so we randomly selected 190 sequences for each of the 690 files to be used in training. This represents 48.7\% of samples used for training. A higher proportion of training samples would imply that some files would be over-represented which could introduce unintended biases in prediction. We choose the same proportion for testing to be able to evaluate the model better given the high heterogeneity of the data leaving only 2.6\% for validation.}

Data encoding differs for the CNN models and the NLP models, so we describe the encoding procedure in the next sections for each type of model.

\subsection*{Convolutional Neural Networks}

We test the performance of four convolutional neural network (CNN) models found in literature \citep{Zeng2016-nu, Nguyen2016-jj, Shadab2020-sb, Trabelsi2019-yt} that have been successful on genomic-based prediction. We assess their performance on their own datasets (when available) and on alternative similar datasets, as well as under incremental modifications of the model characteristics such as data encoding, window size and number of layers. See Table \ref{tab:analyses} for a summary of the performance tests and models.  \revision{For all models, we use the cross-entropy loss.}

\subsubsection*{CNN-Nguyen model \citep{Nguyen2016-jj}.}
\begin{figure}[h!]
    \includegraphics[width=4.75in]{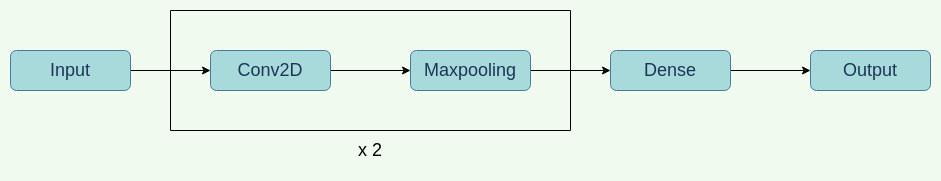}
    \caption{Original CNN-Nguyen (2D) \citep{Nguyen2016-jj}. \revision{The dense layer has a dropout rate of 0.5 to prevent overfitting.}}
    \label{fig:cnn-nguyen}
\end{figure}

\begin{figure}[h!]
    \includegraphics[width=4.75in]{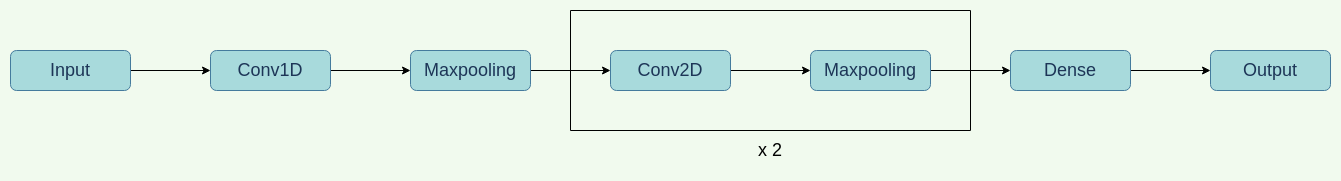}
    \caption{Modified CNN-Nguyen (2D+1D) \citep{Nguyen2016-jj}. \revision{The dense layer has a dropout rate of 0.5 to prevent overfitting.}}
    \label{fig:cnn-nguyen2}
\end{figure}

We implement the simple neural network in \citep{Nguyen2016-jj} as our baseline model (Figure \ref{fig:cnn-nguyen}). The model contains two 2D convolutional layers, each followed by a pooling layer, then the output of the convolutional layers are connected to a fully connected layers. The fully connected layer has a dropout rate of 0.5 to reduce the effect of overfitting. Finally, we use a softmax layer to predict the labels of the input sequences. We denote this original model as CNN-Nguyen2D in the results. In addition to this model, we construct a new model with an extra 1D convolutional layer denoted CNN-Nguyen2D+1D for performance comparison. We compare the performance of the model with a different dimension (1D) and increasing number of layers (Figure \ref{fig:cnn-nguyen2}). \revision{For splice and histone datasets, the batch size is 32 and for motif discovery dataset, the batch size is 512. Kernel size is $(3,3)$ for all 2D convolutional layers and $(1,3)$ for the 1D layer in CNN-Nguyen2D+1D. The number of filters in convolutional layer doubles each time we add a new set of these layers (e.g. 16 filters in the first convolutional layer, 32 in the second, 64 in the third). We use the Adam optimizer with learning rate 0.001 and train for 50 epochs which was assessed to allow sufficient training time for convergence (lack of change in loss over the last few epochs) on all datasets. Early Stopping Callback is not used when training these models as convergence was easily assessed in these cases by studying the loss dynamics.}

\subsubsection*{CNN-Zeng model \citep{Zeng2016-nu}.}
\begin{figure}[h!]
    \includegraphics[width=4.75in]{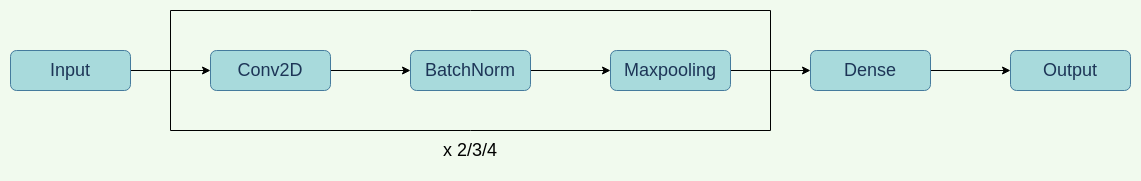}
    \caption{Original CNN-Zeng (2 layers), modified (3,4 layers) \citep{Zeng2016-nu}. \revision{The dense layer has a dropout rate of 0.5 to prevent overfitting.}}
    \label{fig:cnn-zeng}
\end{figure}

We implement the neural network model in \citep{Zeng2016-nu} (Figure \ref{fig:cnn-zeng}) that contains two 2D convolutional layers 
each followed by a batch-normalization and max-pooling layer. The output of the convolutional layers is connected to a fully connected layer. This layer has a dropout rate of 0.5 to prevent overfitting. Finally, a softmax layer is used to predict the class of the input sequence. We denote the original model as CNN-Zeng2 in the results because it has two 2D convolutional layers. We create two new model extensions: CNN-Zeng3 and CNN-Zeng4 with three and four 2D convolutional layers respectively.
To explore the effect of the number of layers, we add 2D convolutional, batch-normalization, and max-pooling layers to the end of the convolutional network (Figure \ref{fig:cnn-zeng}). \revision{For splice and histone datasets, the batch size is 32 and for motif discovery dataset, the batch size is 512. Kernel size is $(3,3)$ for all 2D convolutional layers.} The number of filters in convolutional layer doubles each time we add a new set of these layers (e.g. 16 filters in the first convolutional layer, 32 in the second, 64 in the third). \revision{We use the Adam optimizer with learning rate 0.001 and train for 50 epochs which was assessed to allow sufficient training time for convergence (lack of change in loss over the last few epochs) on all datasets. Early Stopping Callback is not used when training these models as convergence was easily assessed in these cases by studying the loss dynamics.}

\subsubsection*{DeepDBP model \citep{Shadab2020-sb}.}
\begin{figure}[h!]
    \includegraphics[width=4.75in]{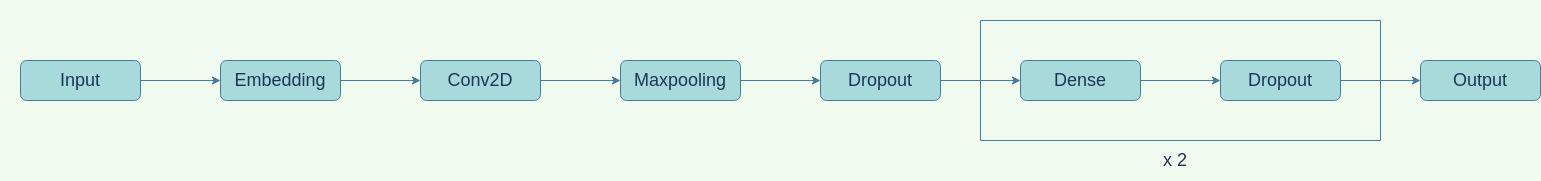}
    \caption{DeepDBP \citep{Shadab2020-sb}}
    \label{fig:deepdbp}
\end{figure}

\revision{Even though the source code of this paper is not well structured and contains many different models that are not properly documented, 
we implemented a model based on the paper description which is what a
domain scientist (like a biomedical researcher) would do.} 
\revision{The model architecture} contains a embedding layer, a convolutional layer, max-pooling layer, followed by fully connected layers and the output layer (Figure \ref{fig:deepdbp}). \revision{For splice and histone datasets, the batch size is 32 and for motif discovery dataset, the batch size is 512. Kernel size is $(1,3)$ for the 1D convolutional layer with 128 filters. We use the Adam optimizer with learning rate 0.001 and train for 50 epochs which was assessed to allow sufficient training time for convergence (lack of change in loss over the last few epochs) on all datasets. Early Stopping Callback is not used when training these models as convergence was easily assessed in these cases by studying the loss dynamics.}
\revision{Unlike CNN-Nguyen, CNN-Zeng and DeepRAM which only have one dropout layer at the dense layer, the DeepDBP model has three dropout layers: dropout -- dense -- dropout -- dense -- dropout -- dense (output) with dropout rate of 0.3.}

\subsubsection*{DeepRAM model \citep{Trabelsi2019-yt}.}
\begin{figure}[h!]
    \includegraphics[width=4.75in]{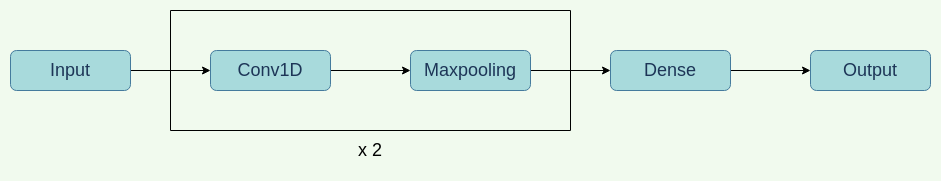}
    \caption{DeepRAM-CNN \citep{Trabelsi2019-yt}. \revision{The dense layer has a dropout rate of 0.5 to prevent overfitting.}}
    \label{fig:deepram-cnn}
\end{figure}

\begin{figure}[h!]
    \includegraphics[width=4.75in]{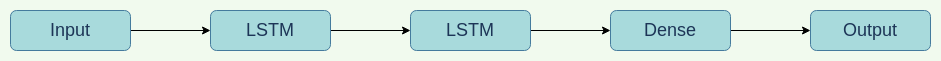}
    \caption{DeepRAM-RNN \citep{Trabelsi2019-yt}. \revision{The dense layer has a dropout rate of 0.5 to prevent overfitting.}}
    \label{fig:deepram-rnn}
\end{figure}

\begin{figure}[h!]
    \includegraphics[width=4.75in]{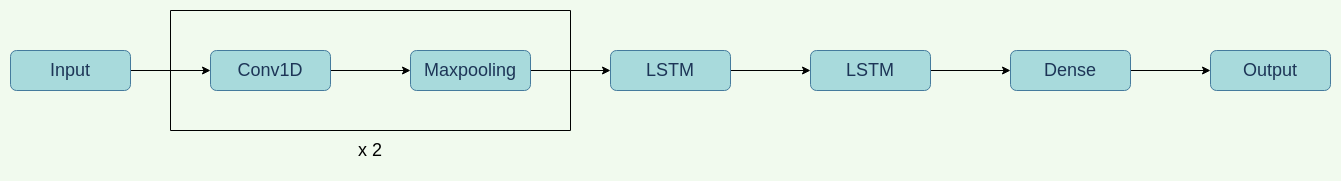}
    \caption{DeepRAM-CNN-RNN \citep{Trabelsi2019-yt}. \revision{The dense layer has a dropout rate of 0.5 to prevent overfitting.}}
    \label{fig:deepram-cnnrnn}
\end{figure}

We implement the three models in \citep{Trabelsi2019-yt}: 1D convolutional neural networks (Figure \ref{fig:deepram-cnn}) denoted DeepRAM-CNN, recurrent neural networks (Figure \ref{fig:deepram-rnn}) denoted DeepRAM-RNN, and a mixture of 1D convolutional and recurrent neural networks (Figure \ref{fig:deepram-cnnrnn}) denoted DeepRAM-CNN-RNN. For convolutional neural networks, we use two 1D convolutional layers, each followed by a max-pooling layer, and finally fully connected layer \revision{(with a dropout rate of 0.5 to prevent overfitting)} and output layer. 
\revision{For splice and histone datasets, the batch size is 32 and for motif discovery dataset, the batch size is 512. Kernel size is $(1,3)$ for the 1D convolutional layer. The number of filters in convolutional layer doubles each time we add a new set of these layers (e.g. 16 filters in the first convolutional layer, 32 in the second). We use the Adam optimizer with learning rate 0.001 and train for 50 epochs which was assessed to allow sufficient training time for convergence (lack of change in loss over the last few epochs) on all datasets. Early Stopping Callback is not used when training these models as convergence was easily assessed in these cases by studying the loss dynamics.}
For recurrent neural networks, we use two LSTM layers followed by fully connected layers and output layer. For the hybrid neural networks, we use two 1D convolutional layers and two Long short-term memory layers, and finally fully connected layers and output layers. We note that the DeepRAM-RNN and the DeepRAM-CNN-RNN models are not entirely CNN models and share many characteristics with the \revision{Natural Language Processing} (NLP) models we will describe next. However, we present these models in this section given that they are all part of the DeepRAM paper \citep{Trabelsi2019-yt} and we follow the comparisons and analyses highlighted in this work.
We compare the changes in performance based on the data encoding as well as comparing the performance of convolutional vs recurrent models.

\subsubsection*{Data encoding.}
For the first three models (CNN-Nguyen, CNN-Zeng and DeepDBP), we use
the same data encoding as in \citep{Nguyen2016-jj} described next. A sliding window of fixed size $k$ \revision{allows us to traverse the sequence focusing on windows of length $k$. The window of length $k$ is a sequence of $k$ nucleotides denoted k-mer. The slide stride is how many nucleotides the window moves to the right as it is traversing the sequence}. At each step, a k-mer is read from the \revision{sequence} and added to the k-mer sequence. For example, if a sequence looks like
``ACTGG", \revision{a window size of 3 with slide stride of 1 would produce the} 3-mers [``ACT", ``CTG", ``TGG"]. The process is similar to how n-grams are created from text \revision{with the k-mer being the word and $k$ being the ``word size"}. After that, one-hot encoding is applied to the k-mers. To also include to spatial information of the sequences, we concatenate the encoding of k-mers within a fixed region size. \revision{For example, for the 3-mers [``ACT", ``CTG", ``TGG"], a region of 2 would imply that we concatenate [``ACT", ``CTG"] to build the 2D encoded data matrix (see Figure 4 in \citep{Nguyen2016-jj} for more details)}. As in \citep{Nguyen2016-jj}, we choose a window size of 3 \revision{with slide stride of 1} and a region of 2. 

For the DeepRAM models, we experiment with two different ways of encoding the sequences. One way is, as described before, to use one-hot encoding with word size 3 and region size 2. The other way is to convert sequences into overlapping k-mers, and embed k-mers into dense vectors using embedding layers. Note that this embedding is different from the one used in the NLP models (described in next section) because of the unit used for encoding. Here, we use the k-mer as the unit for encoding while in \revision{NLP models described next}, we use the nucleotide as the unit.

\subsection*{Natural Language Processing \revision{in conjuction with} Neural Networks \revision{Models for Prediction}}
\label{nlp}

Traditionally, genomic data is stored as a collection of long strings comprised of the four nucleotides: A,C,G,T. 
It is thus intuitive to turn to Natural Language Processing (NLP) theory for solid ways to embed the sequences in a latent space.
Furthermore, NLP methods naturally overcome one of the main drawbacks of CNN models which is the sparsity of the input vectors.

Here, we focus on two widely used NLP tools: doc2vec \citep{le2014distributed, kimothi2016distributed} and Long Short Term Memory (LSTM) \citep{Hochreiter1997, Agarwal2019-pr, kimothi2016distributed}. Both methods share the same objective: represent the input sequence with a low dimensional dense vector yet the specifics differ as is explained below.

We first clarify that the NLP methods are not performing prediction (as the CNN models). Since the purpose of this work is to compare the performance of neural network models on the prediction of phenotypes from genomes, we need to add a neural network model to the NLP model that will perform the prediction of labels (Figure \ref{fig:simple-nn}).  
Table \ref{tab:analyses2} presents a summary of the performance tests and models.

\subsubsection*{LSTM-layer model.}
\begin{figure}[h!]
    \includegraphics[width=4.75in]{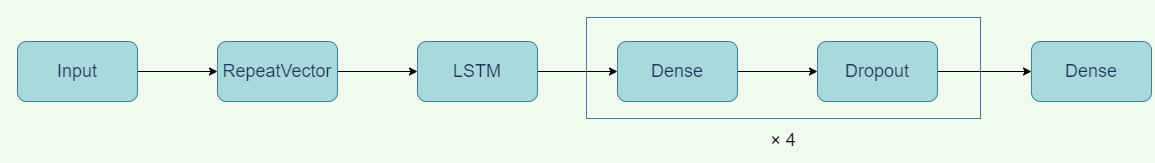}
    \caption{LSTM-layer}
    \label{fig:lstm-layer}
\end{figure}

We implement the neural network model (Figure \ref{fig:lstm-layer}) that contains, after the input layer, an embedding layer followed by an LSTM layer with size of 30 for both datasets. There are four dense layers with size decaying by a factor of 2 (128-64-32-16). There is one dropout layer between any two dense layers with dropout rate of 0.2. With this model, we study the changes in performance when using different optimizers: Adam and SGD. 
\revision{For the Adam optimizer, we use a learning rate of 0.001 and for SGD optimizer, we used a learning rate of 0.01.}
We use Early Stopping Callback on both Adam and SGD optimizers \revision{with a maximum number of epochs set at 4000 for the splice and the histone data, and 200 for the motif discovery data. The patience parameter (the threshold to stop the training if the loss stops decreasing further after a certain number of epochs) is set at 100 for both optimizers for the splice data, at 100 and 400 for Adam and SGD respectively for the histone data, and at 10 for both optimizers for the motif discovery data. Changes in the patience parameter are due to differences in speed between the two optimizers when training.}
\revision{For the splice data, training with the Adam optimizer stopped early at 289 epochs while training with SGD optimizer 
stopped early at 2872 epochs. For the histone data, training with the Adam optimizer stopped early at 154 epochs and at 526 for the SGD optimizer. Finally, for the motif discovery data, training stopped early at 51 epochs for the Adam optimizer, but training
reached the maximum number of epochs allowed (200) for the SGD optimizer bringing into question the convergence of such training.}

\subsubsection*{doc2vec+NN model \citep{le2014distributed, kimothi2016distributed}.}
The nature of the doc2vec sequence representation as a semantic vector preserves similarity of sequences in terms of frequency and location of n-grams.
We apply the distributed memory mode (DV-PM) as in \citep{le2014distributed, kimothi2016distributed}, and then, we use the simple fully connected neural network in Figure \ref{fig:simple-nn} containing two dense layer with size shrinking by a factor of 2 with a dropout layer in between.
We study the effect of embedding size in the performance of the model. \revision{For all instances of this model, we use the SGD optimizer with learning rate of 0.01 and momentum of 0.9. We use Early Stopping Callback with maximum number of iterations allowed as 1000 for the splice and histone data, and 400 for the motif discovery data. The patience parameter is set at 50 for the splice data, 30 for the histone data and 10 for the motif discovery data. Again, changes in the maximum number of iterations and patience parameter are due to the speed of training for different sample sizes. Training stopped early on all instances of the model. For the splice data, training stopped early at 119, 143, 166, and 264 epochs for the four embedding sizes used (50, 100, 150, and 200). For the histone data, training stopped early at 114, 177, 61, and 118 for the four embedding sizes used (50, 100, 150, and 200). For the motif discovery data, training stopped early at 36, 39, 58, and 11 for the four embedding sizes used (50, 100, 150, and 200).}

\subsubsection*{LSTM-AE+NN model \citep{Agarwal2019-pr}.}
\begin{figure}[h!]
    \includegraphics[scale=0.35]{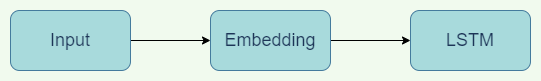}
    \caption{LSTM-AE (encoder) \citep{Agarwal2019-pr}}
    \label{fig:lstm-ae-encoder}
\end{figure}

\begin{figure}[h!]
    \includegraphics[scale=0.35]{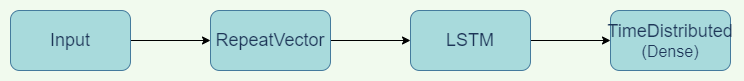}
    \caption{LSTM-AE (decoder) \citep{Agarwal2019-pr}}
    \label{fig:lstm-ae-decoder}
\end{figure}

\begin{figure}[h!]
    \includegraphics[scale=0.35]{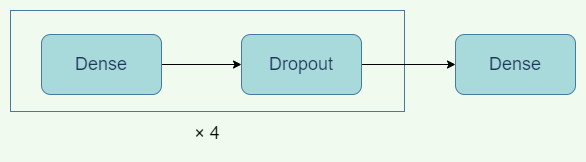}
    \caption{Simple NN for prediction after LSTM-AE (encoder) and doc2vec}
    \label{fig:simple-nn}
\end{figure}

A LSTM autoencoder model (LSTM-AE) aims to represent a sequence by a dense vector that can be converted back to the original sequence. Indeed, LSTM-AE is comprised of two parts: an encoder network (Figure \ref{fig:lstm-ae-encoder}) that compresses the original sequence into a low dimensional dense vector, and a decoder network (Figure \ref{fig:lstm-ae-decoder}) that converts the vector back to the original sequence. 
The encoder reads as input an encoded DNA sequence and outputs a dense vector as the embedding for this sequence whose length is a hyper parameter to tune. 
The decoder reads as input the dense vector produced by the encoder and produces a reconstructed sequence. 
The accuracy of the autoencoder is measured by comparing the reconstructed sequence to the original sequence.
We implement a LSTM-AE following \citep{Agarwal2019-pr} based on the description in their publication given that no reproducible script was available.
The LSTM-AE model is trained to achieve maximum reconstruction accuracy of the sequences.
Then, since LSTM-AE is not performing classification, we add a simple fully connected neural network (Figure \ref{fig:simple-nn}) containing two dense layer with size shrinking by a factor of 2 with a dropout layer in between for the prediction of class labels. The size of the first dense layer is adjusted, as a rule of thumb, to match 1 to 4 times the embedding dimension. We denote this model as LSTM-AE+NN.
We note that only the weights corresponding to the simple fully connected neural network are optimized for classification which is different to the LSTM-layer model whose embedding indeed changes during training. 
We highlight that the LSTM-layer and LSTM-AE models differ on how an embedding is evaluated. The embedding produced by the LSTM-layer model aims to better classify the sequences into the right category while the embedding produced by the LSTM-AE model aims to better capture the sequence itself. 
We study the effect of the batch size in the performance of the model.
\revision{For the LSTM-AE training, we use the Adam optimizer with
  learning rate of 0.001 while for the training of the simple NN for
  prediction, we use the SGD optimizer with momentum of 0.9 and with
  learning rate of 0.01 for the splice and motif discovery data, and
  0.001 for the histone data. We use Early Stopping Callback on all
  cases with 2000, 4000 and 200 maximum iterations allowed for splice,
  histone and motif discovery data respectively for the LSTM-AE
  training, and 1000, 1500 and 500 maximum iterations allowed for
  splice, histone and motif discovery data respectively for the simple
  NN training. In terms of the patience parameter, we set it at 100,
  200 and 10 for the splice, histone and motif discovery data
  respectively for both LSTM-AE and simple NN training. The LSTM-AE
  training stopped early in almost all cases: 1) for the splice data,
  at epoch 1474, 424, and 1005 for the three batch sizes used (32,
  256, and 1024 respectively); 2) for the histone data, at epoch 549,
  422, and 646 for the three batch sizes used (32, 256, and 1024
  respectively), and 3) for the motif discovery data, at epoch 78 and
  195 for batch sizes 32 and 256. For this data, training reached the
  maximum number of iterations allowed (200) for the case of batch
  size of 1024 bringing into question the convergence of this
  case. The training of the simple NN stopped early in all cases: 1)
  for the splice data, at epoch 123, 114, and 200 for the three batch
  sizes used (32, 256, and 1024 respectively); 2) for the histone data, at epoch 212, 805, and 999 for the three batch sizes used (32, 256, and 1024 respectively), and 3) for the motif discovery data, at epoch 185, 146, and 62 for the three batch sizes used (32, 256, and 1024 respectively).}


\revision{We summarize the training details for the NLP models in Table \ref{tab:training}. We note that since the training of the CNN was simpler (50 epochs in all cases), we do not need a summarizing table for the training of the CNN models.}

\subsubsection*{Data encoding.}
For the LSTM-layer and the LSTM-AE models, we use the same data
encoding as described next. Each nucleotide is converted to a label number. For example, [``A", ``C", ``G", ``T"], are encoded as [3, 2, 1, 0] in descending lexicographical order.
The LSTM-layer is crucial given the intractable growth in dimension of the input vector. That is, a sequence containing 6000 nucleotides would be represented by a sequence of 6000 numbers.
For the doc2vec model, 
we encode the sequences based on 3-mers with \revision{slide stride} of 1.
For example, for the ``ACTGG" sequence, the 3-mers are [``ACT", ``CTG", ``TGG"]. 
We construct a dictionary with all the 3-mers in the training set. 
While it is unlikely for 3-mers in the test set to not appear in the dictionary, we categorize these instances as out-of-vocabulary (OOV) with a unique encoding.

\section*{Results}

\subsection*{The role of dimension and number of layers on CNN}

\begin{figure}[h]
    \centering
    \includegraphics[scale=0.175]{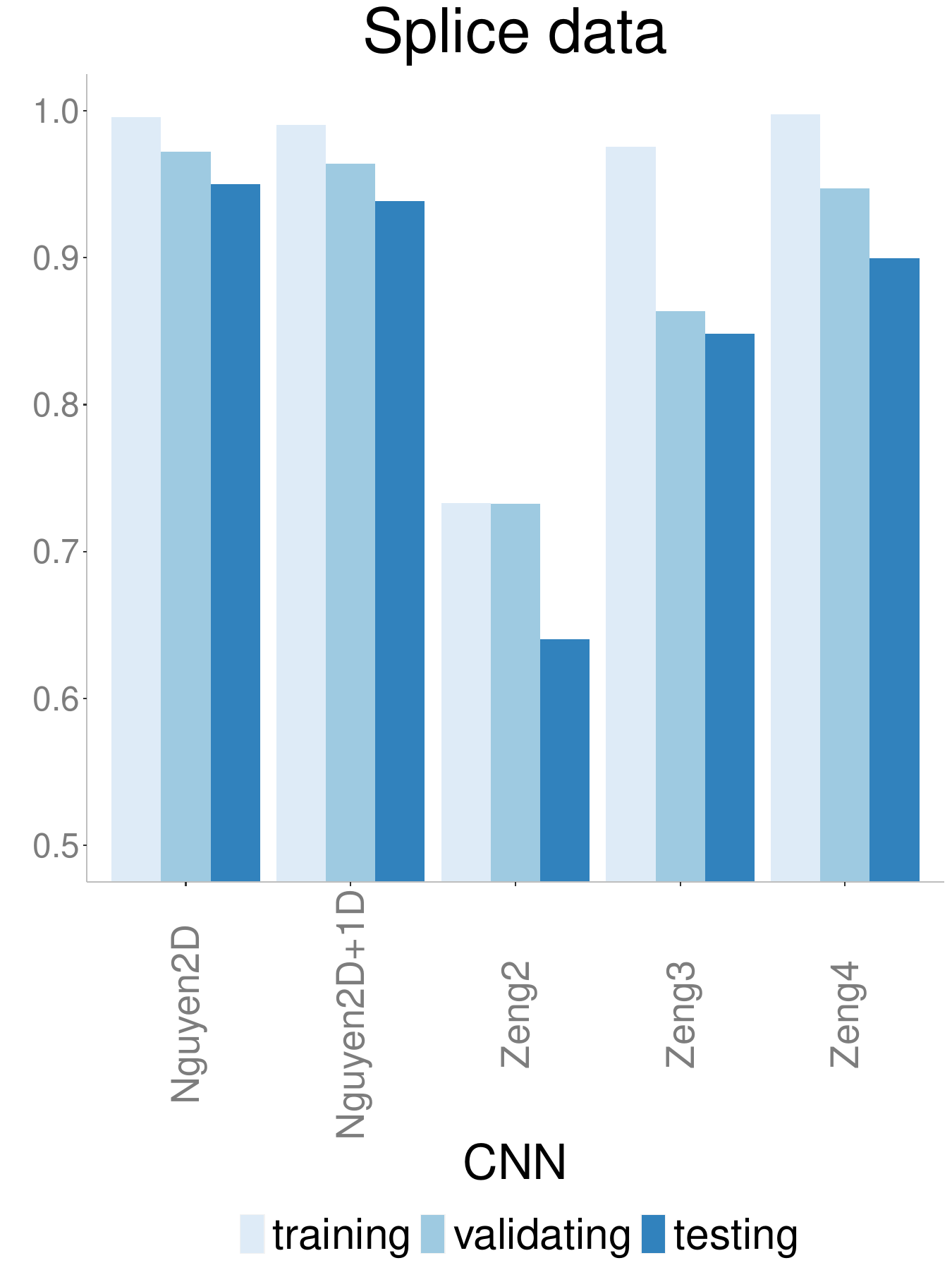}
    \includegraphics[scale=0.175]{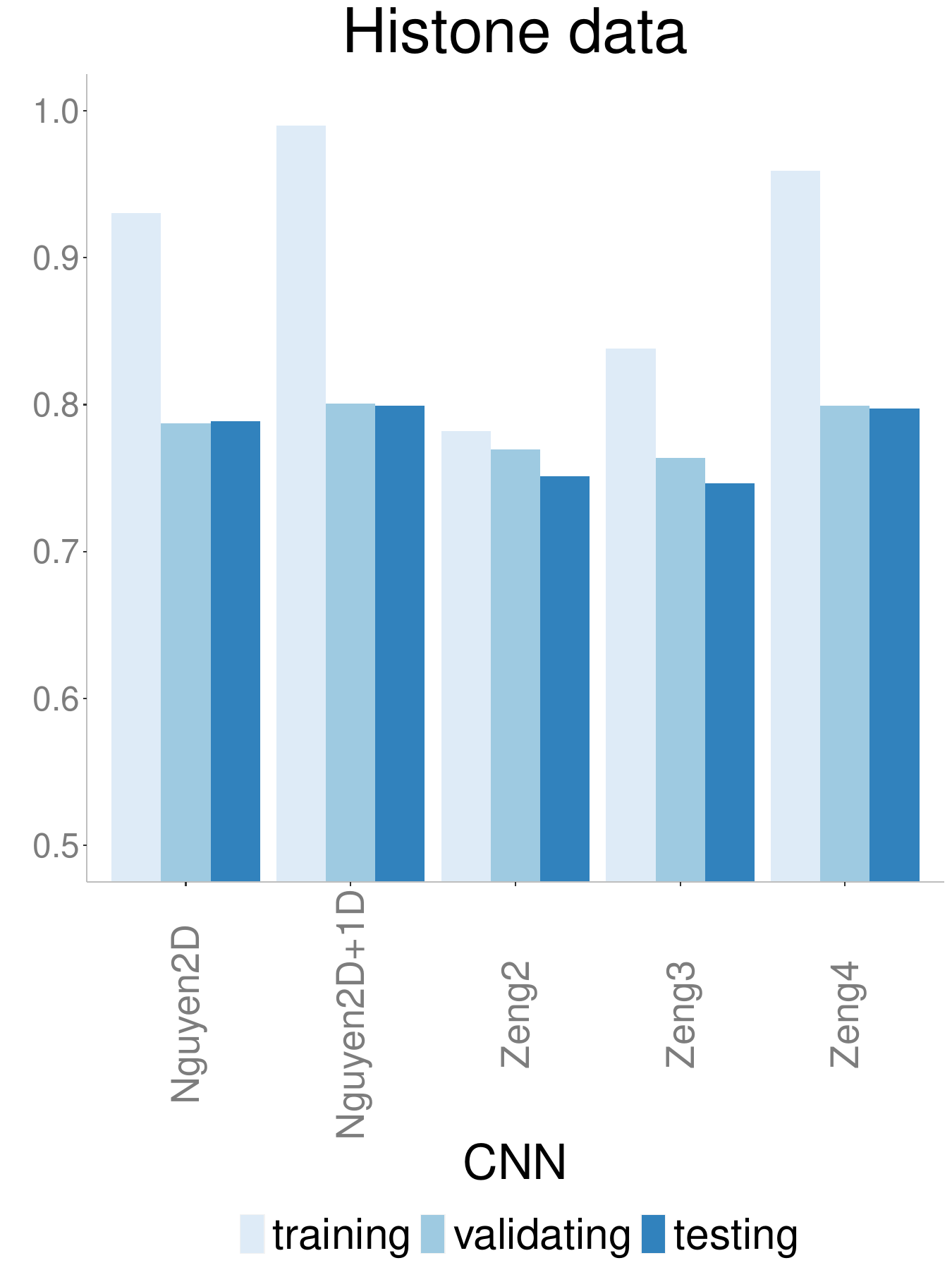}
    \includegraphics[scale=0.175]{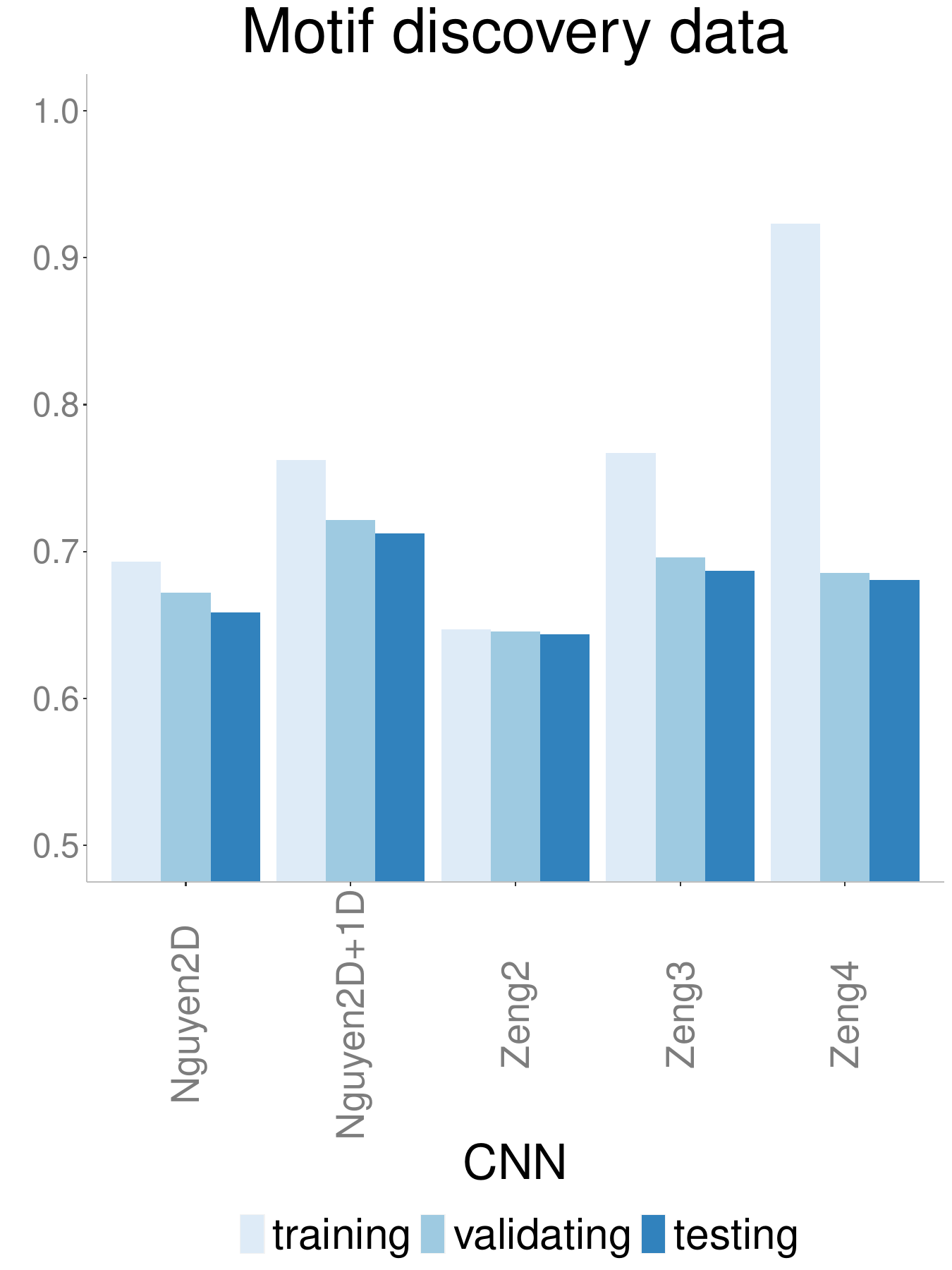}
    \caption{Accuracy of CNN models of increasing number of layers on three datasets of increasing size. Nguyen2D corresponds to the original CNN model in \citep{Nguyen2016-jj}, while Nguyen2D+1D corresponds to the same model with an extra 1D convolutional layer. Similarly, Zeng2 corresponds to the original model in \citep{Zeng2016-nu} which has two 2D convolutional layers while Zeng3 and Zeng4 correspond to models with three and four 2D convolutional layers respectively. There is an inverse relationship between accuracy and data size with the largest dataset (motif discovery) having the lowest accuracy overall. Adding one layer with different dimension (CNN-Nguyen2D+1D) improves the accuracy slightly, but more layers of the same dimension (CNN-Zeng3 with 3 layers and CNN-Zeng4 with 4 layers) only increase the overfitting.}
    \label{fig:dim-layers-acc}
\end{figure}

Figure \ref{fig:dim-layers-acc} shows the training, validating and testing accuracy of the CNN models when varying the number of layers for the three datasets. Nguyen2D corresponds to the original CNN model in \citep{Nguyen2016-jj}, while Nguyen2D+1D corresponds to the same model with an extra 1D convolutional layer. Similarly, Zeng2 corresponds to the original model in \citep{Zeng2016-nu} which has two 2D convolutional layers while Zeng3 and Zeng4 correspond to models with three and four 2D convolutional layers respectively.

For the smallest dataset (splice), all models have a testing accuracy higher than 80\% which is similar to what is reported in the original CNN-Nguyen paper \citep{Nguyen2016-jj} (88.9\%) except for the original model in CNN-Zeng \citep{Zeng2016-nu}. Adding more layers improves the performance of the CNN-Zeng model, but not the CNN-Nguyen model. There is not any strong evidence of overfitting in any of the models in the splice data. 

For the medium size dataset (histone), all models have a similar testing accuracy (slightly below 80\%). In this dataset, adding more layers merely increases the training accuracy and thus, the overfitting. Lastly, for the largest dataset (motif discovery), all testing accuracies are below 70\%. There is a slight improvement in the CNN-Nguyen model when adding one more 1D layer, yet for the case of CNN-Zeng, more layers only increase the training accuracy and thus, the overfitting. We compare the performance of the CNN-Zeng model with and without regularization in the Appendix.

To sum up, accuracy decreases with data size with the largest data having the lowest reported accuracy. In addition, adding more layers to a CNN model does increase accuracy for smaller datasets, but it appears to only increase overfitting on larger datasets. This assertion is counterintuitive as overfitting is thought to be the result of parameter-rich models on small size data. In our analyses, overfitting indeed appears as a result of more complex models (more layers) yet only on larger datasets. \revision{It is important to note that this atypical performance could be due to the distinct data partition chosen for the motif discovery data (48-3-48 in contrast to a standard 70-15-15). For this dataset, we prioritized a equal contribution to the training samples from each of the 690 input files in order to prevent unintended bias in predictions caused by heterogeneity in the sequences. This choice is not meant to be perfect and can create another set of complications (e.g., the unexpected decreased accuracy). Future work should address implications in prediction due to data partition choices when faced with highly heterogeneous datasets.}

\begin{figure}[h]
    \centering
    \includegraphics[scale=0.3]{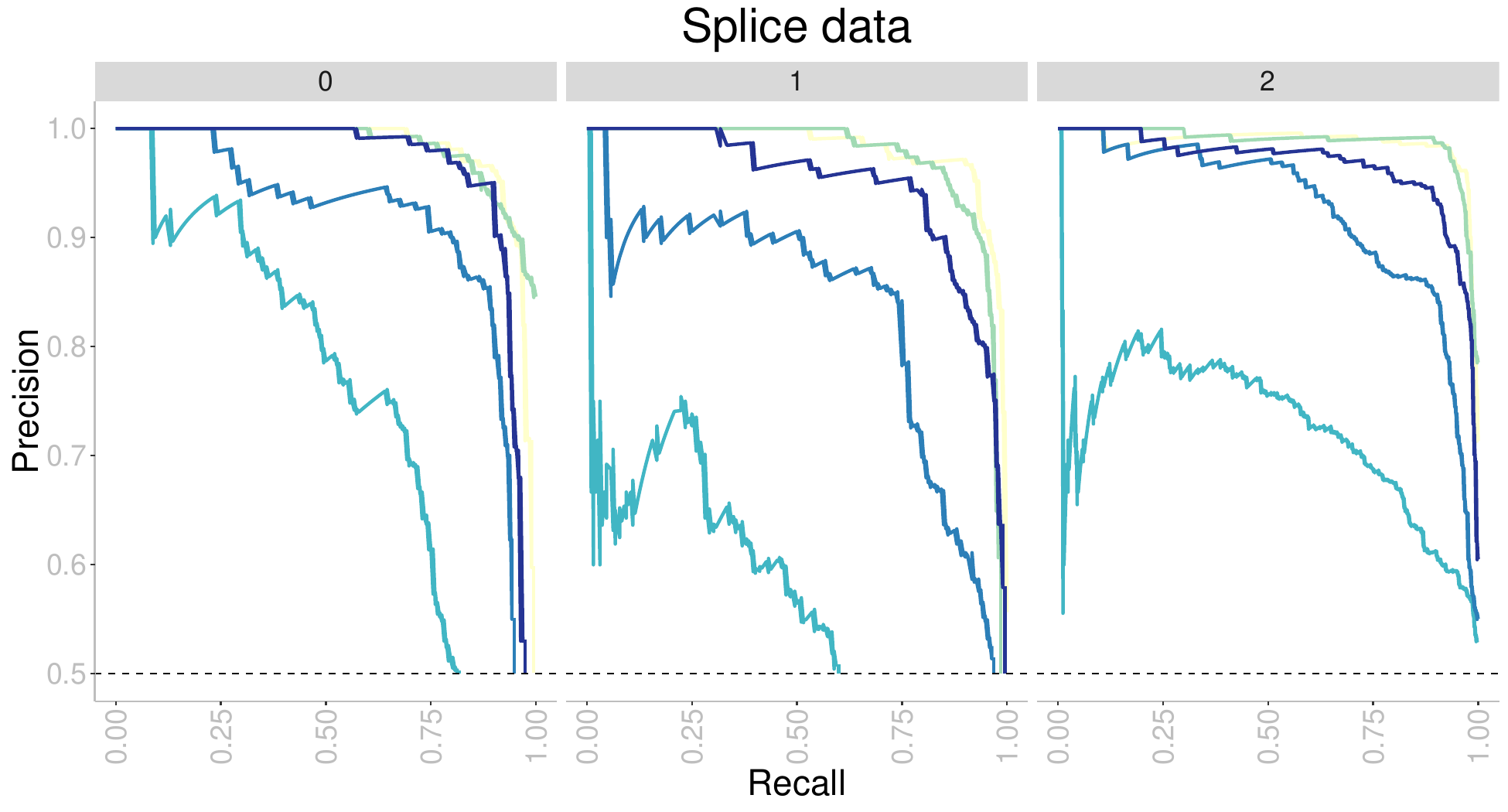}\\[0.25cm]
    \includegraphics[scale=0.25]{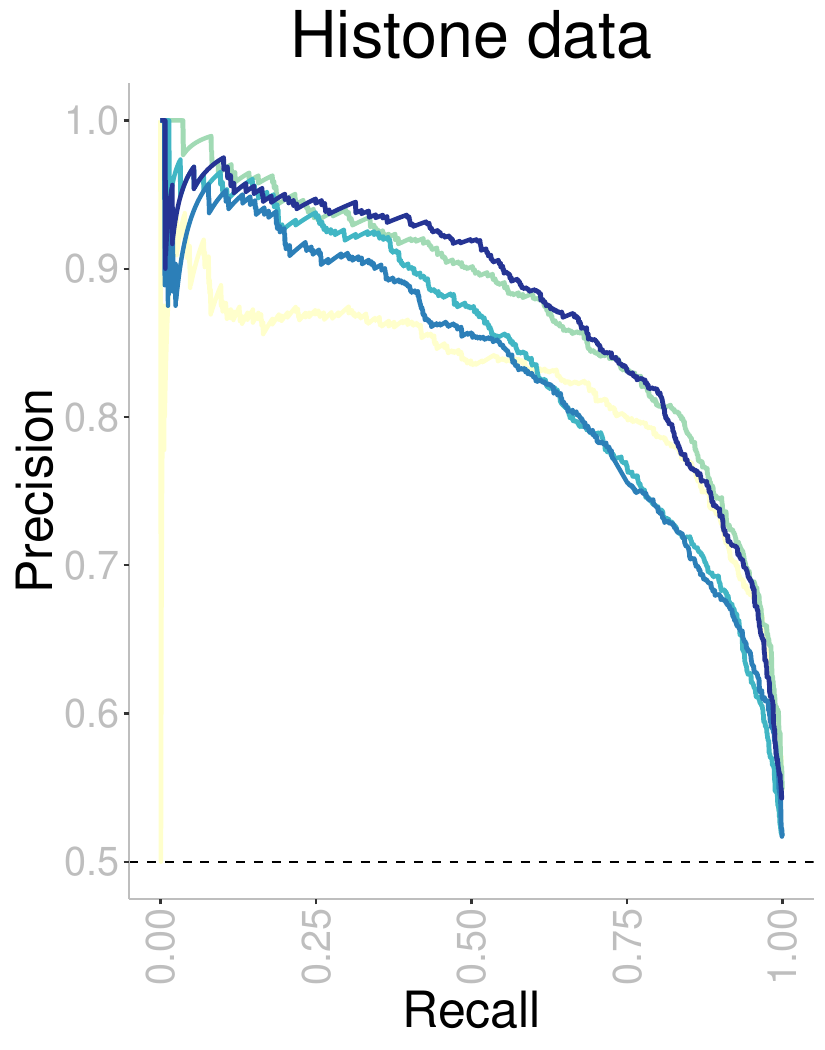}
    \includegraphics[scale=0.25]{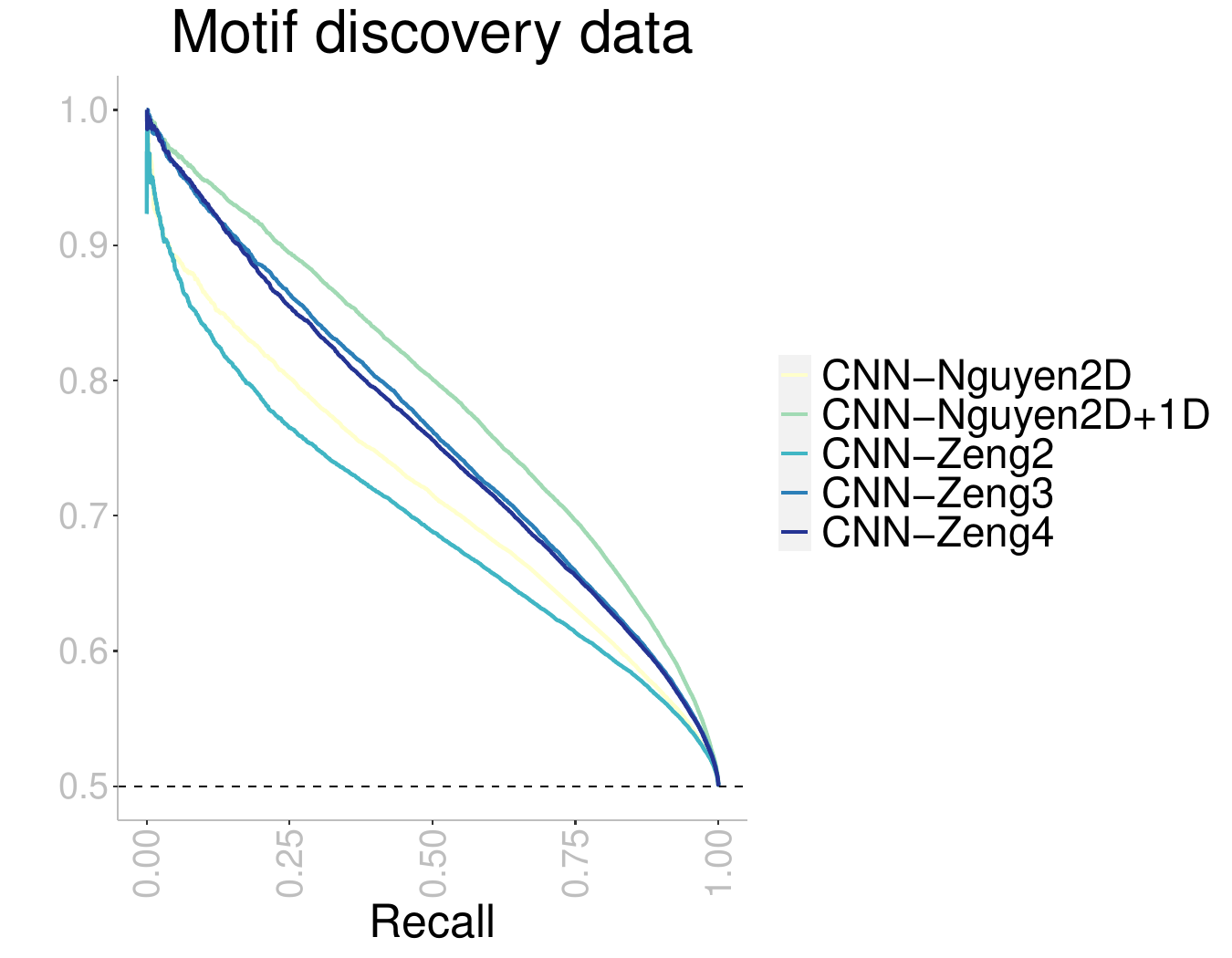}
    \caption{Precision-recall curves of CNN models of increasing number of layers on the three datasets of increasing size. The higher the curve, the better performance with a horizontal dashed line to represent random prediction. An ideal precision-recall curve would cross the (1,1) point. The splice data has three panels since precision-recall curves assume binary classification and the splice dataset has three classes (0, 1, 2). Each panel corresponds to prediction one class vs the other two combined. Nguyen2D corresponds to the original CNN model in \citep{Nguyen2016-jj}, while Nguyen2D+1D corresponds to the same model with an extra 1D convolutional layer. Similarly, Zeng2 corresponds to the original model in \citep{Zeng2016-nu} which has two 2D convolutional layers while Zeng3 and Zeng4 correspond to models with three and four 2D convolutional layers respectively. The CNN-Nguyen models outperform the other models across datasets.}
    \label{fig:dim-layers-pr}
\end{figure}

\begin{figure}[h]
    \centering
    \includegraphics[scale=0.3]{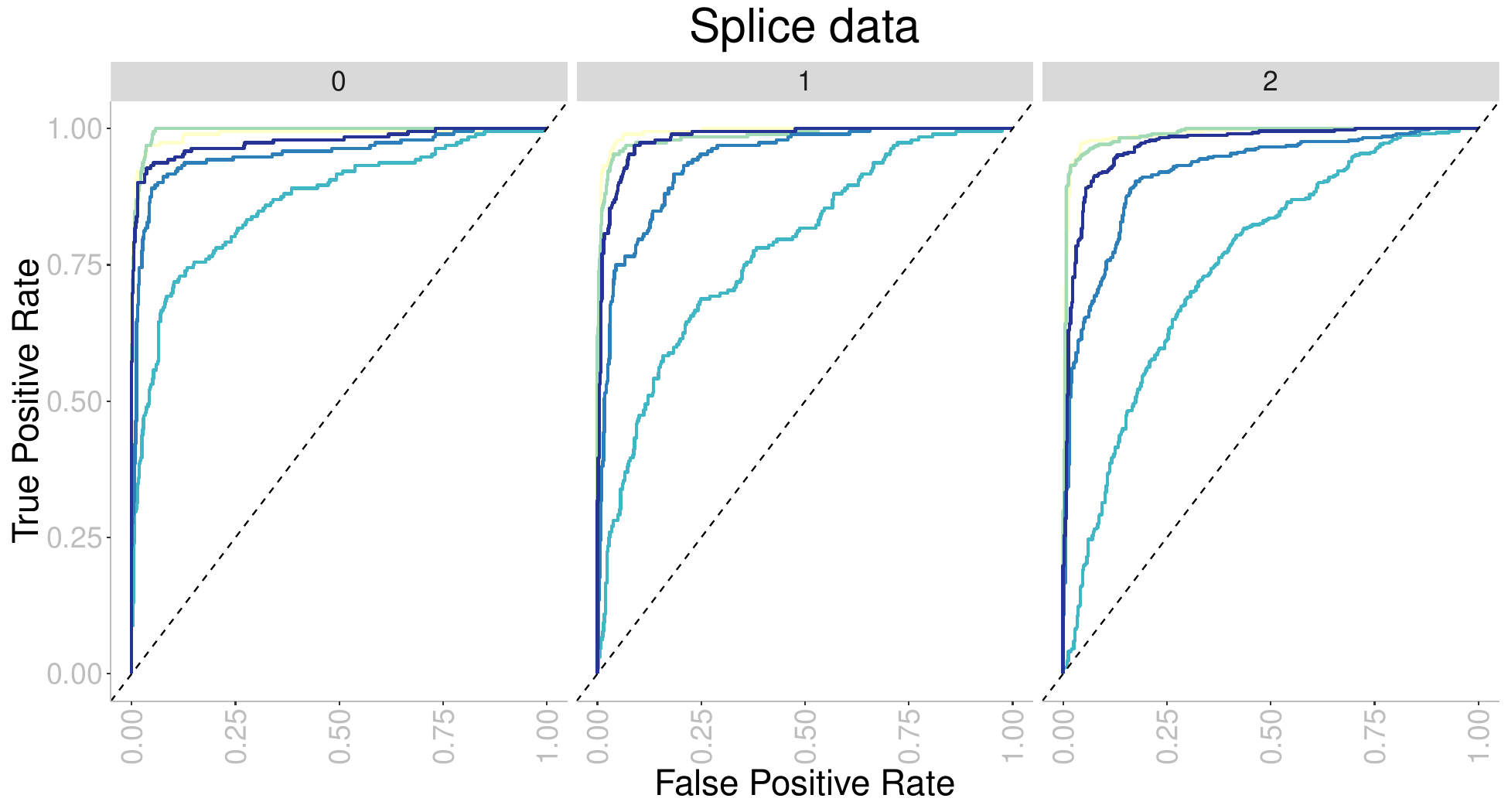}\\[0.25cm]
    \includegraphics[scale=0.25]{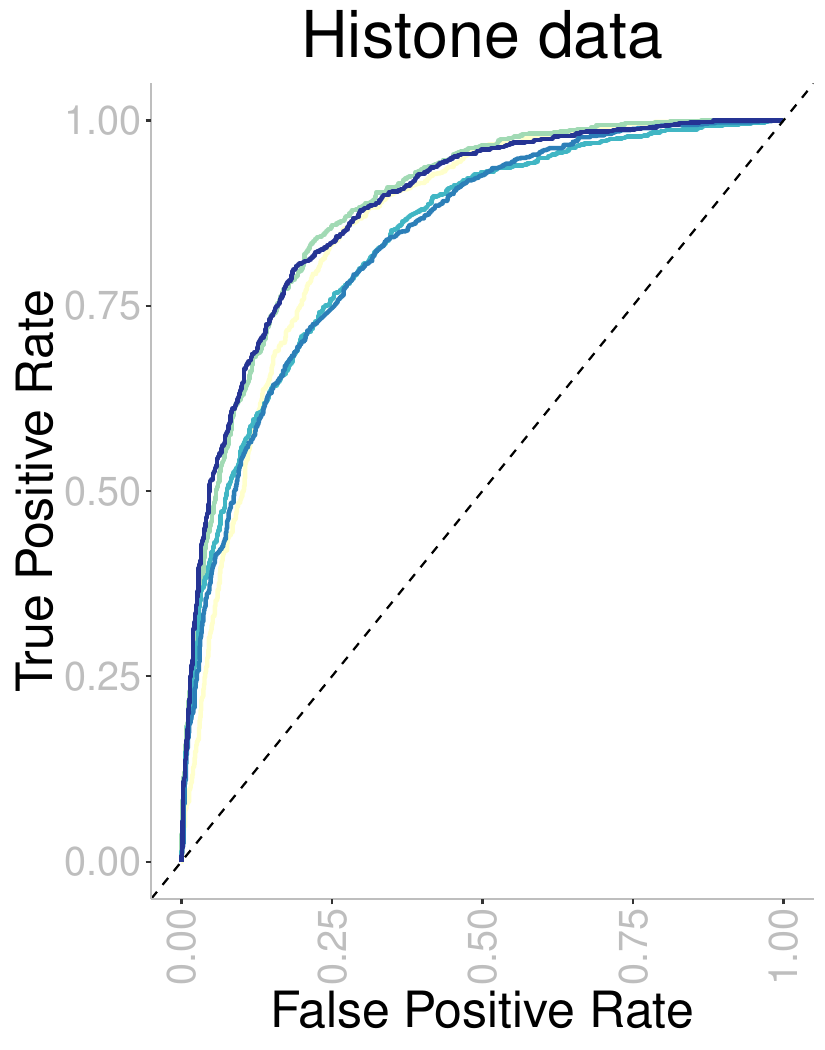}
    \includegraphics[scale=0.25]{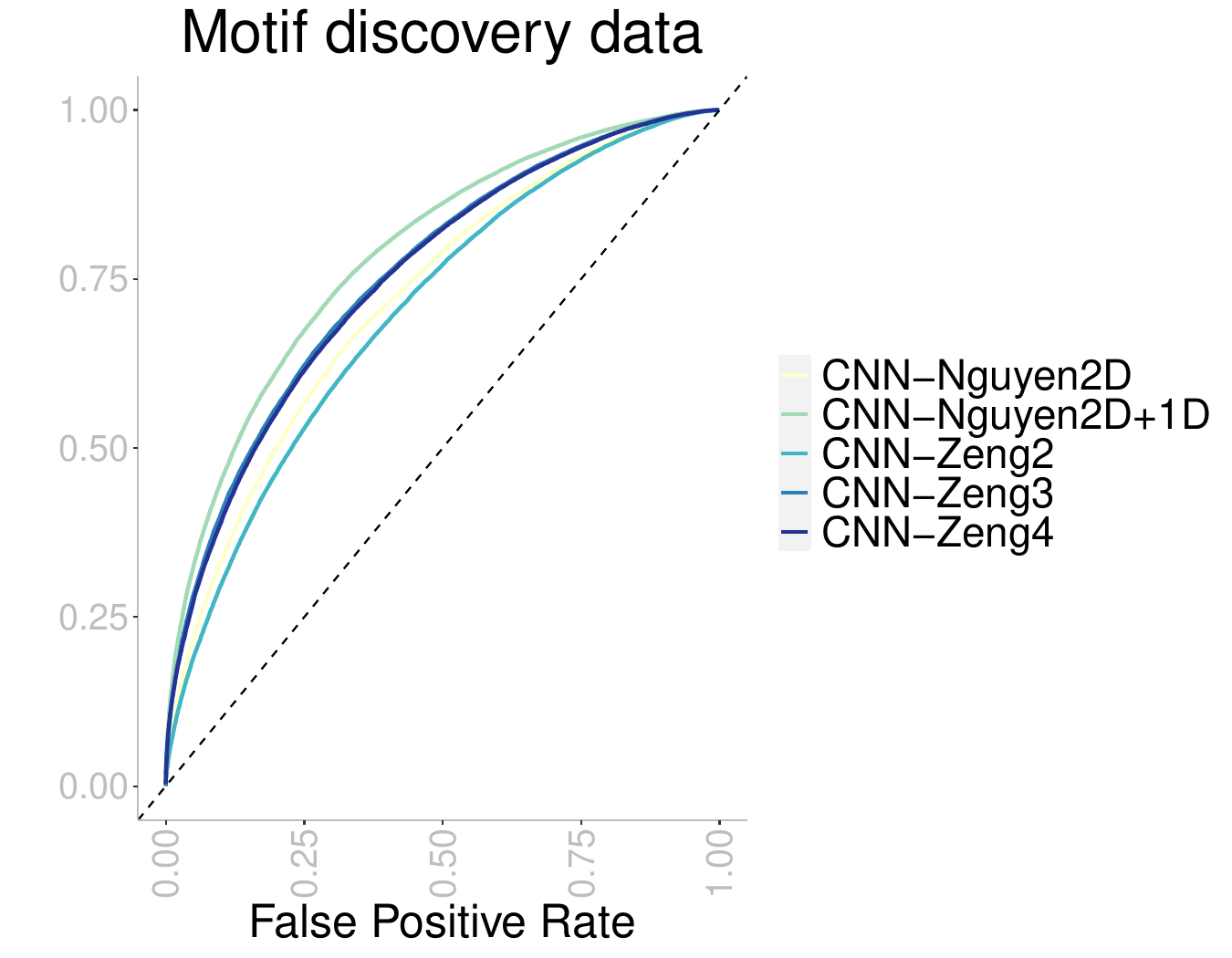}
    \caption{ROC curves of CNN models of increasing number of layers on the three datasets of increasing size. The higher the curve, the better performance with a 45$^{\circ}$ dashed line to represent random prediction. The splice data has three panels since ROC curves assume binary classification and the splice dataset has three classes (0, 1, 2). Each panel corresponds to prediction one class vs the other two combined. Nguyen2D corresponds to the original CNN model in \citep{Nguyen2016-jj}, while Nguyen2D+1D corresponds to the same model with an extra 1D convolutional layer. Similarly, Zeng2 corresponds to the original model in \citep{Zeng2016-nu} which has two 2D convolutional layers while Zeng3 and Zeng4 correspond to models with three and four 2D convolutional layers respectively. The CNN-Nguyen models outperform the other models across datasets.}
    \label{fig:dim-layers-roc}
\end{figure}

Finally, we investigate the precision-recall curves of the models in Figure \ref{fig:dim-layers-pr}. The CNN-Nguyen models outperform those in CNN-Zeng across datasets with the original CNN-Zeng2 displaying the worse performance. This behavior is confirmed with the ROC curves in Figure \ref{fig:dim-layers-roc}.

\subsection*{The role of data encoding}

\begin{figure}[h]
    \centering
    \includegraphics[scale=0.175]{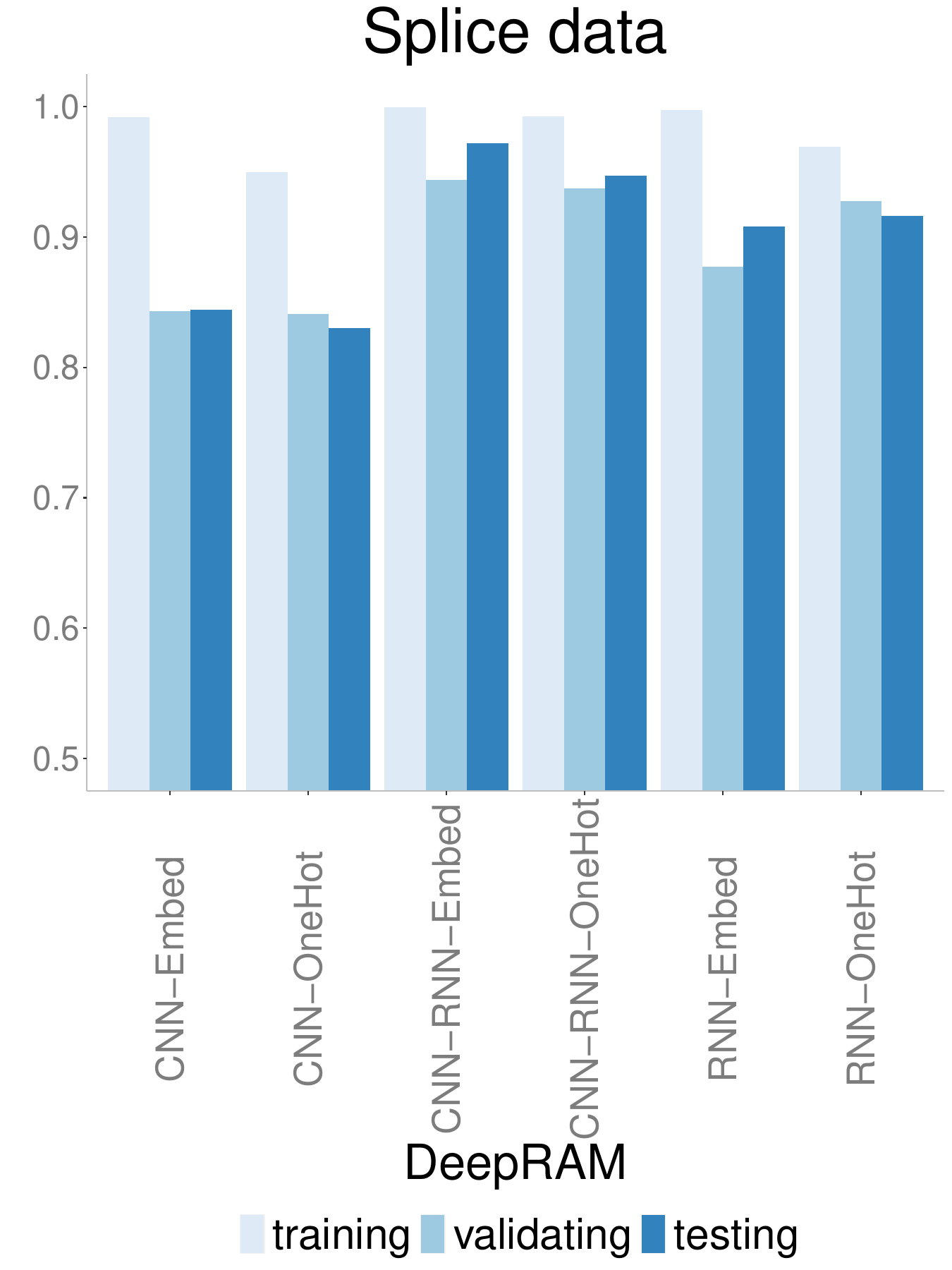}
    \includegraphics[scale=0.175]{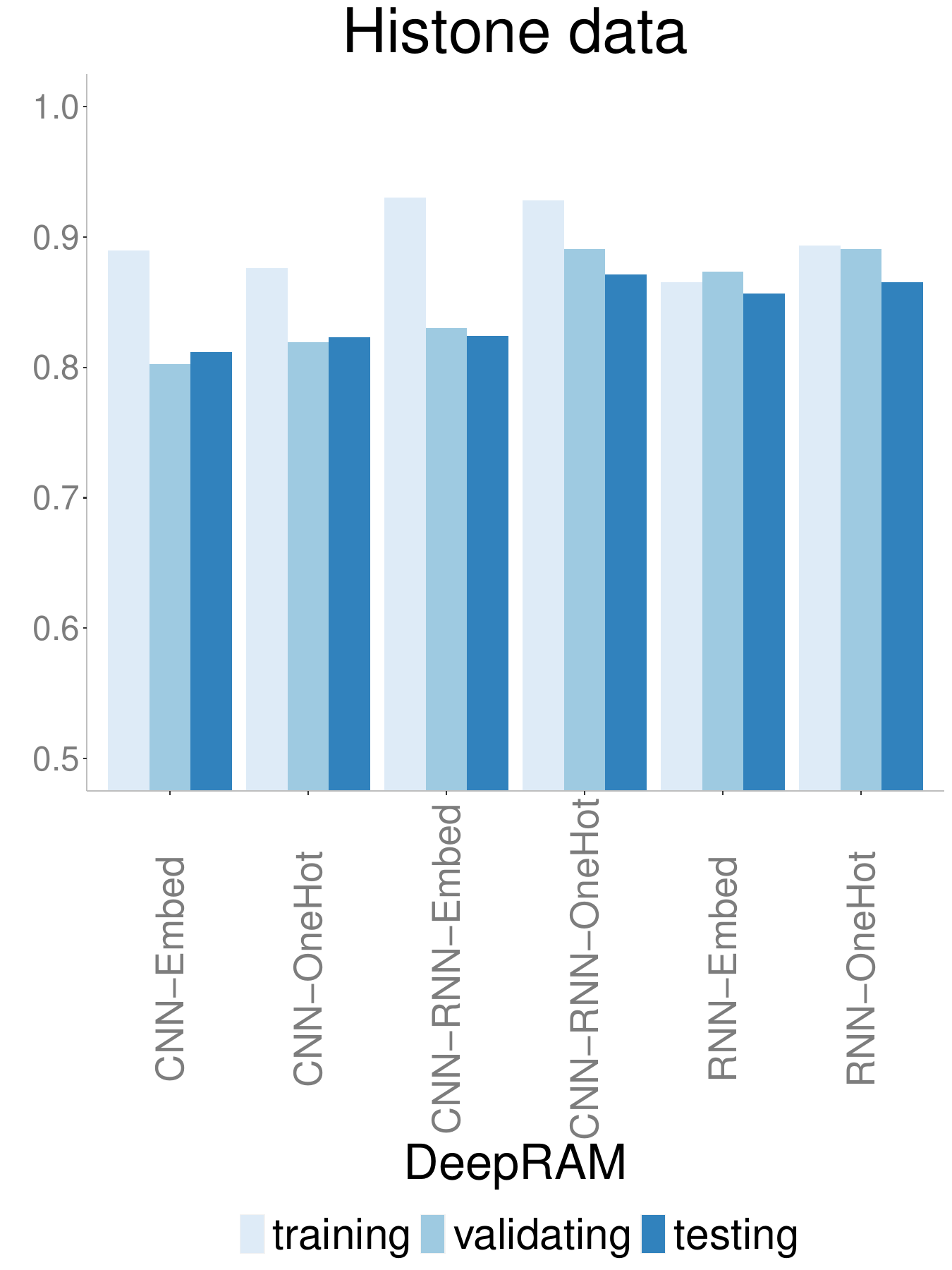}
    \includegraphics[scale=0.175]{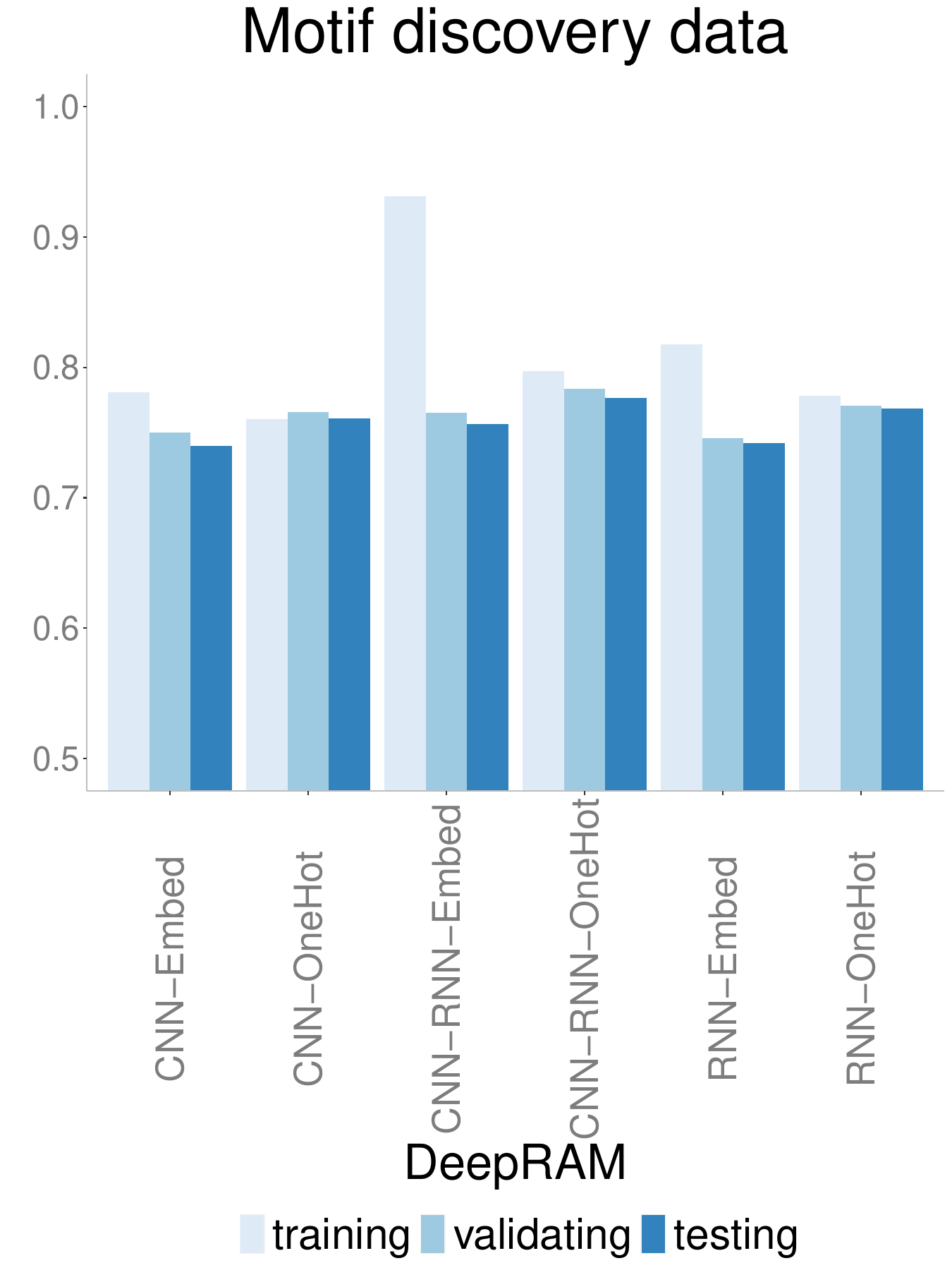}
    \caption{Accuracy of DeepRAM models \citep{Trabelsi2019-yt} with two different data encoding schemes: one-hot encoding (OneHot) and embedding layer (Embed) on three datasets of increasing size. CNN corresponds to the convolutional model, RNN corresponds to the recurrent model and CNN-RNN corresponds to the combined model. Accuracy decreases with data size, and all models display a similar behavior on the different data encoding schemes.}
    \label{fig:embedding-acc}
\end{figure}

\begin{figure}[h]
    \centering
    \includegraphics[scale=0.3]{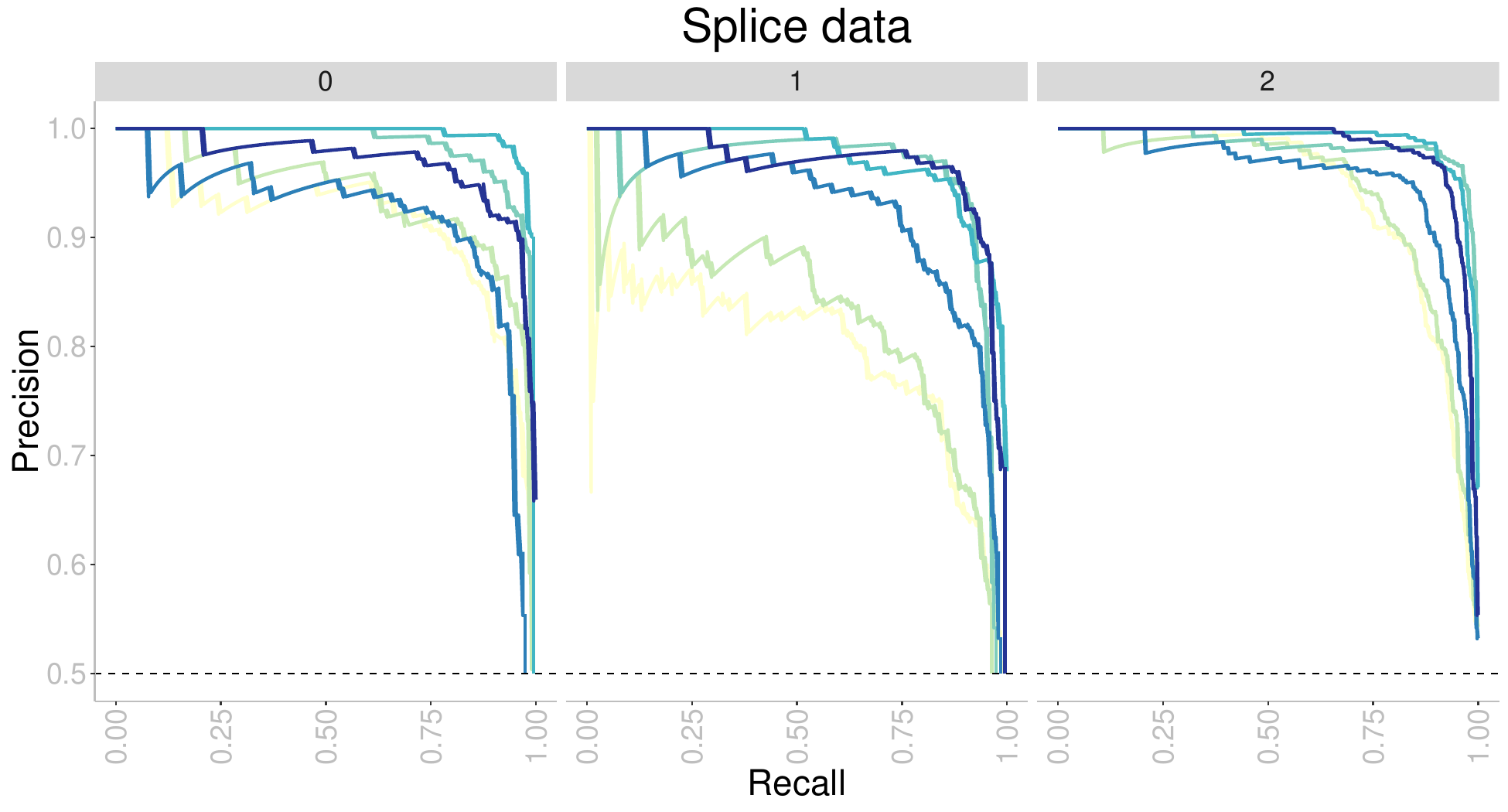}\\[0.25cm]
    \includegraphics[scale=0.25]{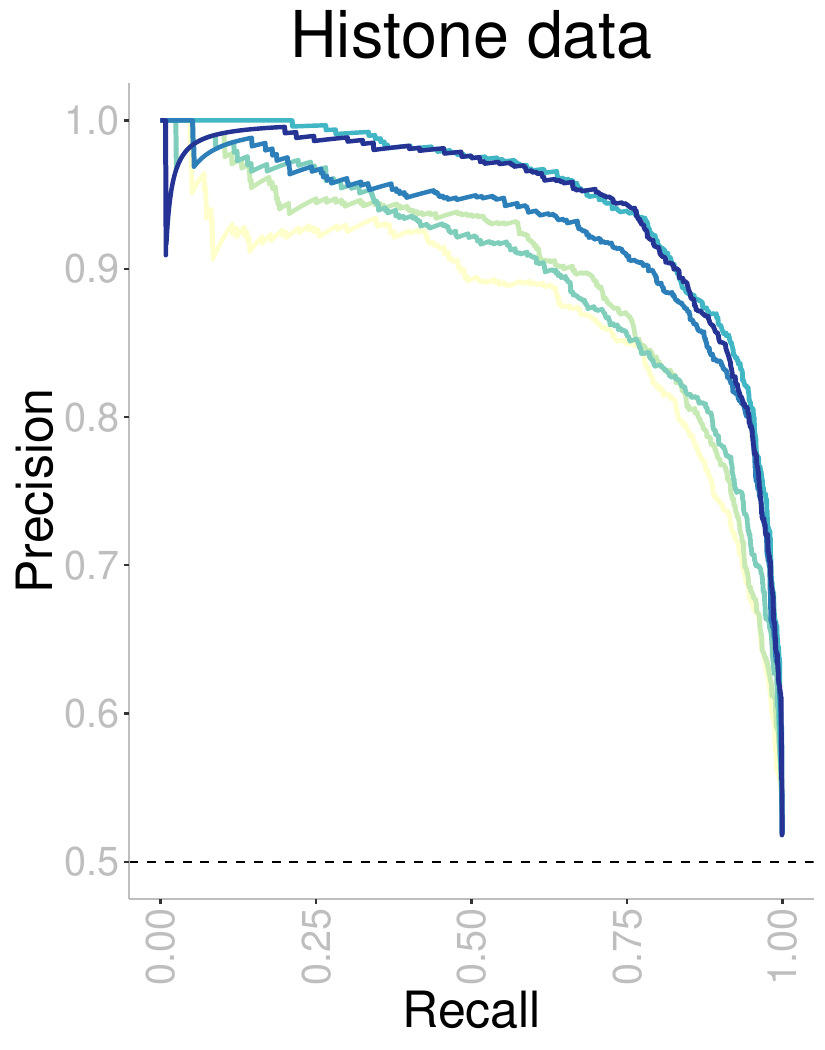}
    \includegraphics[scale=0.25]{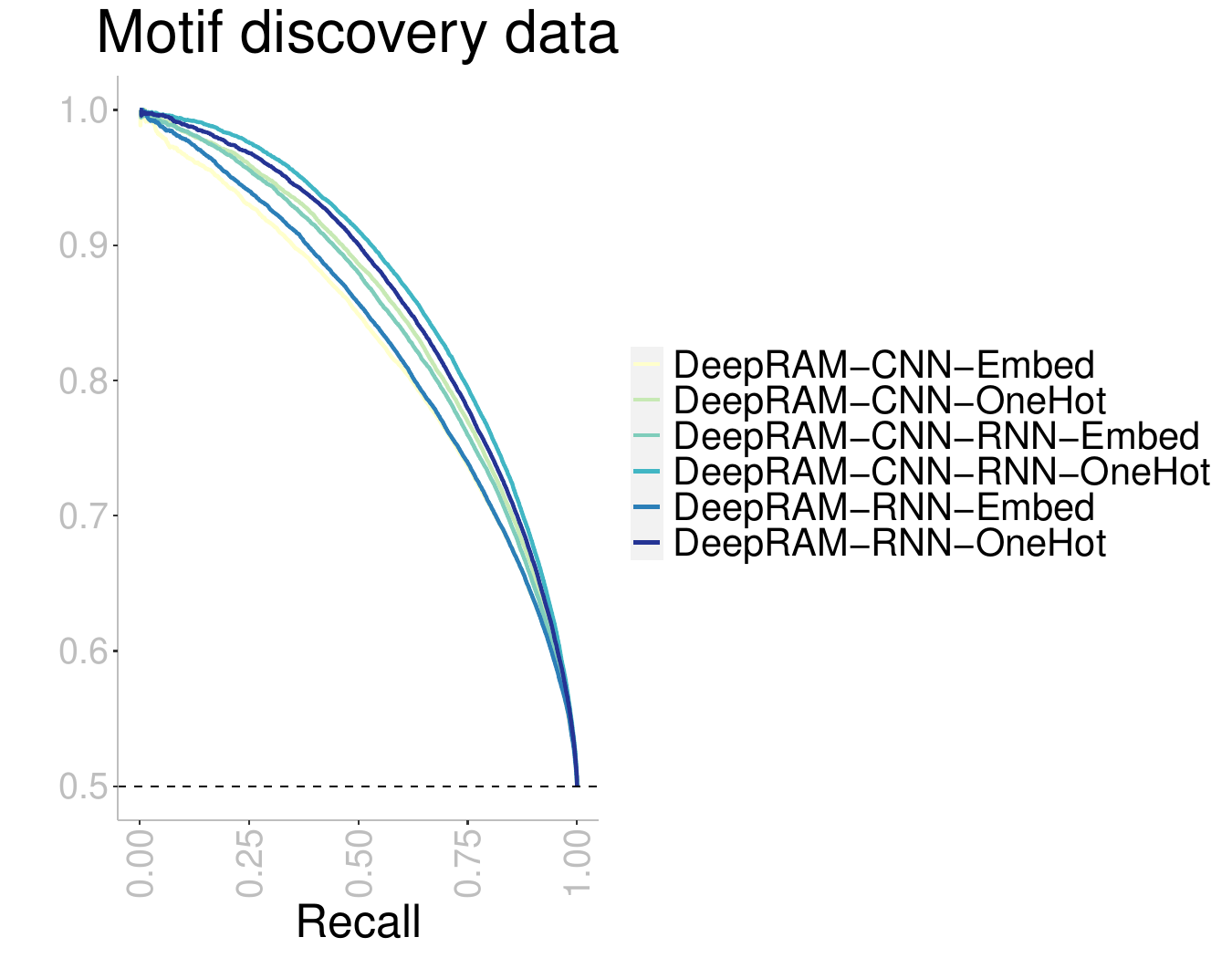}
    \caption{Precision-recall curves of DeepRAM models \citep{Trabelsi2019-yt} with two different data encoding schemes: one-hot encoding (OneHot) and embedding layer (Embed) on three datasets of increasing size. The higher the curve, the better performance with a horizontal dashed line to represent random prediction. An ideal precision-recall curve would cross the (1,1) point. The splice data has three panels since precision-recall curves assume binary classification and the splice dataset has three classes (0, 1, 2). Each panel corresponds to prediction one class vs the other two combined. CNN corresponds to the convolutional model, RNN corresponds to the recurrent model and CNN-RNN corresponds to the combined model. Accuracy decreases with data size, and all models display a similar behavior on the different data encoding schemes.}
    \label{fig:embedding-pr}
\end{figure}

\begin{figure}[h]
    \centering
    \includegraphics[scale=0.3]{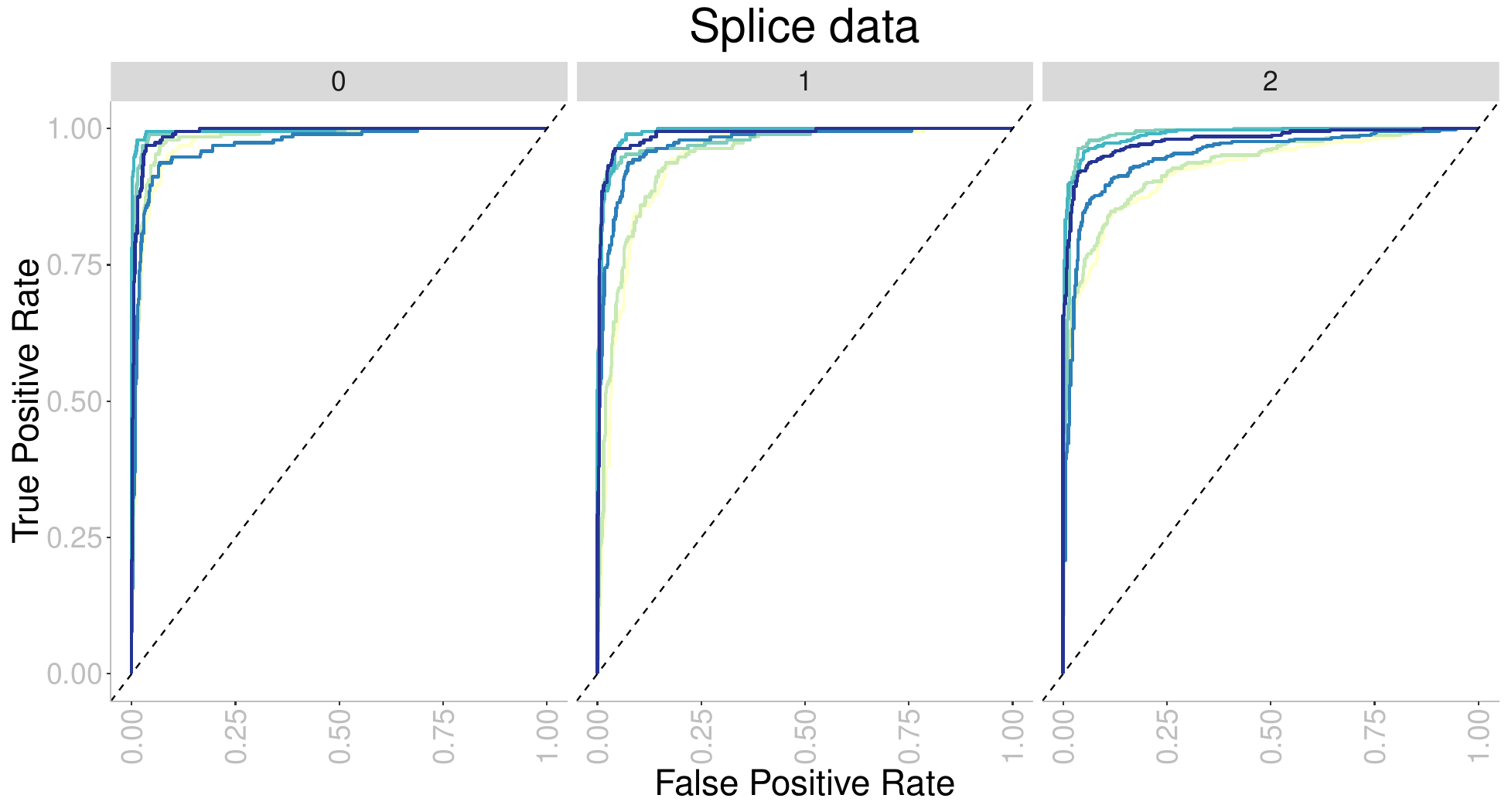}\\[0.25cm]
    \includegraphics[scale=0.25]{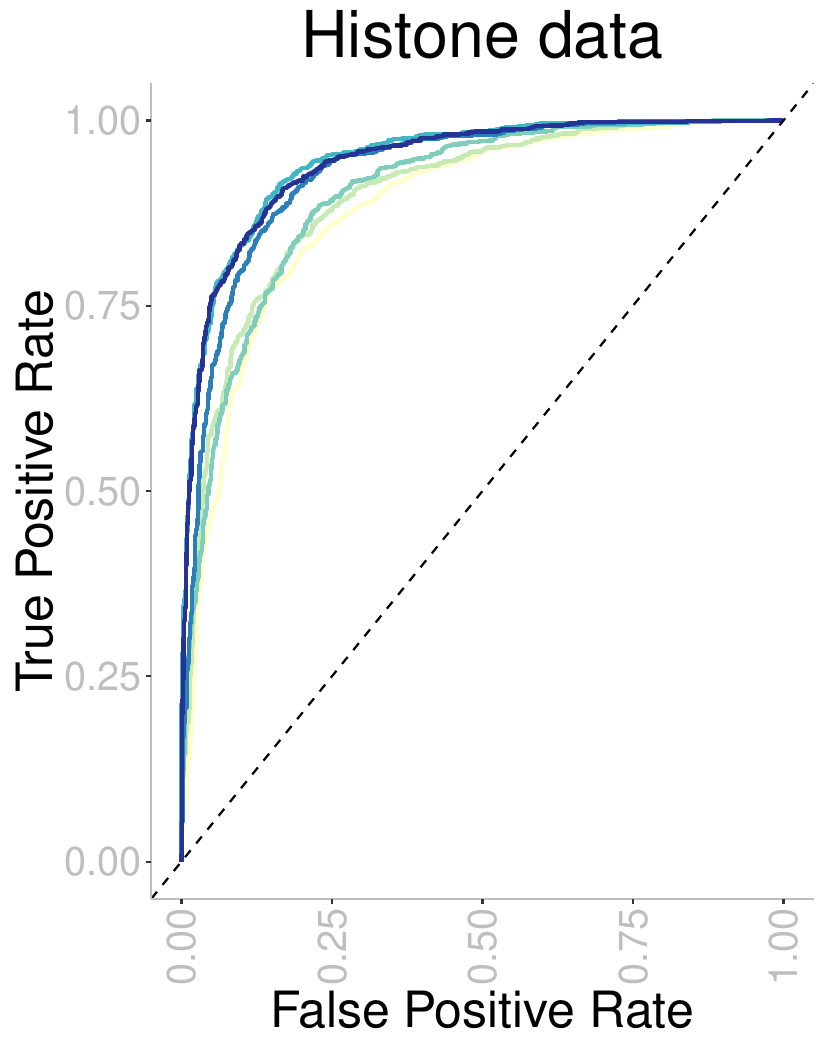}
    \includegraphics[scale=0.25]{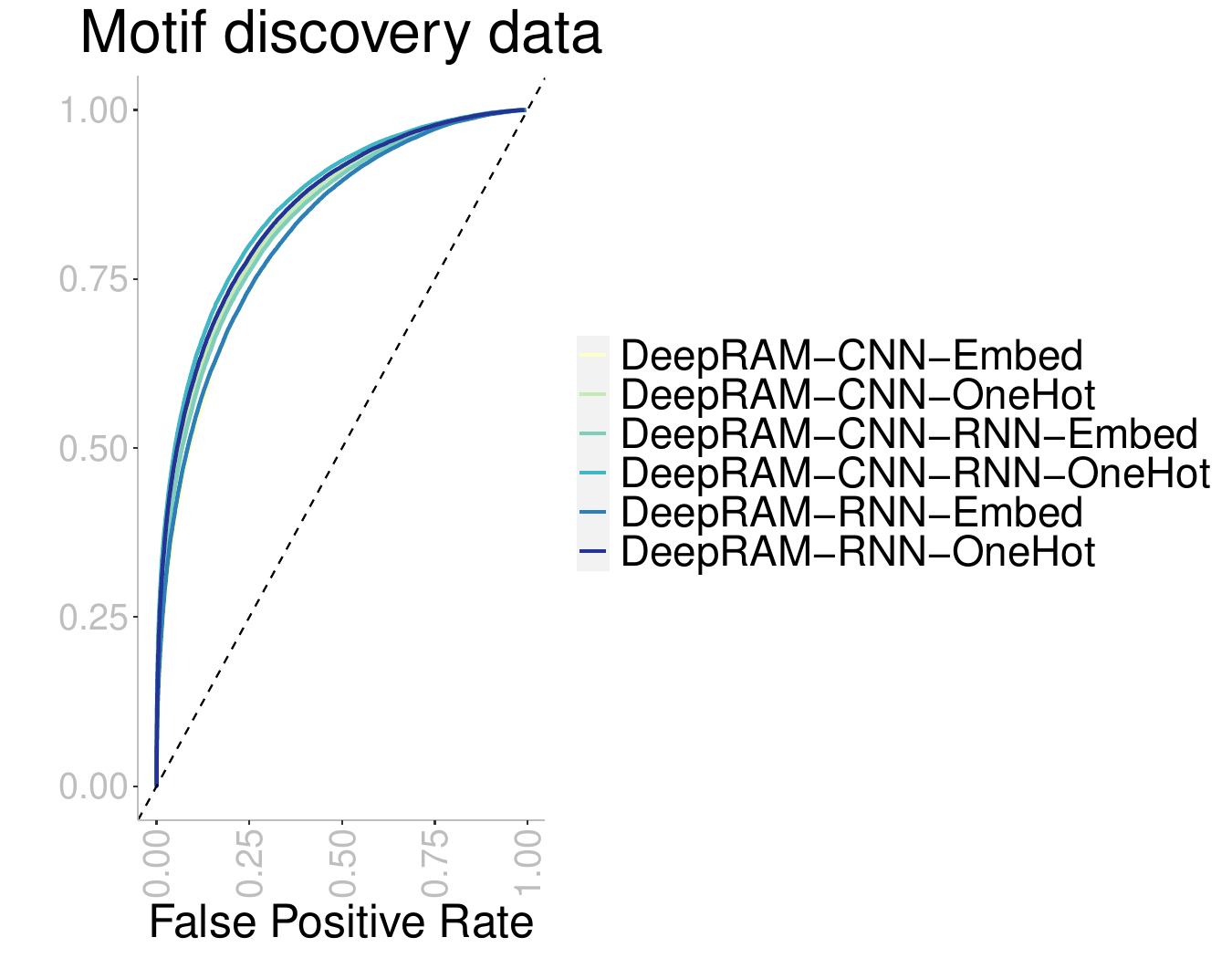}
    \caption{ROC curves of DeepRAM models \citep{Trabelsi2019-yt} with two different data encoding schemes: one-hot encoding (OneHot) and embedding layer (Embed) on three datasets of increasing size. The higher the curve, the better performance with a 45$^{\circ}$ dashed line to represent random prediction. The splice data has three panels since ROC curves assume binary classification and the splice dataset has three classes (0, 1, 2). Each panel corresponds to prediction one class vs the other two combined. CNN corresponds to the convolutional model, RNN corresponds to the recurrent model and CNN-RNN corresponds to the combined model. Accuracy decreases with data size, and all models display a similar behavior on the different data encoding schemes.}
    \label{fig:embedding-roc}
\end{figure}

It appears that the type of data encoding (one-hot encoding vs embedding layer) does not have a strong influence on the performance of the DeepRAM models in \citep{Trabelsi2019-yt}. Figure \ref{fig:embedding-acc} shows the accuracy of the models which is lowest overall for the largest dataset (motif discovery) yet there is not a clear difference across models or data encoding types. 
The combined model (CNN-RNN) seems to slightly outperform the other models and this behavior is also apparent in the precision-recall curves (Figure \ref{fig:embedding-pr}) and in the ROC curves (Figure \ref{fig:embedding-roc}). However, care must be taken in that the
combined model with embedding layer (CNN-RNN-Embed) seems to overfit in the motif discovery data while the one-hot encoding version of the same model does not show overfitting, so it appears that one-hot encoding should be preferred.

Importantly, the behavior of DeepRAM seems to translate well across datasets. Accuracy lies between 80\% and 90\% for the smallest dataset (splice) and around 75\% for the largest dataset (motif discovery). As a point of comparison, the accuracy presented in original DeepRAM paper \citep{Trabelsi2019-yt} ranges from 83.6\% to 99.4\% on data from 83 ChIP-seq experiments in the ENCODE project.


\subsection*{The role of the optimizer}

\begin{figure}[h]
    \centering
    \includegraphics[scale=0.175]{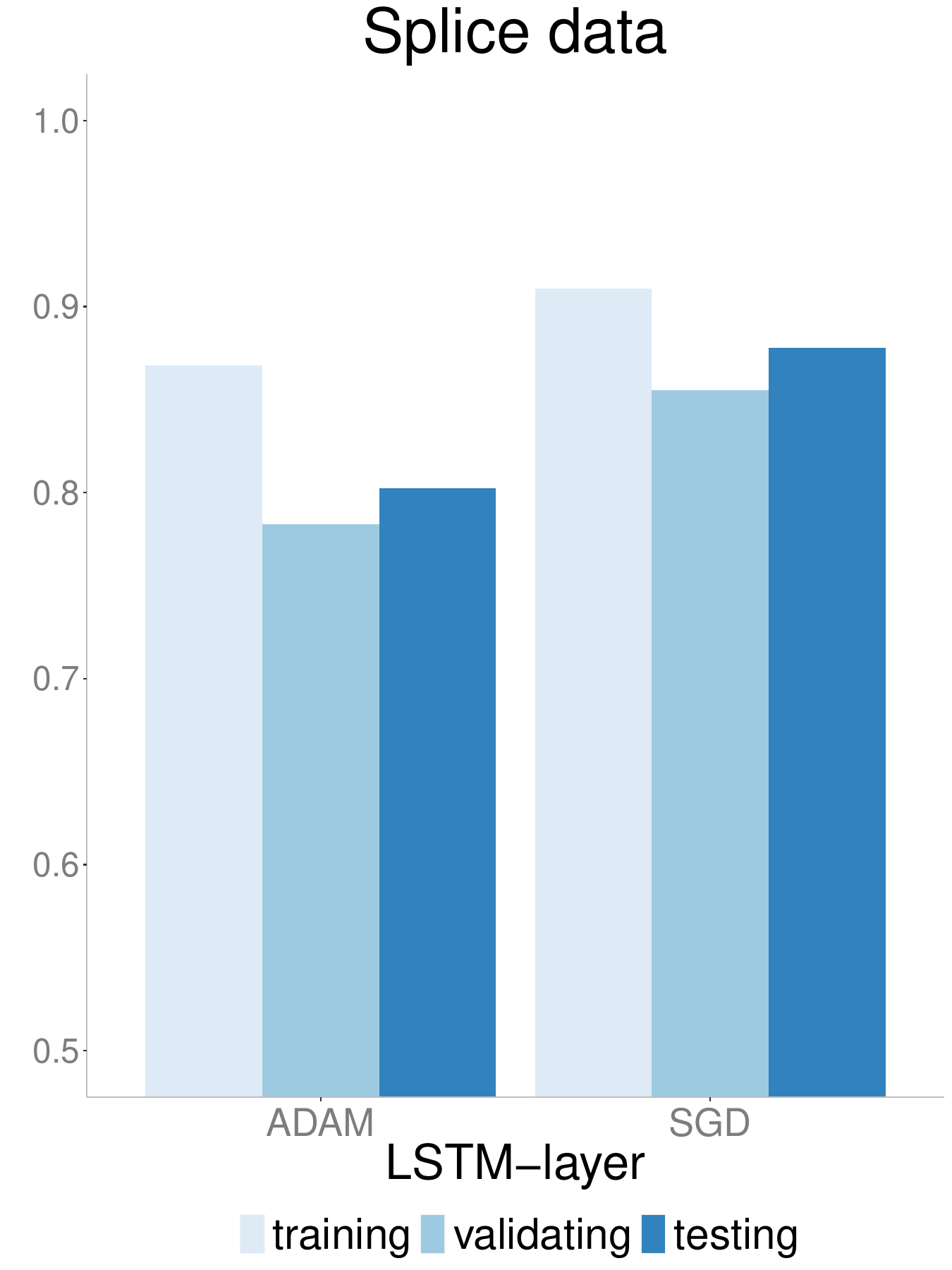}
    \includegraphics[scale=0.175]{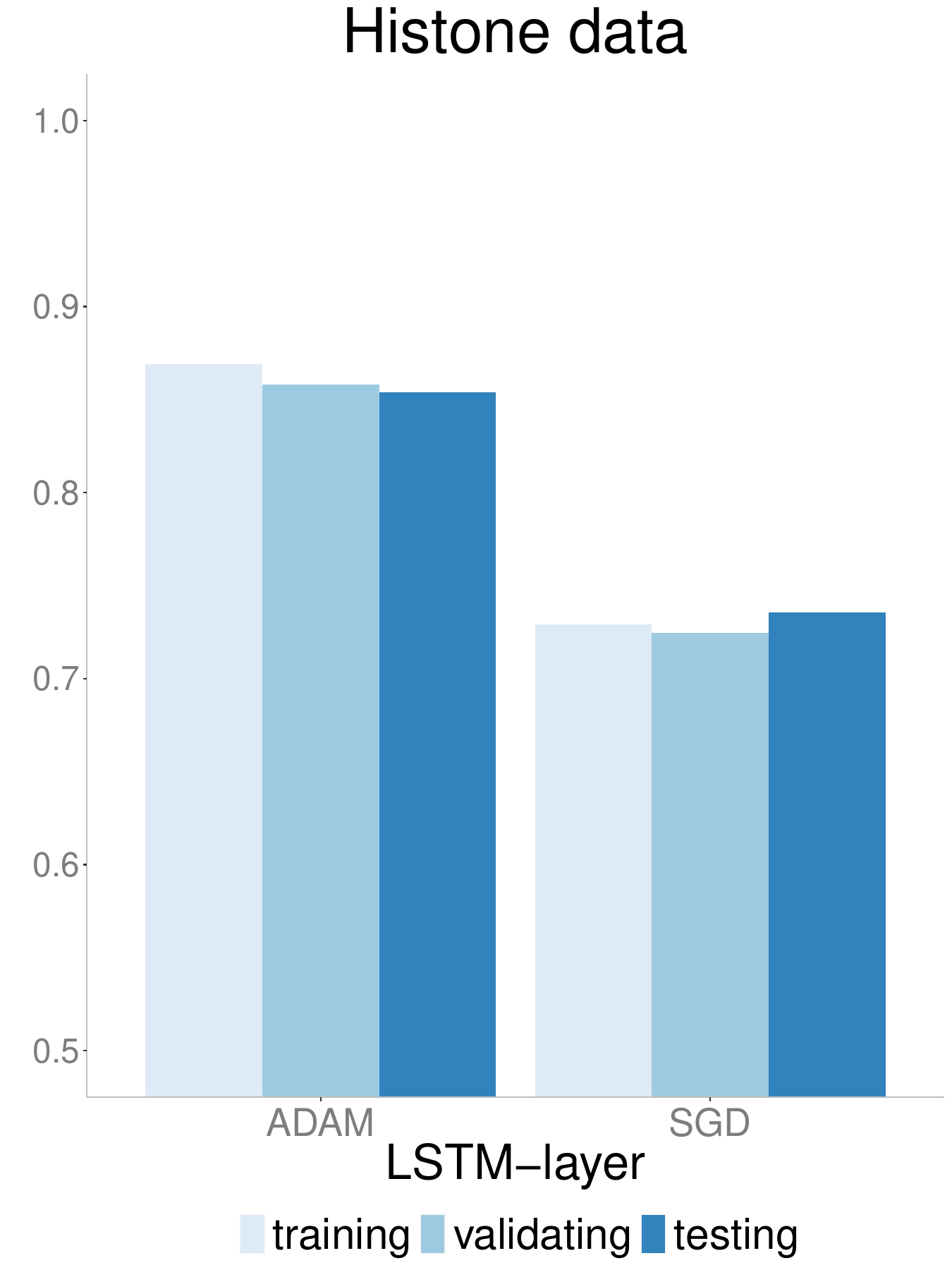}
    \includegraphics[scale=0.175]{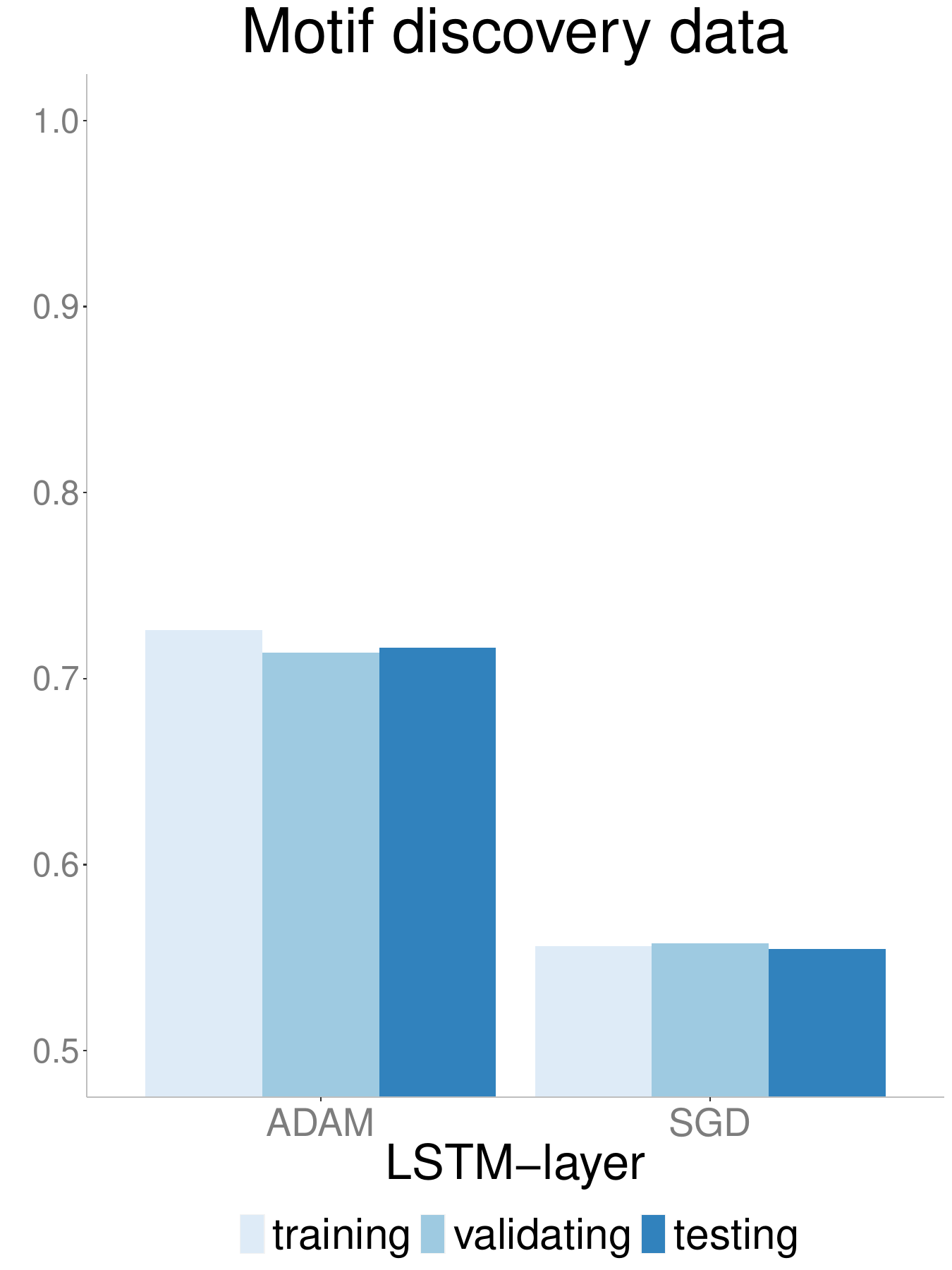}
    \caption{Accuracy of LSTM-layer model with two different optimizers: ADAM and SGD on three datasets of increasing size. There is no evidence of overfitting with this model on any of the datasets. In addition, the best optimizer varies with SGD outperforming ADAM for the smallest dataset (splice) and ADAM outperforming SGD on the other two datasets. It is widely accepted that SGD performs better in terms of finding global optima. However, due to its low speed, it can get stuck in one plateau too long.}
    \label{fig:optimizer-acc}
\end{figure}

\begin{figure}[h]
    \centering
    \includegraphics[scale=0.3]{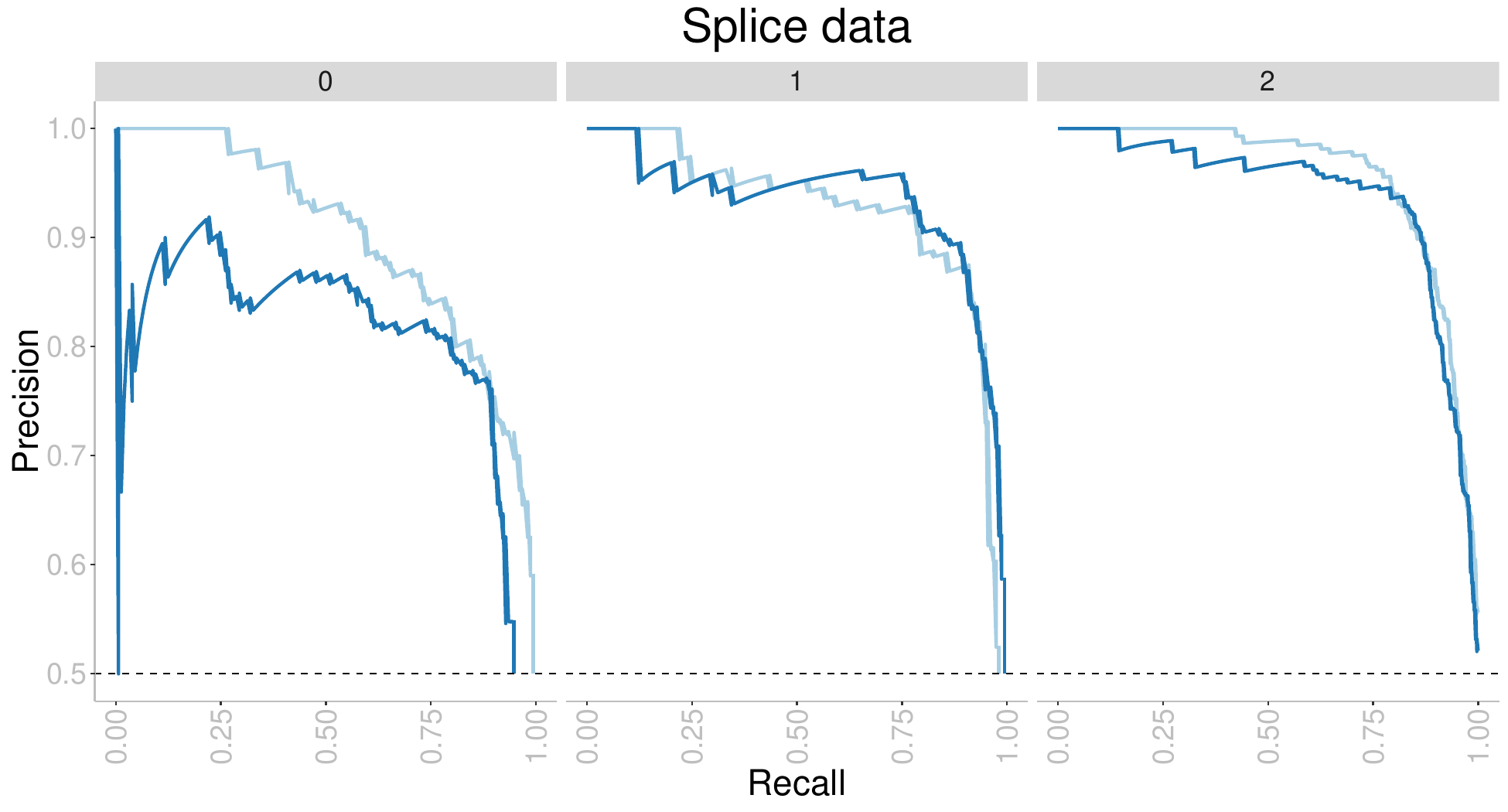}\\[0.25cm]
    \includegraphics[scale=0.25]{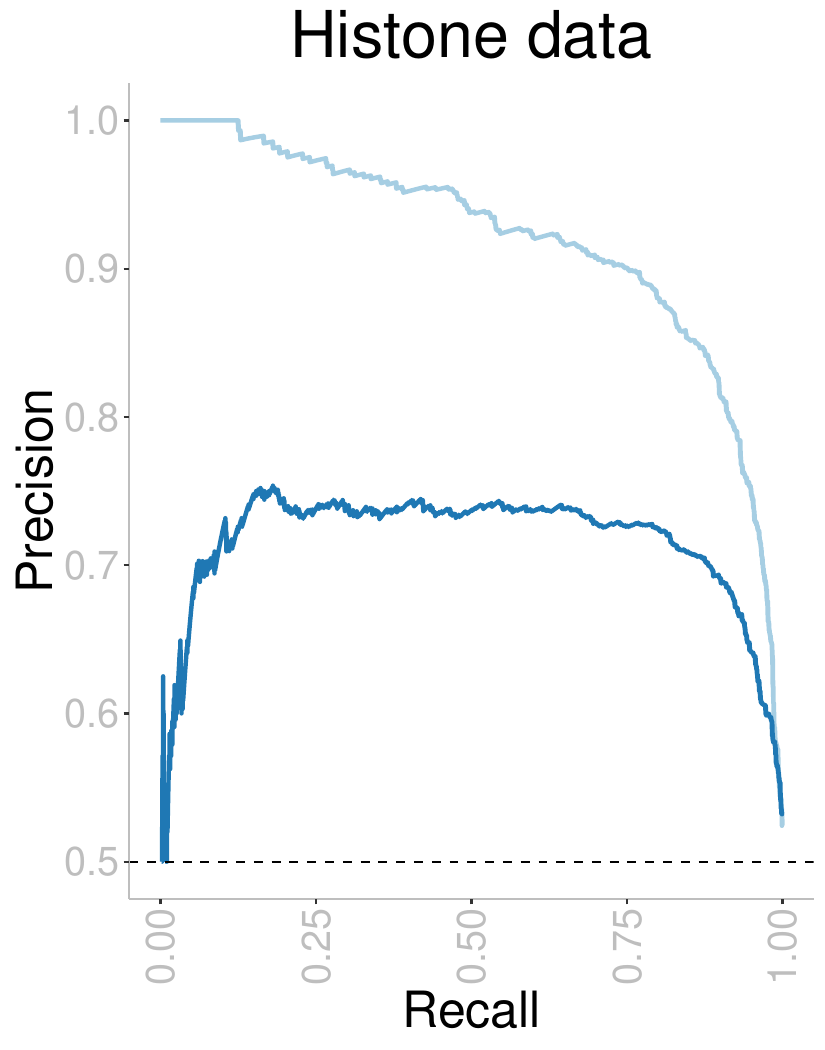}
    \includegraphics[scale=0.25]{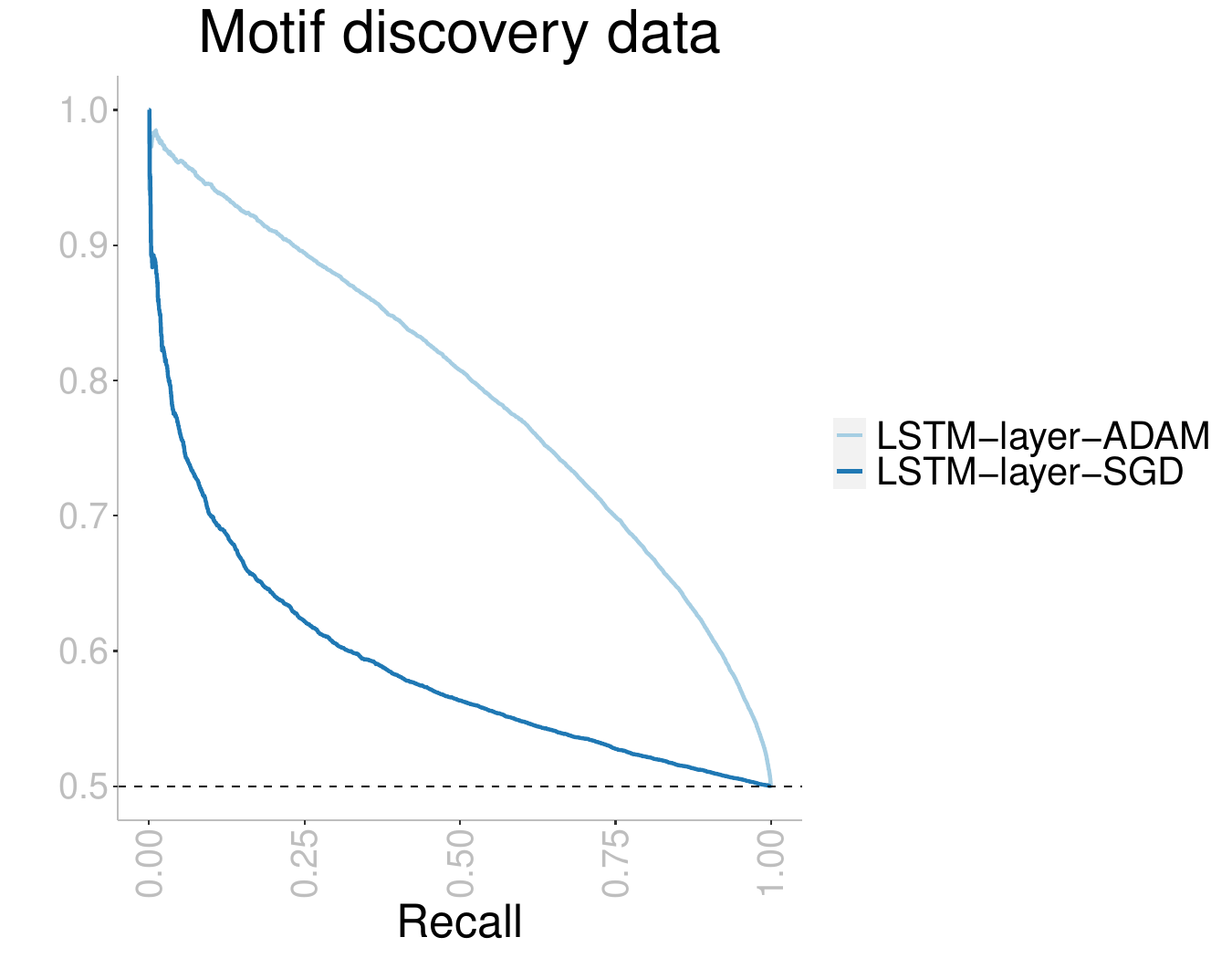}
    \caption{Precision-recall curves of LSTM-layer model with two different optimizers: ADAM and SGD on three datasets of increasing size. The higher the curve, the better performance with a horizontal dashed line to represent random prediction. An ideal precision-recall curve would cross the (1,1) point. The splice data has three panels since precision-recall curves assume binary classification and the splice dataset has three classes (0, 1, 2). Each panel corresponds to prediction one class vs the other two combined. The best optimizer varies with SGD outperforming ADAM for the smallest dataset (splice) and ADAM outperforming SGD on the other two datasets. It is widely accepted that SGD performs better in terms of finding global optima. However, due to its low speed, it can get stuck in one plateau too long.}
    \label{fig:optimizer-pr}
\end{figure}

\begin{figure}[h]
    \centering
    \includegraphics[scale=0.3]{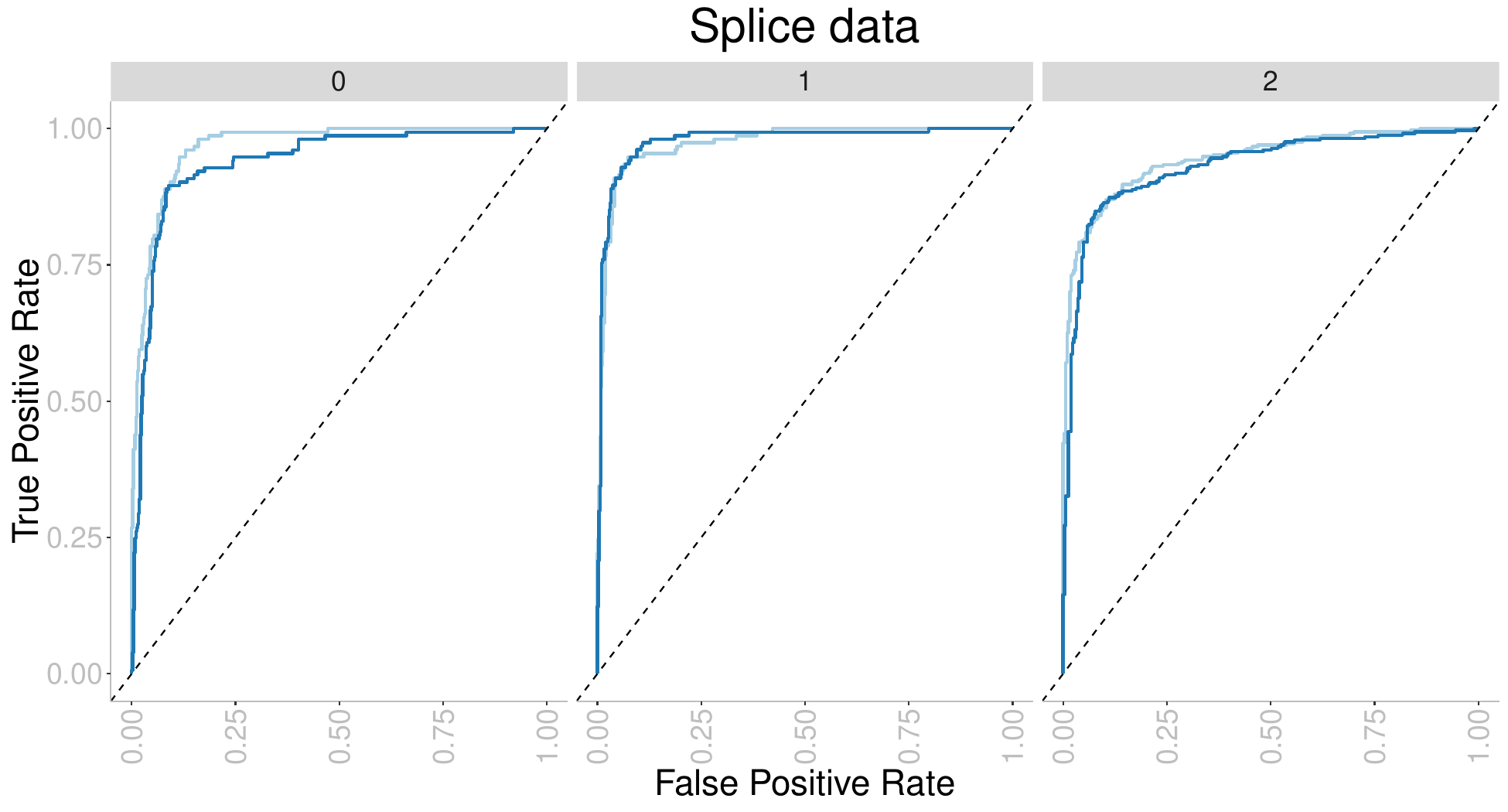}\\[0.25cm]
    \includegraphics[scale=0.25]{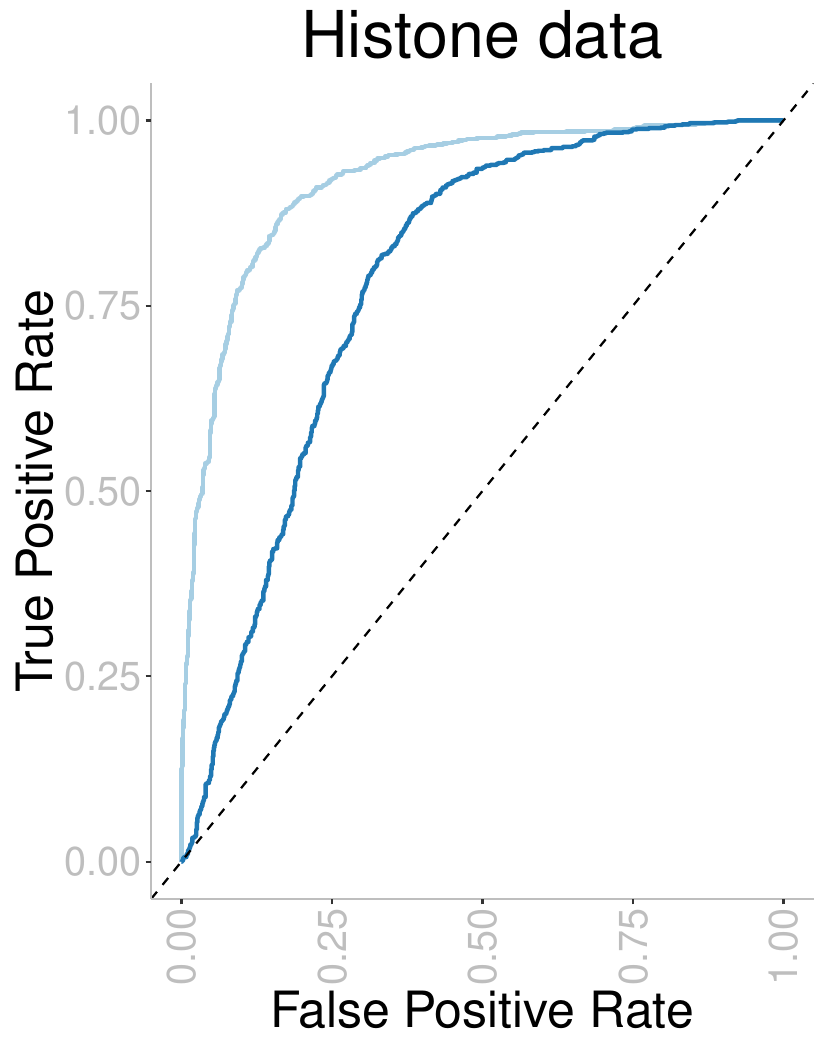}
    \includegraphics[scale=0.25]{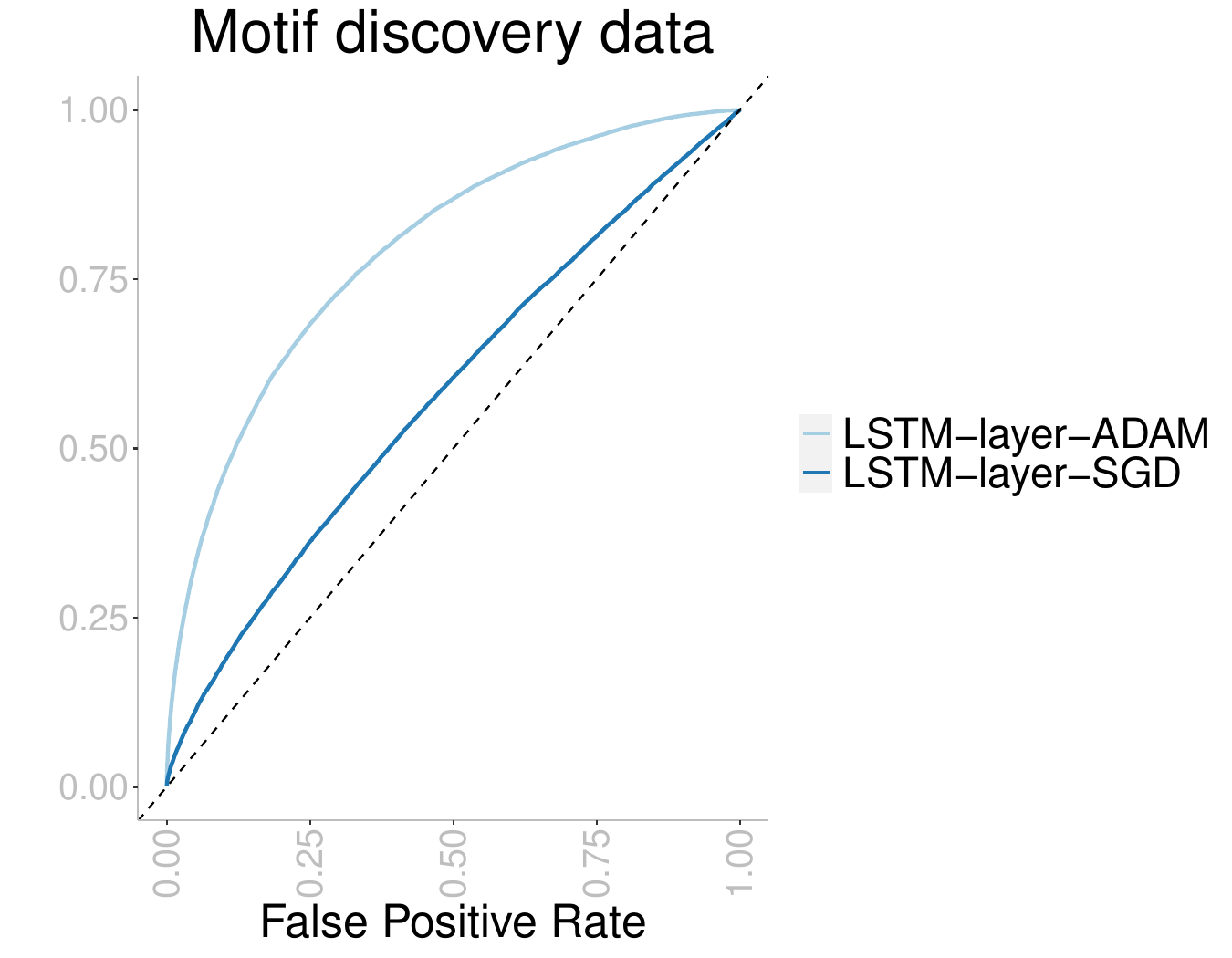}
    \caption{ROC curves of LSTM-layer model with two different optimizers: ADAM and SGD on three datasets of increasing size. The higher the curve, the better performance with a 45$^{\circ}$ dashed line to represent random prediction. The splice data has three panels since ROC curves assume binary classification and the splice dataset has three classes (0, 1, 2). Each panel corresponds to prediction one class vs the other two combined. The best optimizer varies with SGD outperforming ADAM for the smallest dataset (splice) and ADAM outperforming SGD on the other two datasets. It is widely accepted that SGD performs better in terms of finding global optima. However, due to its low speed, it can get stuck in one plateau too long.}
    \label{fig:optimizer-roc}
\end{figure}

The SGD optimizer outperforms the Adam optimizer on the LSTM-layer model for the smallest dataset (splice) while Adam outperforms SGD for the two larger datasets (histone and motif discovery). See Figure \ref{fig:optimizer-acc} for accuracy, Figure \ref{fig:optimizer-pr} for precision-recall curves and Figure \ref{fig:optimizer-roc} for ROC curves.
While \citep{reddi2019convergence} has already discussed the convergence issues of the Adam optimizer, we also need to note that the difference in performance can be due to differences with the Early Stopping Callback patience parameter. It is widely accepted that SGD performs better in terms of finding global optima. However, due to its low speed, it can get stuck in one plateau too long.
We note that the comparison of optimizer behavior has ignited multiple studies. For a more comprehensive investigation on the role of optimizers in neural network models, see \citep{nado2021large}.

\subsection*{The role of the embedding size}

\begin{figure}[h]
    \centering
    \includegraphics[scale=0.175]{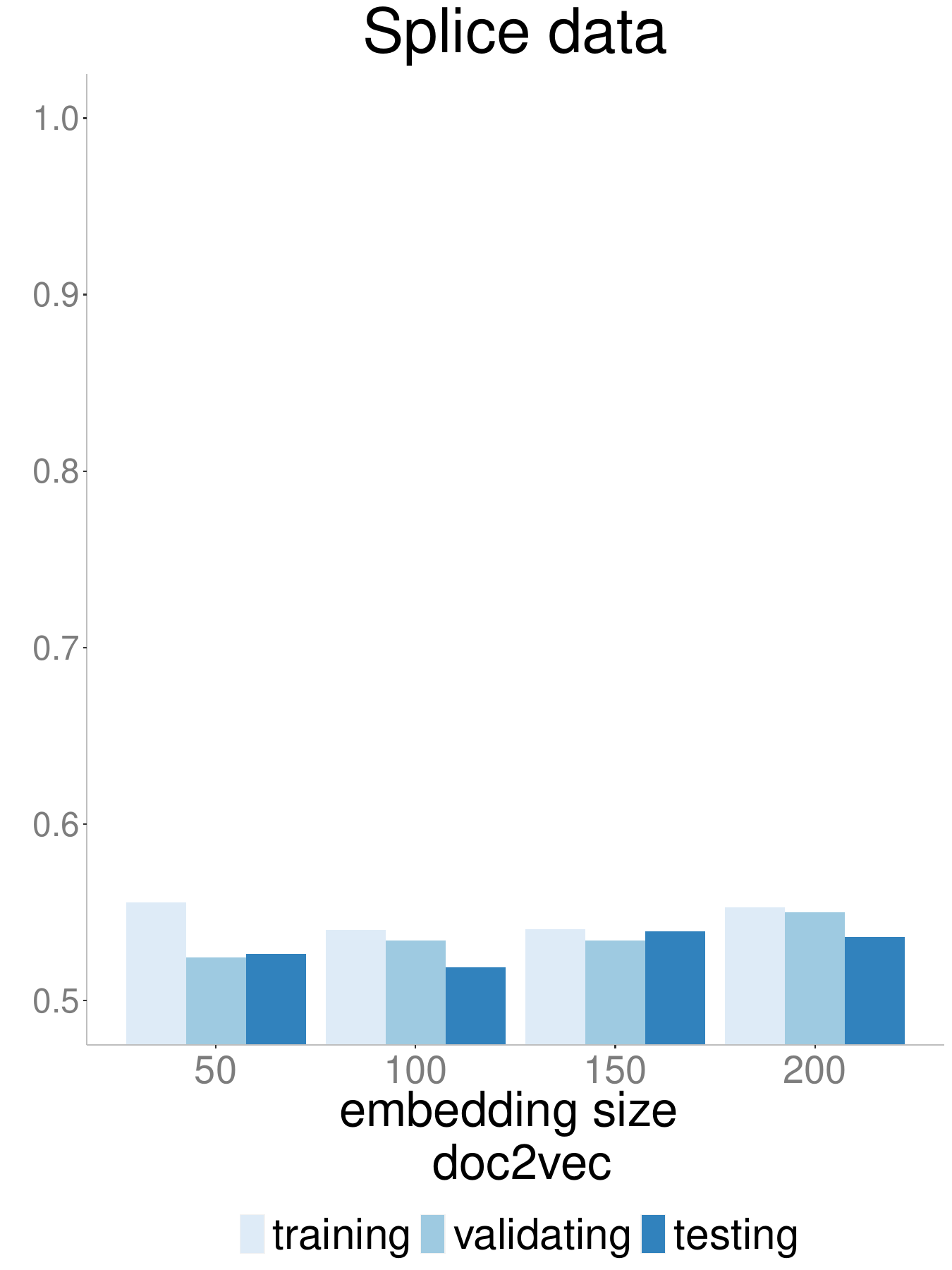}
    \includegraphics[scale=0.175]{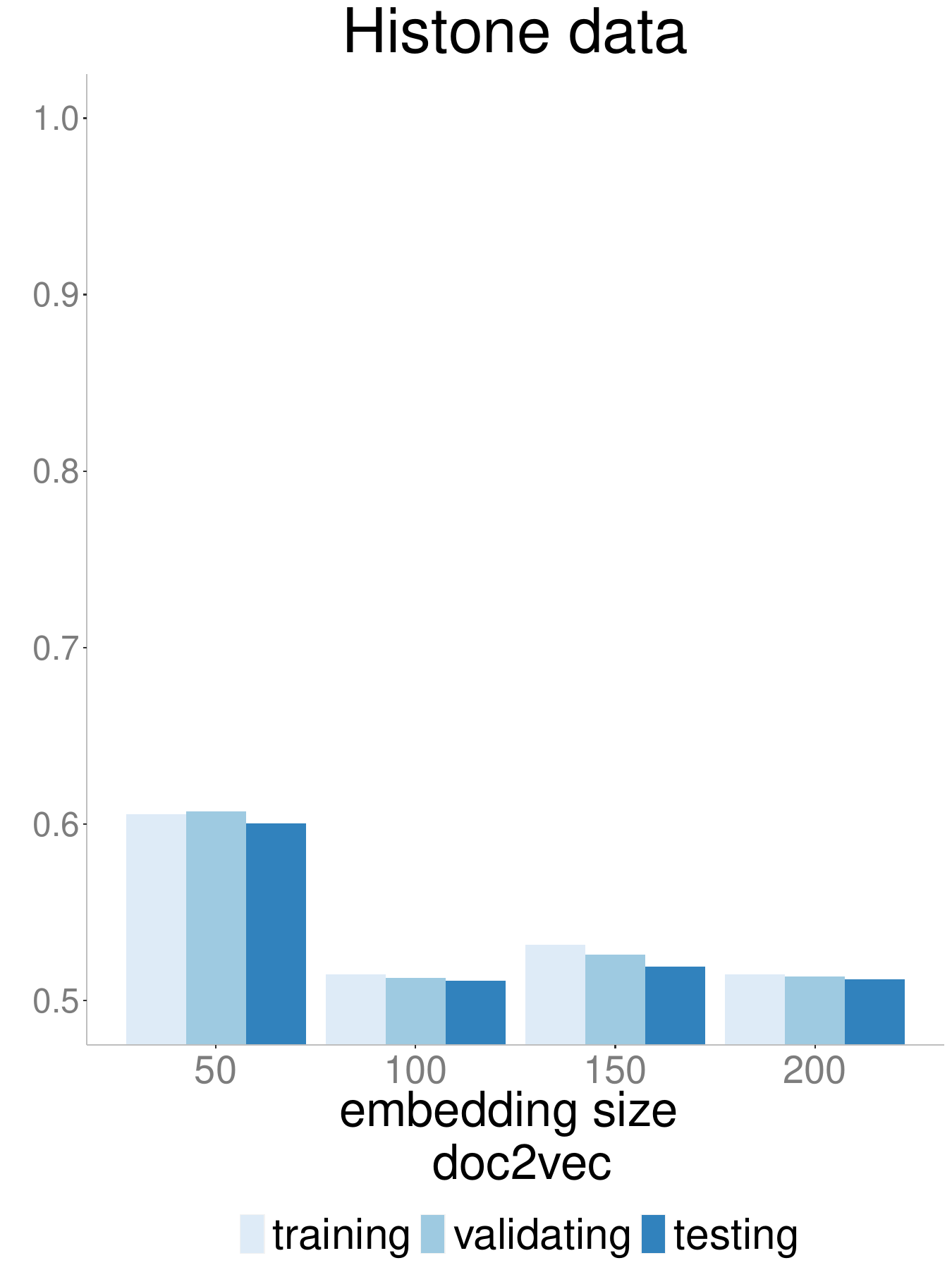}
    \includegraphics[scale=0.175]{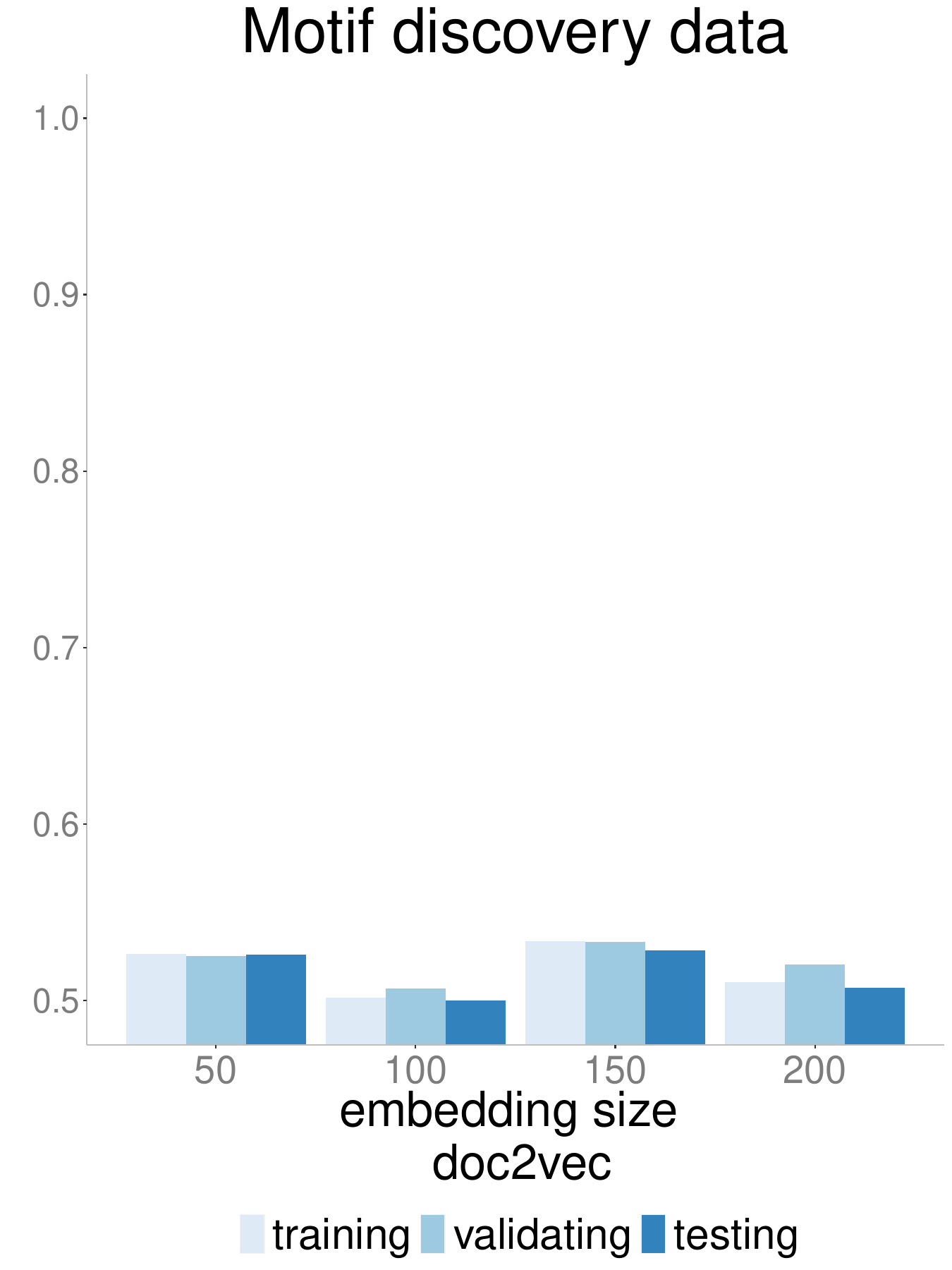}
    \caption{Accuracy of doc2vec+NN model with four different embedding sizes (50, 100, 150, 200) on three datasets of increasing size. The performance is poor in all cases with accuracy barely exceeding 50\% across datasets.}
    \label{fig:embed-size-acc}
\end{figure}

Of all the neural network models compared in this work, the doc2vec
version performs the worse with accuracy barely exceeding 50\% (Figure
\ref{fig:embed-size-acc}). The size of embedding does not appear to
have a strong influence on the accuracy, and if anything, it appears
to slightly decrease accuracy as the size of embedding increases for
some datasets (e.g. histone). The poor performance of the doc2vec
models is evident in the precision-recall curves (Appendix) and the
ROC curves as well (Appendix). This behavior contradicts the results
of the original work of doc2vec on sequences
\citep{kimothi2016distributed} which reported 97\% specificity (true
negative rate), 93\% sensitivity (true positive rate or recall), and
95\% accuracy for binary classification (as in histone and motif
discovery data) and 83\% precision, 81.5\% sensitivity, 81\% accuracy
for multiclass classification (as in splice data). The lack of
congruence could be due to lack of robustness of the model across
datasets, but more likely can be explained by the length of the
sequences.  While the original study has an average length of 425 bp
with sequences as long as 22,152 bp, the sequences used here have
length 60, 101 and 500 bp.

\subsection*{The role of batch size}

\begin{figure}[h]
    \centering
    \includegraphics[scale=0.175]{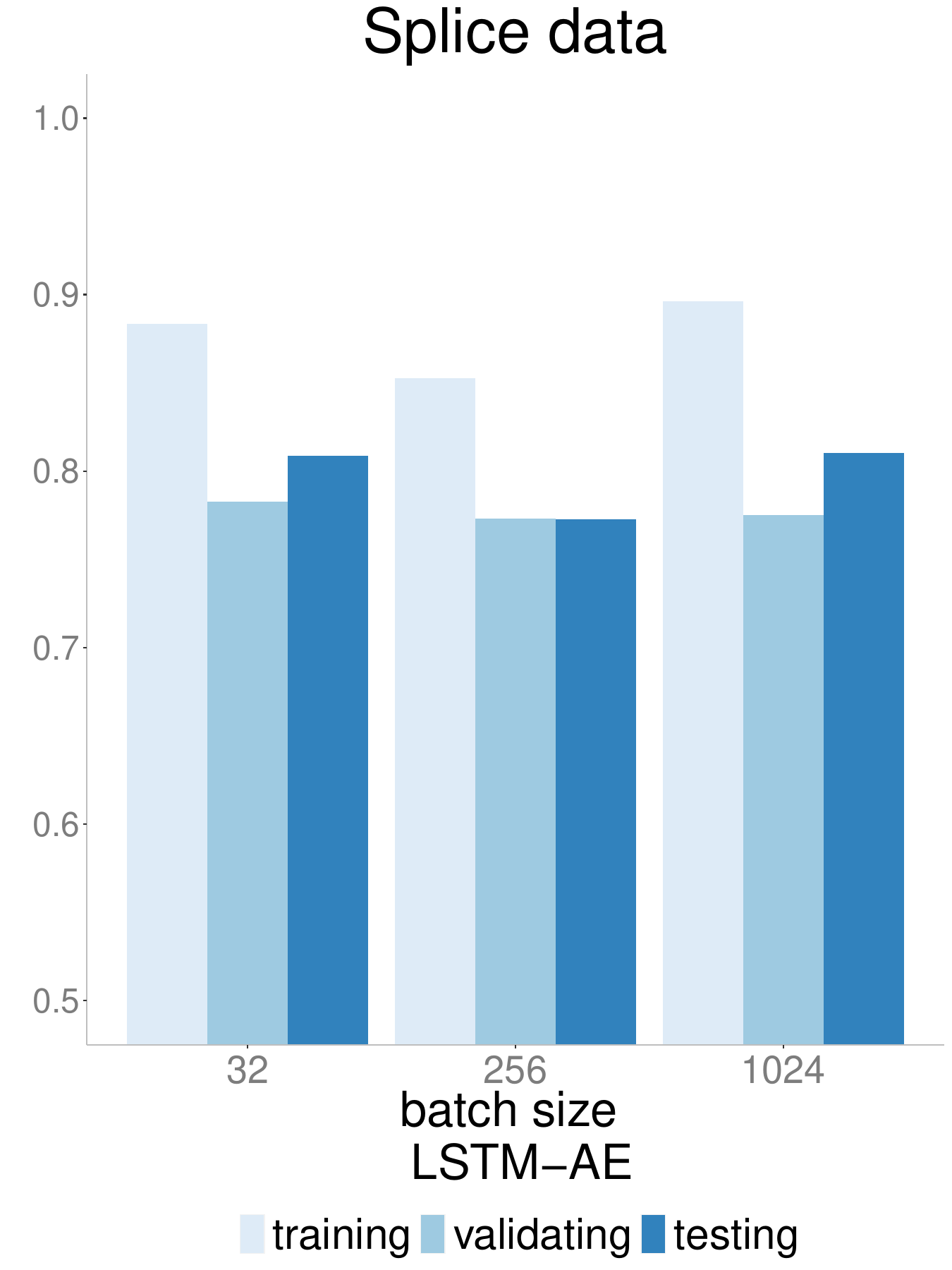}
    \includegraphics[scale=0.175]{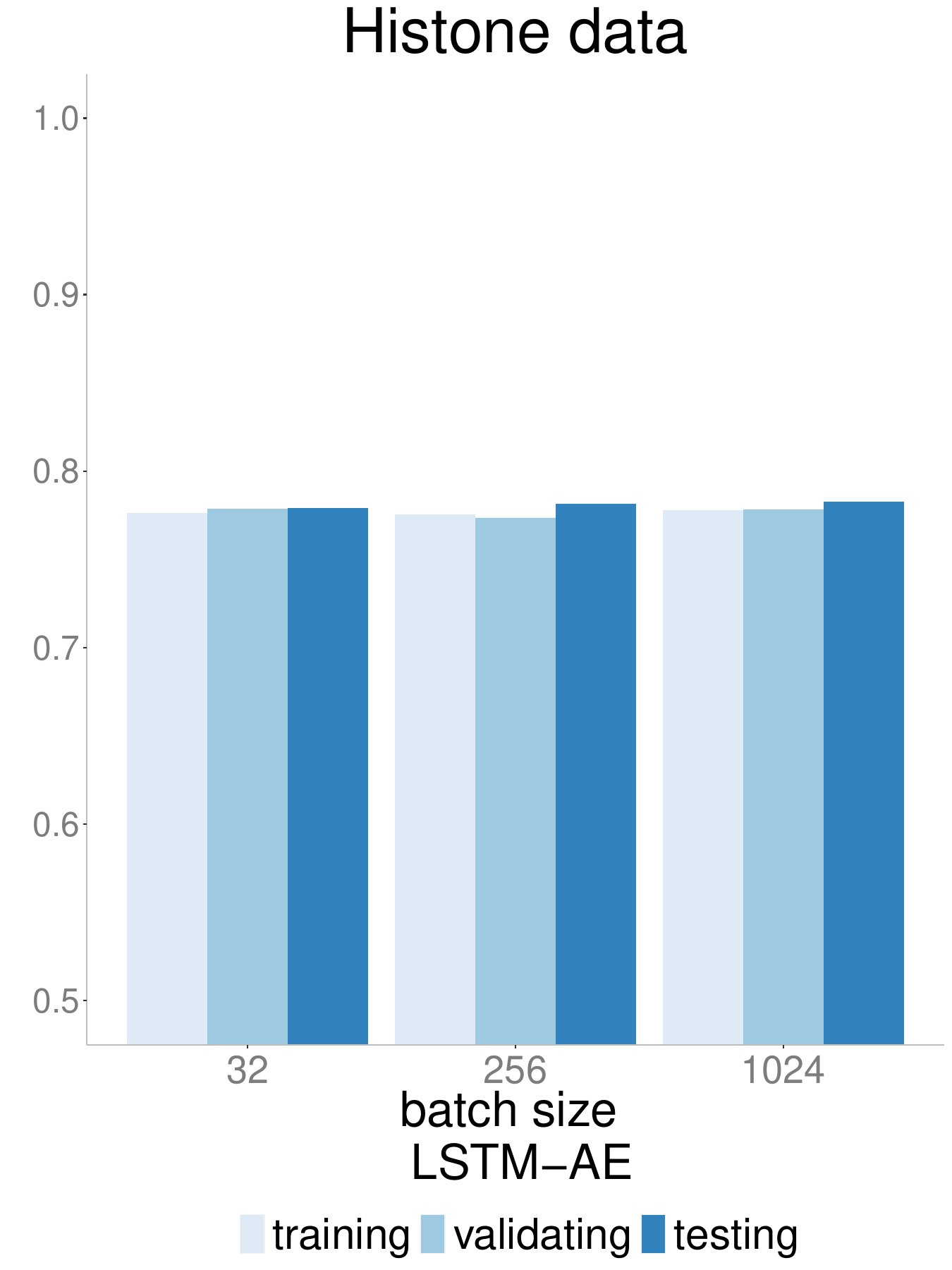}
    \includegraphics[scale=0.175]{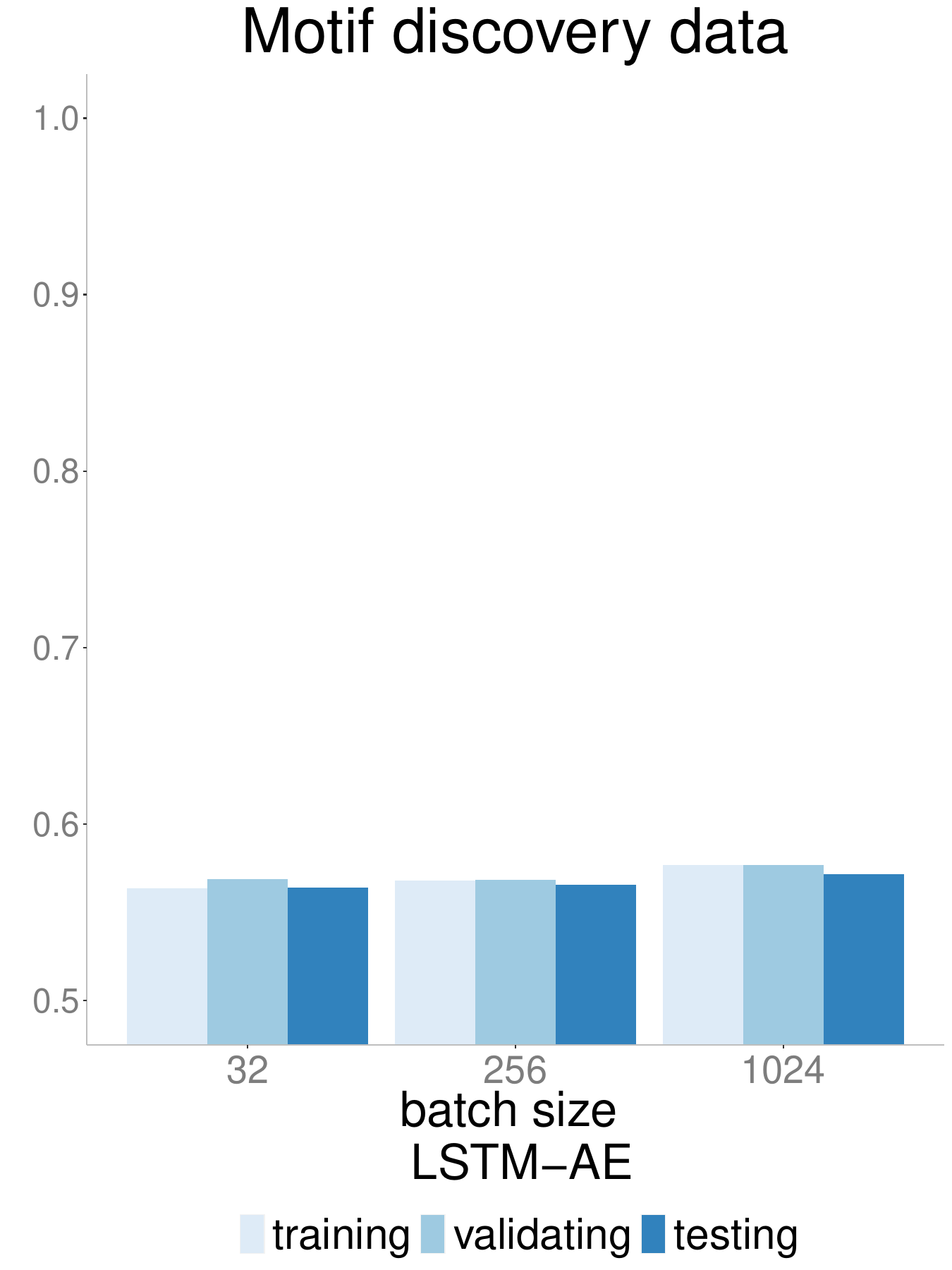}
    \caption{Accuracy of LSTM-AE+NN model \citep{Agarwal2019-pr} with three different batch sizes (32, 256 and 1024) on three datasets of increasing size. There is no evidence of overfitting with this model, and accuracy seems to decrease as the data size increases with the largest data (motif discovery) having the smallest accuracy (barely above 50\%).}
    \label{fig:batch-size-acc}
\end{figure}

\begin{figure}[h]
    \centering
    \includegraphics[scale=0.3]{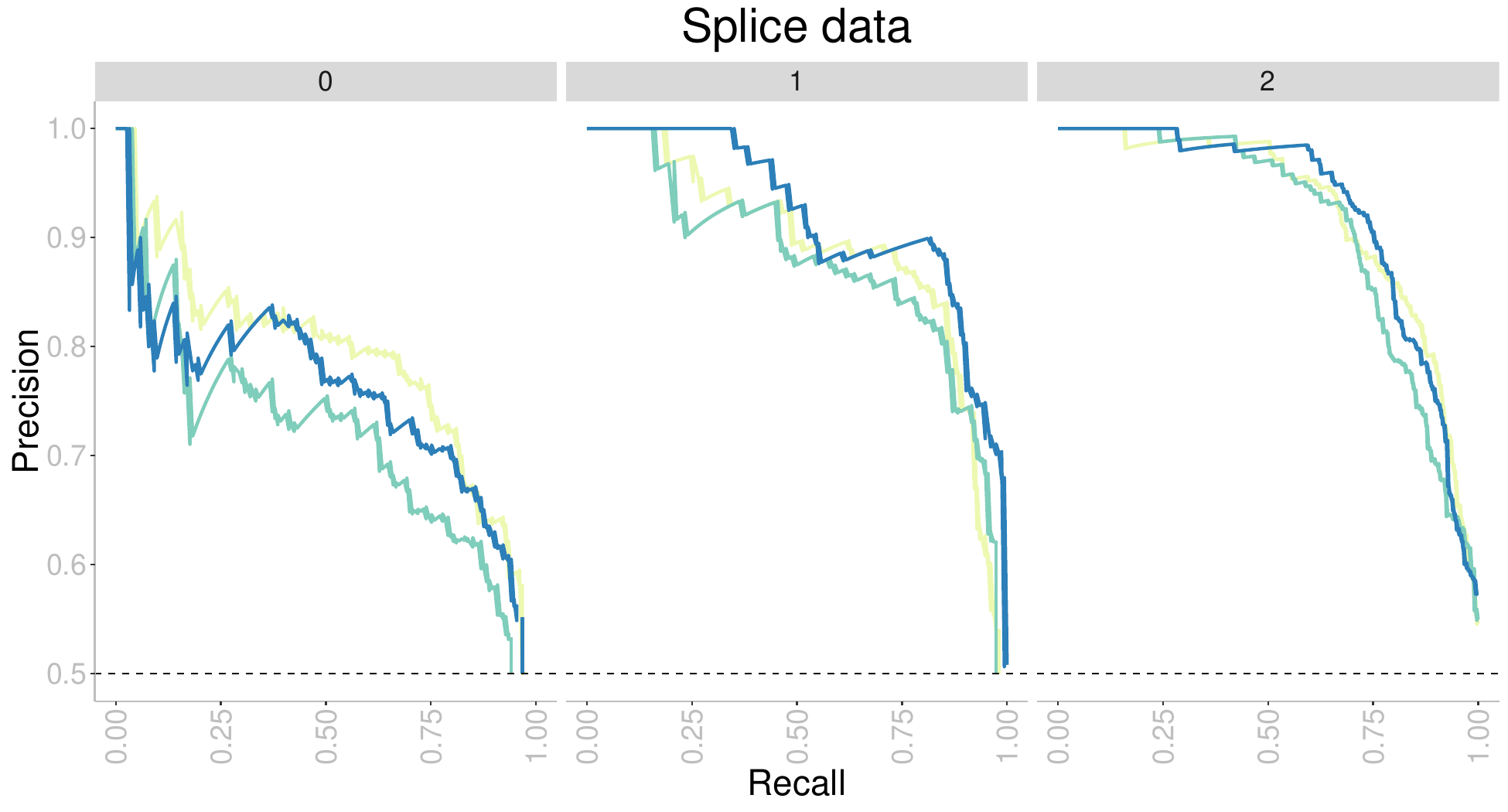}\\[0.25cm]
    \includegraphics[scale=0.25]{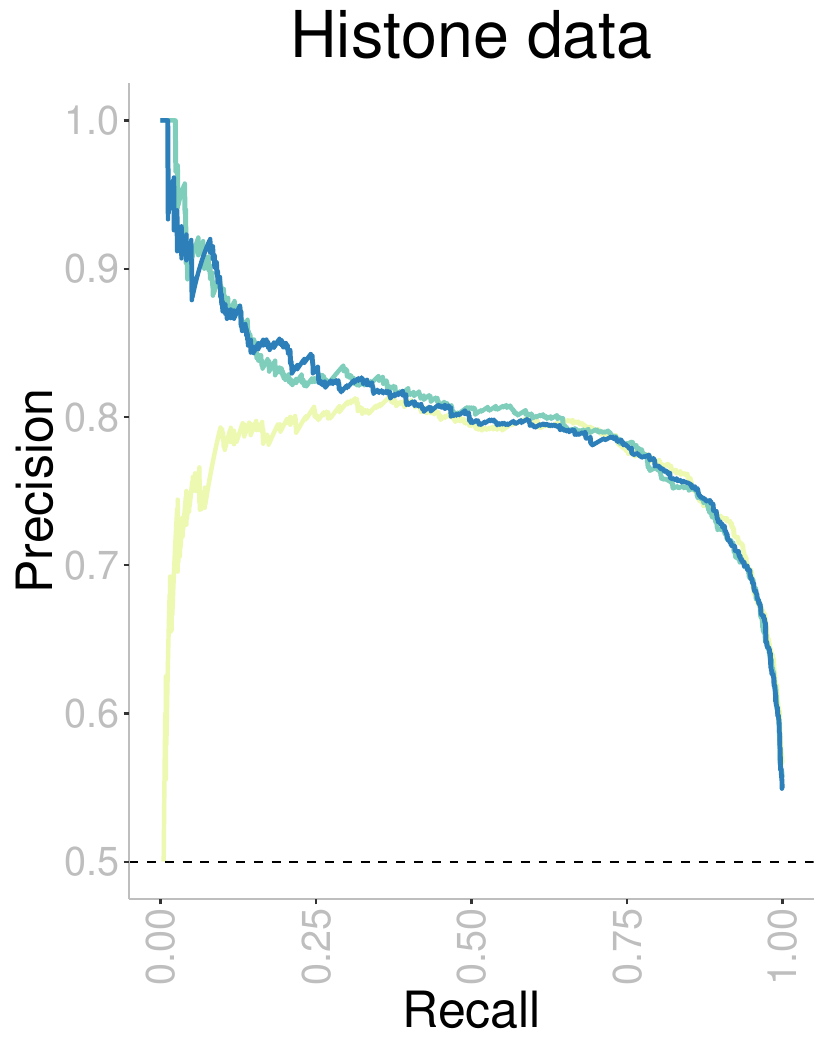}
    \includegraphics[scale=0.25]{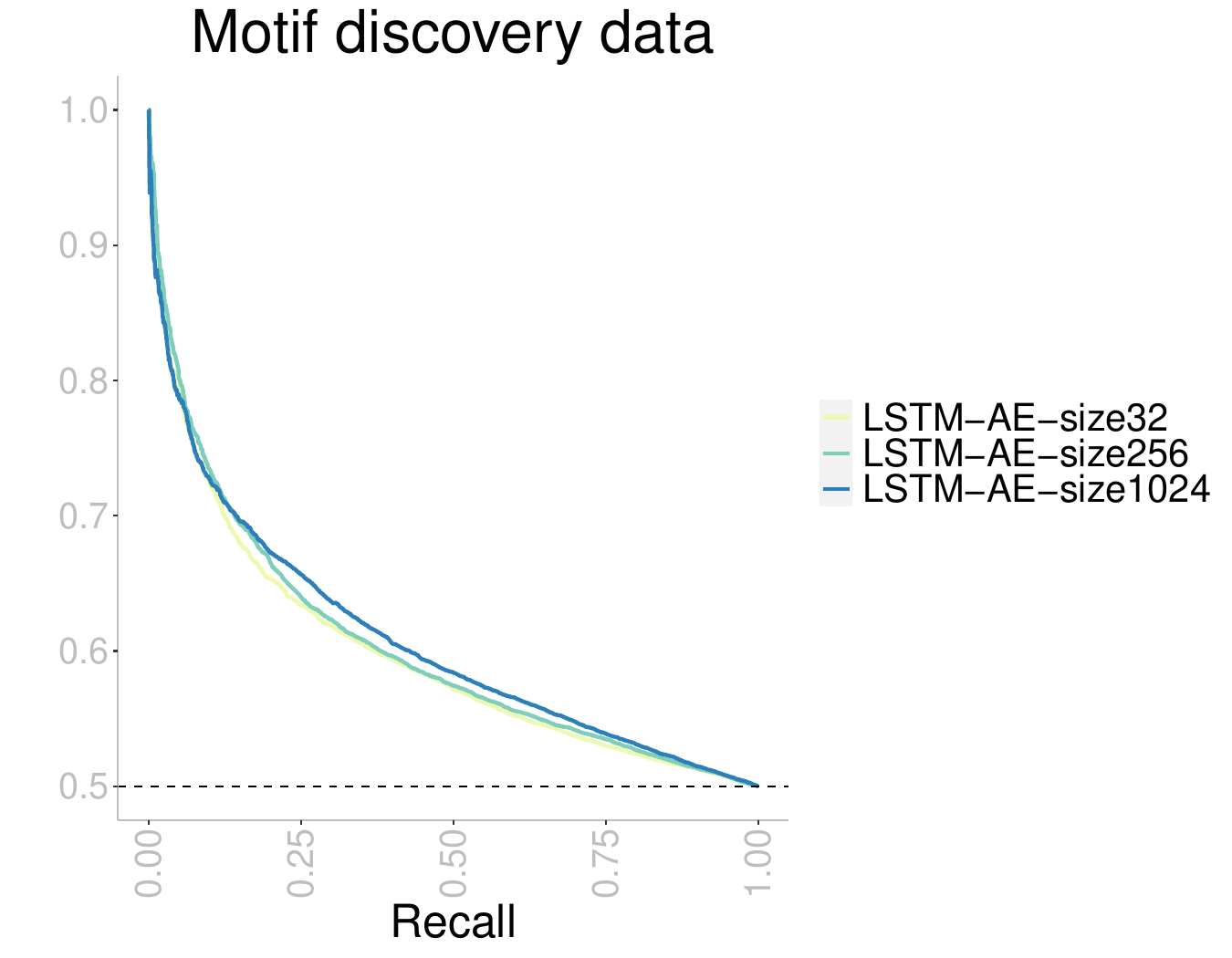}
    \caption{Precision-recall curves of LSTM-AE+NN model \citep{Agarwal2019-pr} with three different batch sizes (32, 256 and 1024) on three datasets of increasing size. The higher the curve, the better performance with a horizontal dashed line to represent random prediction. An ideal precision-recall curve would cross the (1,1) point. The splice data has three panels since precision-recall curves assume binary classification and the splice dataset has three classes (0, 1, 2). Each panel corresponds to prediction one class vs the other two combined. Unlike other models, there is a clear distinction in the class 0 prediction performance of this model compared to other classes in the splice data. It appears that class 0 is harder to predict with a lower recall for a given precision value compared to the other classes.}
    \label{fig:batch-size-pr}
\end{figure}

\begin{figure}[h]
    \centering
    \includegraphics[scale=0.3]{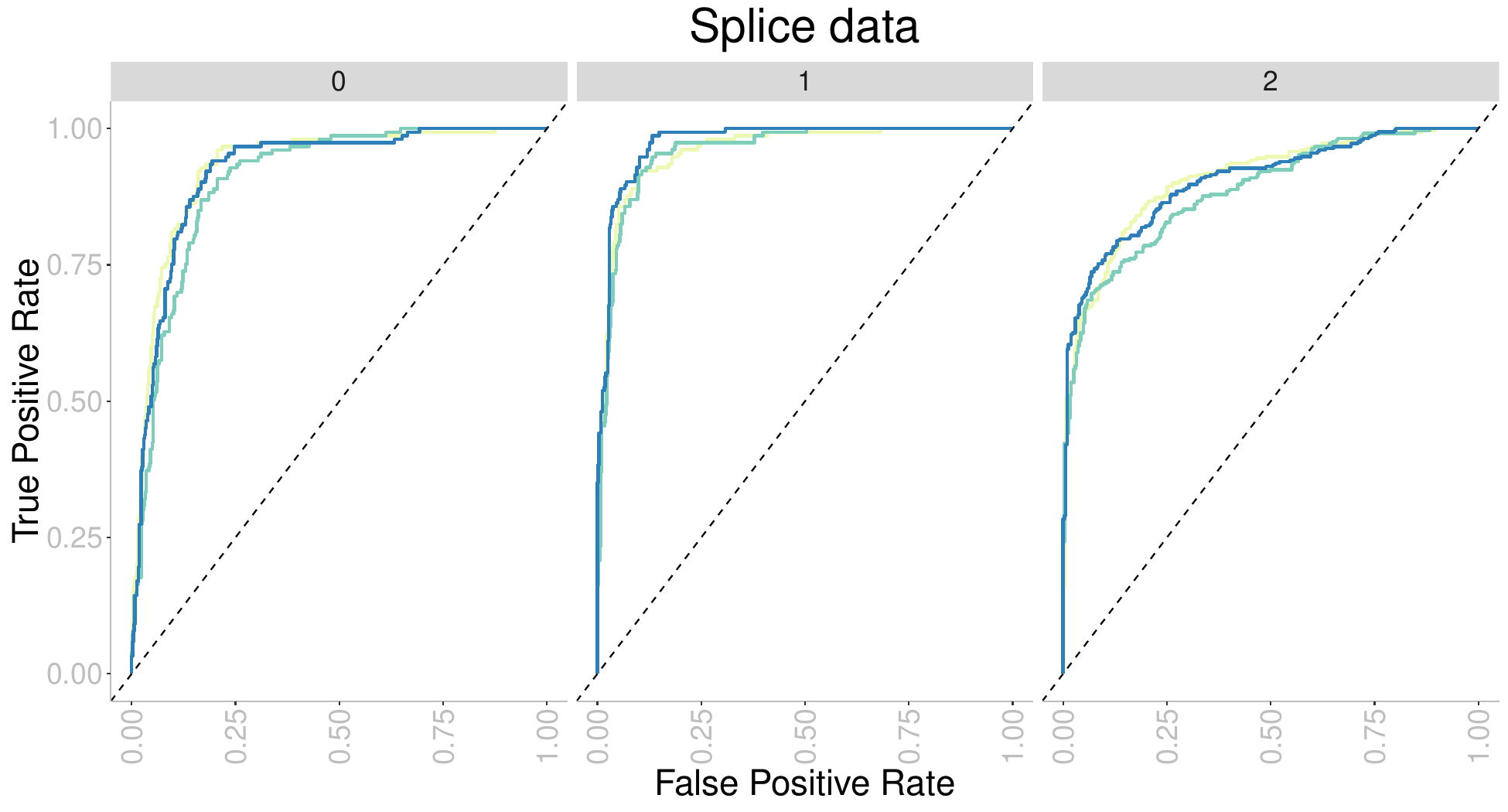}\\[0.25cm]
    \includegraphics[scale=0.25]{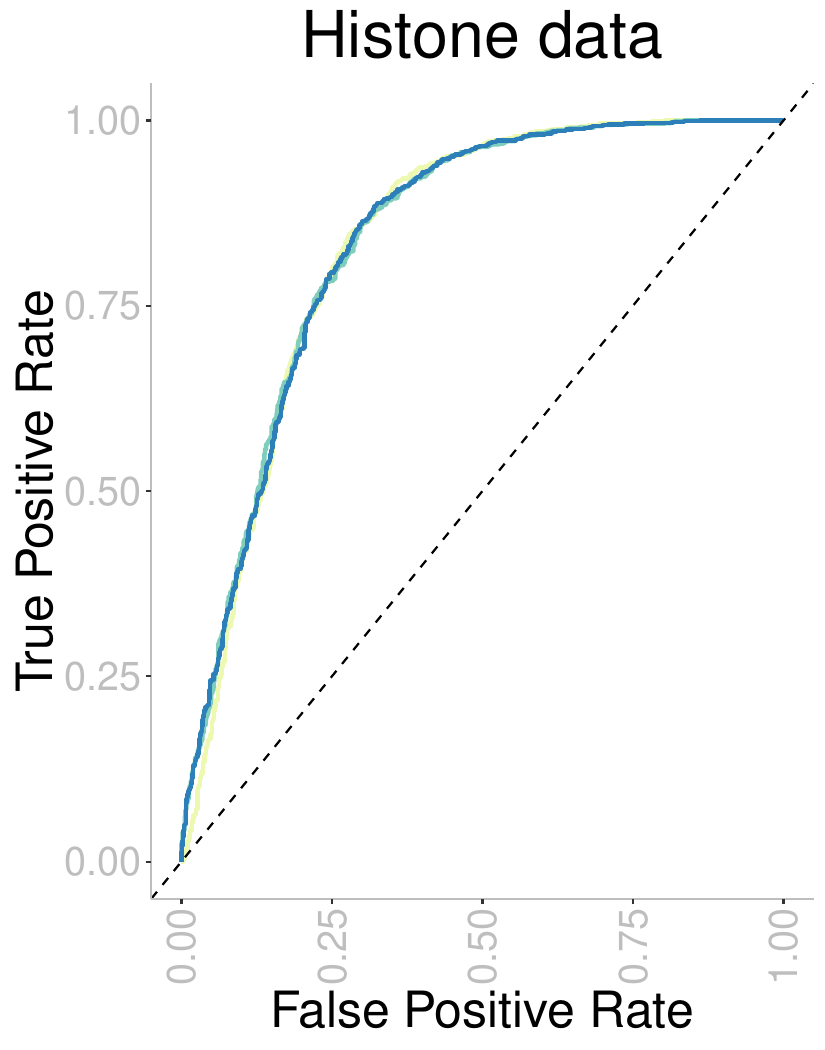}
    \includegraphics[scale=0.25]{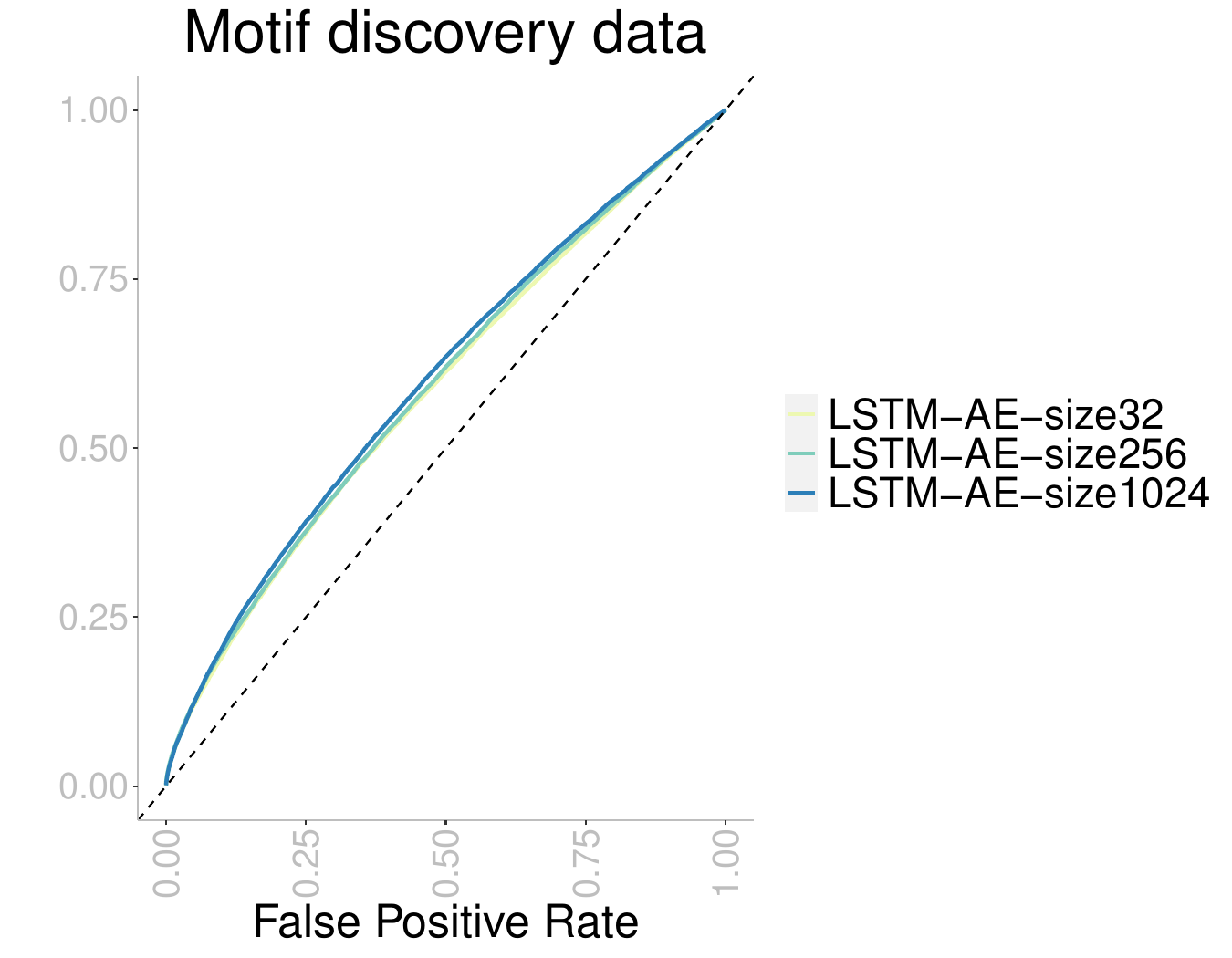}
    \caption{ROC curves of LSTM-AE+NN model \citep{Agarwal2019-pr} with three different batch sizes (32, 256 and 1024) on three datasets of increasing size. The higher the curve, the better performance with a 45$^{\circ}$ dashed line to represent random prediction. The splice data has three panels since ROC curves assume binary classification and the splice dataset has three classes (0, 1, 2). Each panel corresponds to prediction one class vs the other two combined. Again, we see no differences with respect to batch size and ROC curves close to the 45$^{\circ}$ dashed line (random prediction) for the motif discovery data.}
    \label{fig:batch-size-roc}
\end{figure}

Batch size has zero impact on the accuracy of the LSTM-AE model \citep{Agarwal2019-pr} (Figure \ref{fig:batch-size-acc}) with all three batch sizes (32, 256 and 1024) showing the same accuracy levels for a given dataset. Accuracy is instead affected by the size of the data with the largest dataset (motif discovery) barely exceeding 50\%. Also, this model appears to be robust to overfitting across datasets. 
The same conclusions can be drawn from the precision-recall curves (Figure \ref{fig:batch-size-pr}) and the ROC curves (Figure \ref{fig:batch-size-roc}). In the precision-recall curve it stands out that the class 0 in the splice data is harder to be predicted with his model compared to the other classes.

\subsection*{Overall comparison of models}

Among of all options, we select the models with highest testing accuracy for each of the categories (listed in Table \ref{tab:best}): CNN-Nguyen \citep{Nguyen2016-jj}, CNN-Zeng \citep{Zeng2016-nu}, DeepRAM \citep{Trabelsi2019-yt}, doc2vec, LSTM-AE \citep{Agarwal2019-pr} and LSTM-layer and for each of the three datasets (splice, histone and motif discovery). We also add the model in DeepDBP \citep{Shadab2020-sb} to the comparison.

\begin{figure}[h]
    \centering
    \includegraphics[scale=0.175]{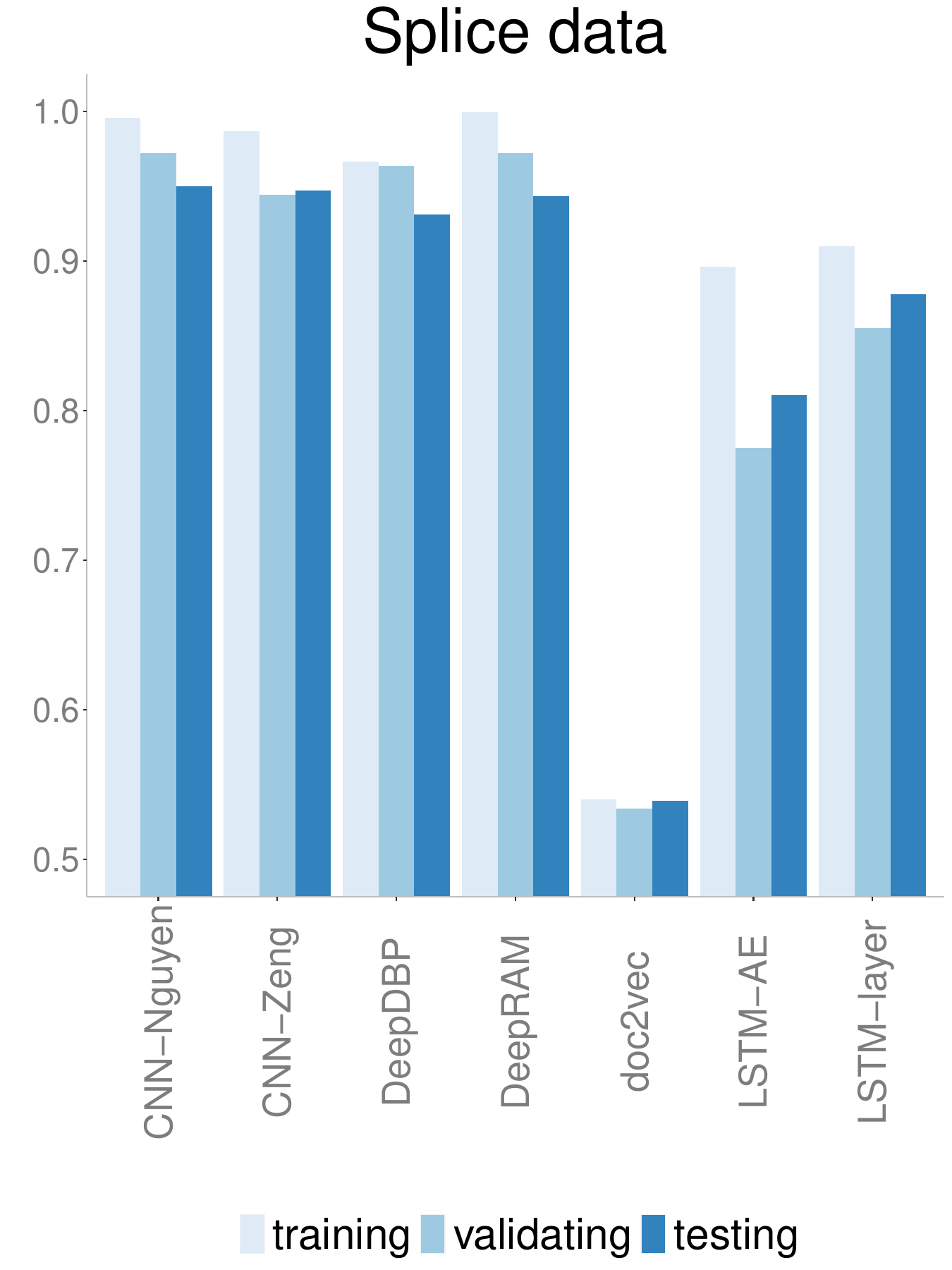}
    \includegraphics[scale=0.175]{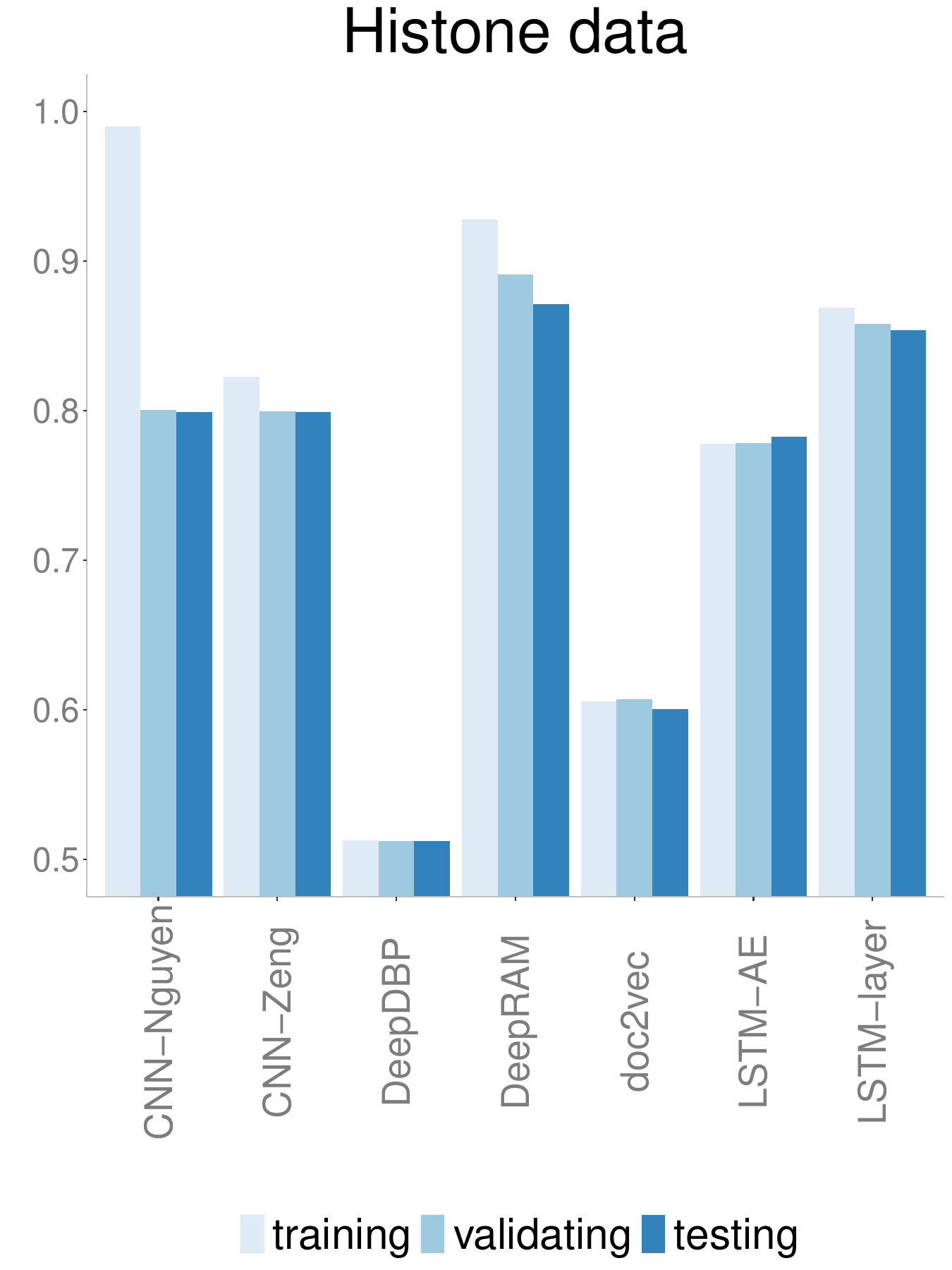}
    \includegraphics[scale=0.175]{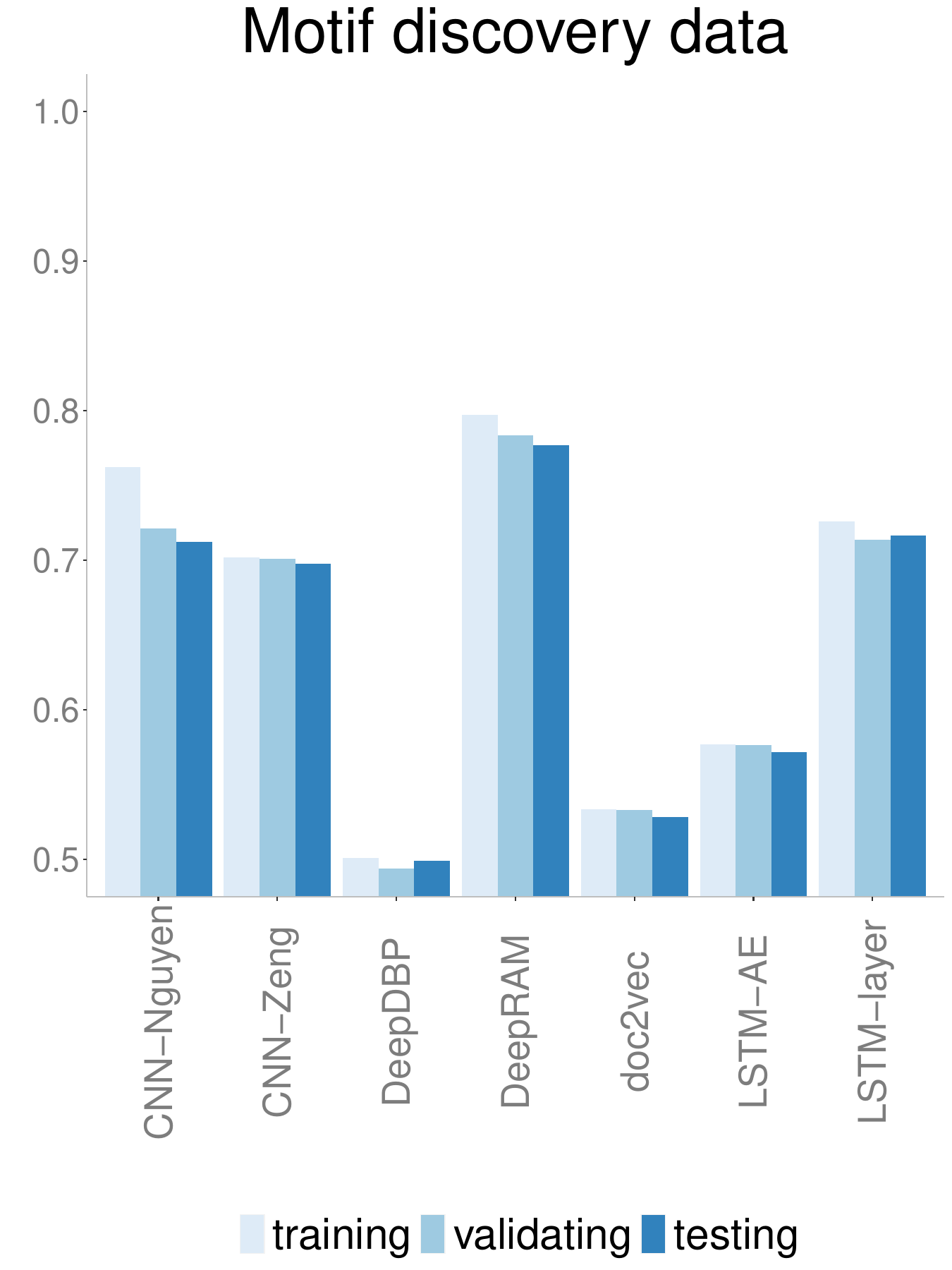}
    \caption{Accuracy of the models with highest testing accuracy among all comparisons (Table \ref{tab:best}) across datasets. DeepDBP \citep{Shadab2020-sb} shows the worst robustness across datasets, while DeepRAM \citep{Trabelsi2019-yt} shows both the best accuracy and robustness across datasets. Overfitting does not appear to be an issue except for CNN-Nguyen on the histone data.}
    \label{fig:best-acc}
\end{figure}

Regarding accuracy (Figure \ref{fig:best-acc}), first, we note that
the behavior of DeepDBP is not robust across datasets with accuracy
levels never exceeding 55\% for the histone and motif discovery data
while the reported accuracy on the original paper
\citep{Shadab2020-sb} was 84.31\% for a data of sample size of 1075 sequences.
It appears that the performance of DeepDBP is highly dependent on the specifics of the data at hand.

Next, we notice that doc2vec+NN behaves poorly with accuracy levels
barely exceeding 50\% in all three datasets. We reiterate that this
poor performance could be due to the short length of the sequences
used here. Overall, DeepRAM outperforms all other models across datasets which has the added strength of robustness given that the accuracy levels are not far from the accuracy levels reported in the original DeepRAM paper \citep{Trabelsi2019-yt} (88.9\%). Both CNN models (CNN-Nguyen and CNN-Zeng) perform well across datasets albeit less accurately than DeepRAM. Overfitting does not appear to be a relevant factor among these models, except for CNN-Nguyen on the histone data.

\begin{figure}[h]
    \centering
    \includegraphics[scale=0.3]{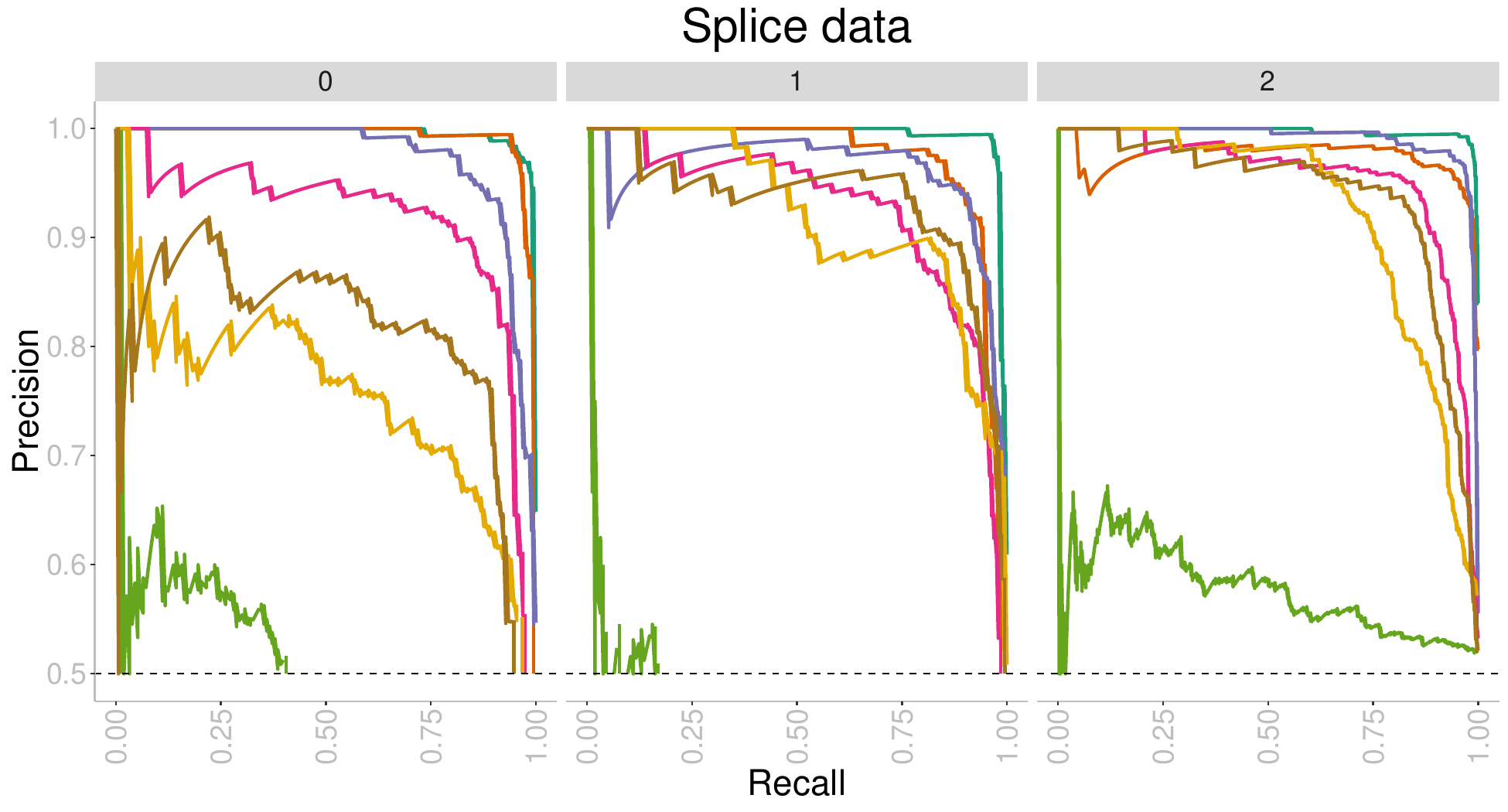}\\[0.25cm]
    \includegraphics[scale=0.25]{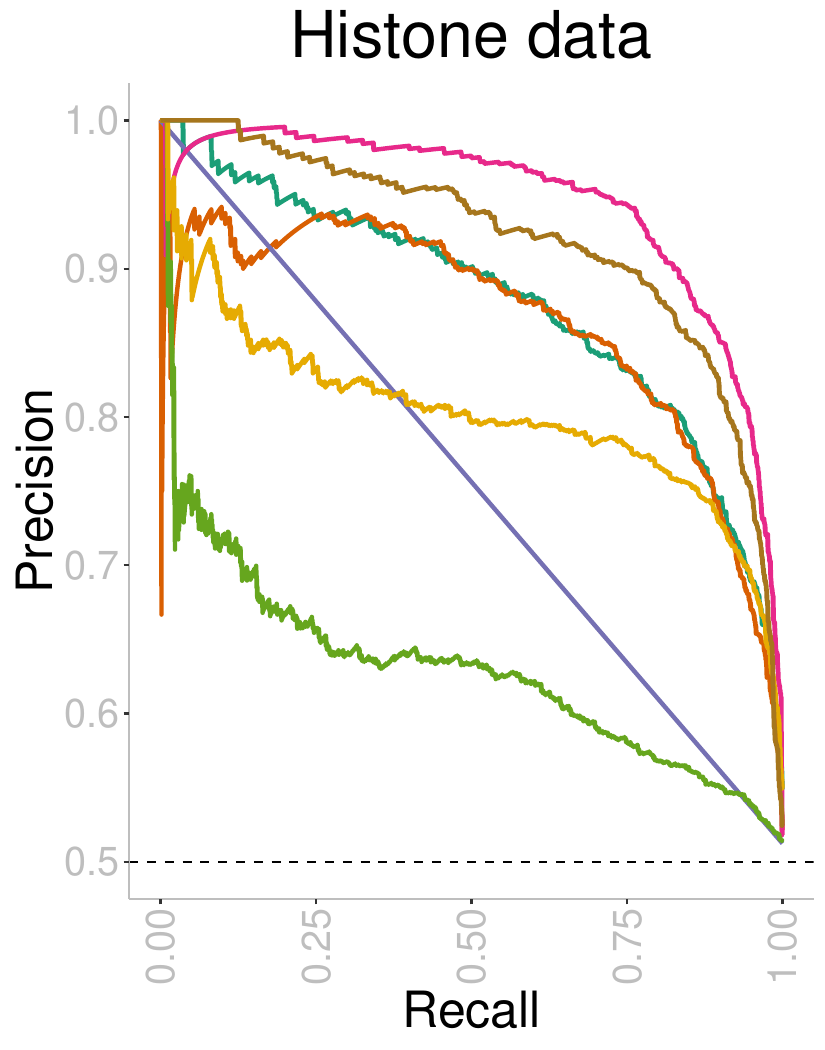}
    \includegraphics[scale=0.25]{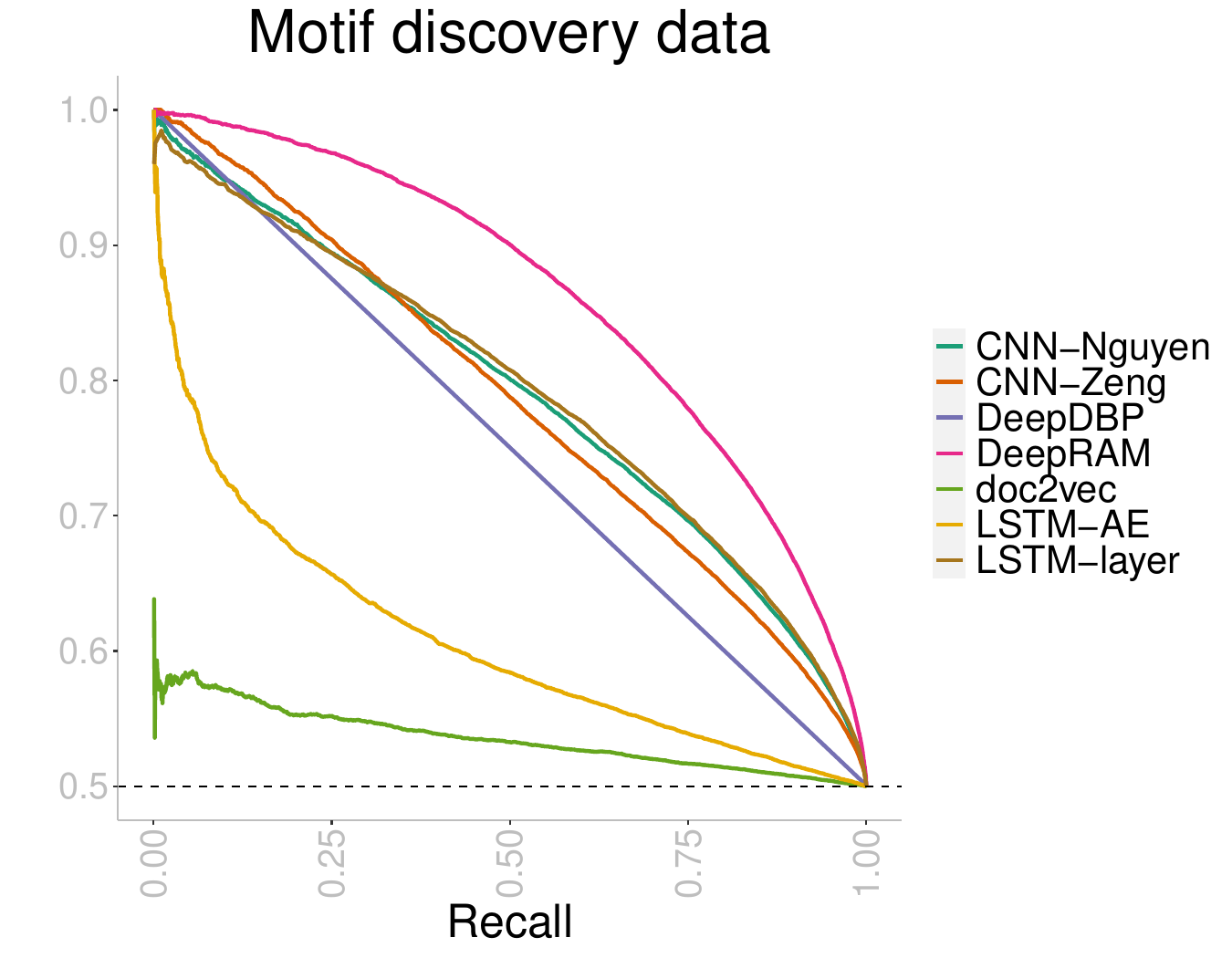}
    \caption{Precision-recall curves of CNN models of increasing number of layers on the three datasets of increasing size. The higher the curve, the better performance with a horizontal dashed line to represent random prediction. An ideal precision-recall curve would cross the (1,1) point. The splice data has three panels since precision-recall curves assume binary classification and the splice dataset has three classes (0, 1, 2). Each panel corresponds to prediction one class vs the other two combined. DeepRAM outperforms all models in the histone and motif discovery datasets, and behaves well on the splice data. The CNN models (Nguyen and Zeng) outperform all models on the splice data.}
    \label{fig:best-pr}
\end{figure}

\begin{figure}[h]
    \centering
    \includegraphics[scale=0.3]{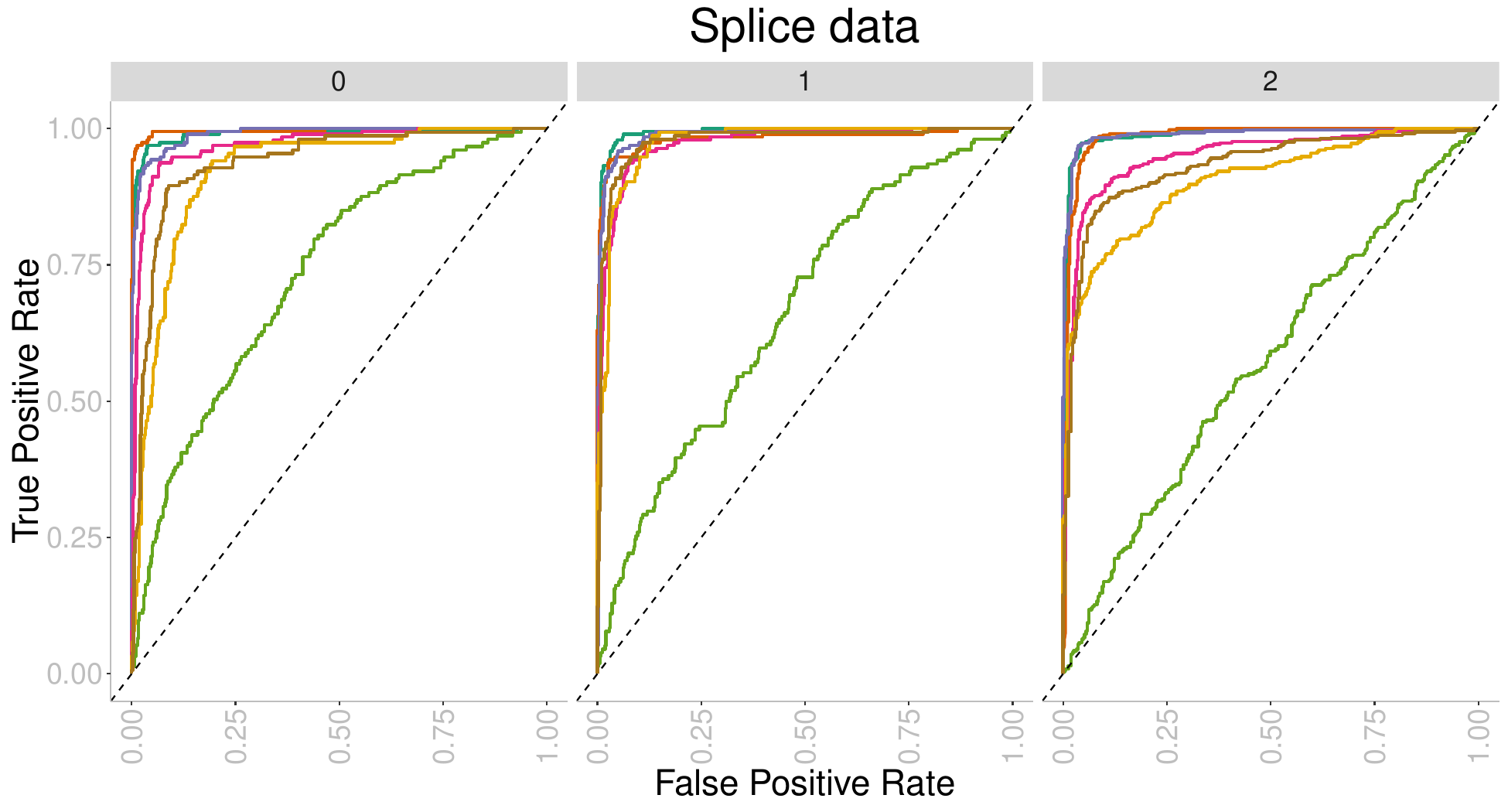}\\[0.25cm]
    \includegraphics[scale=0.25]{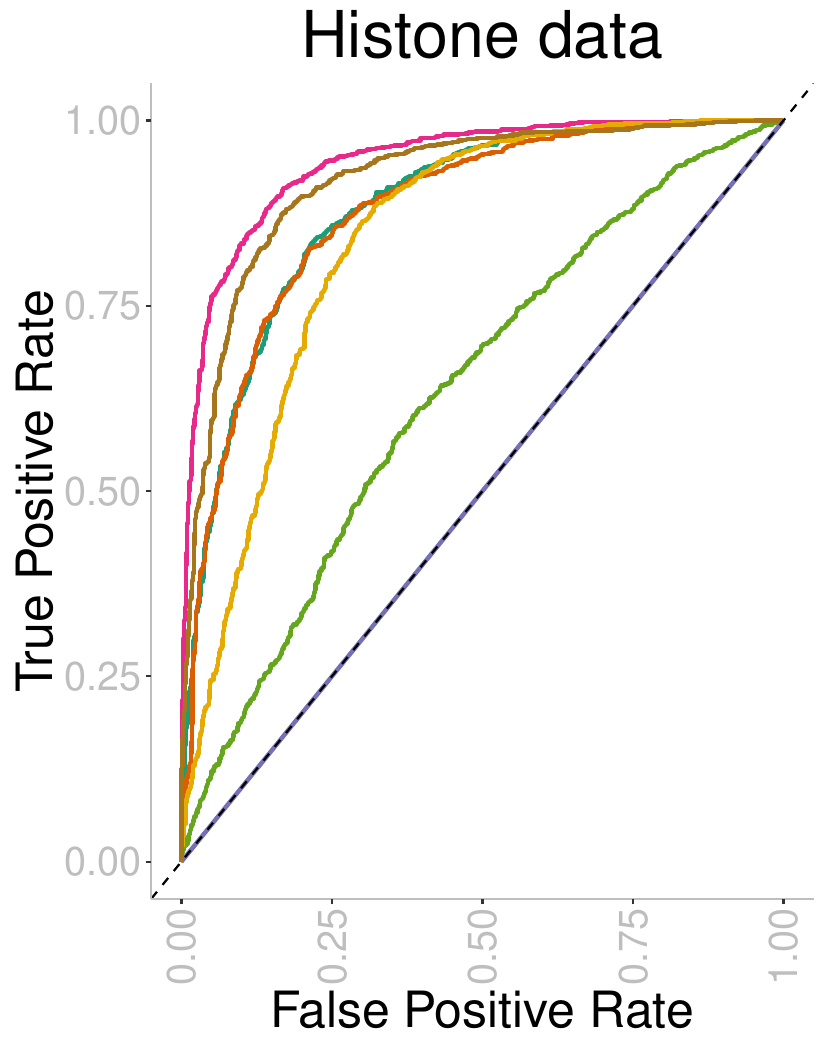}
    \includegraphics[scale=0.25]{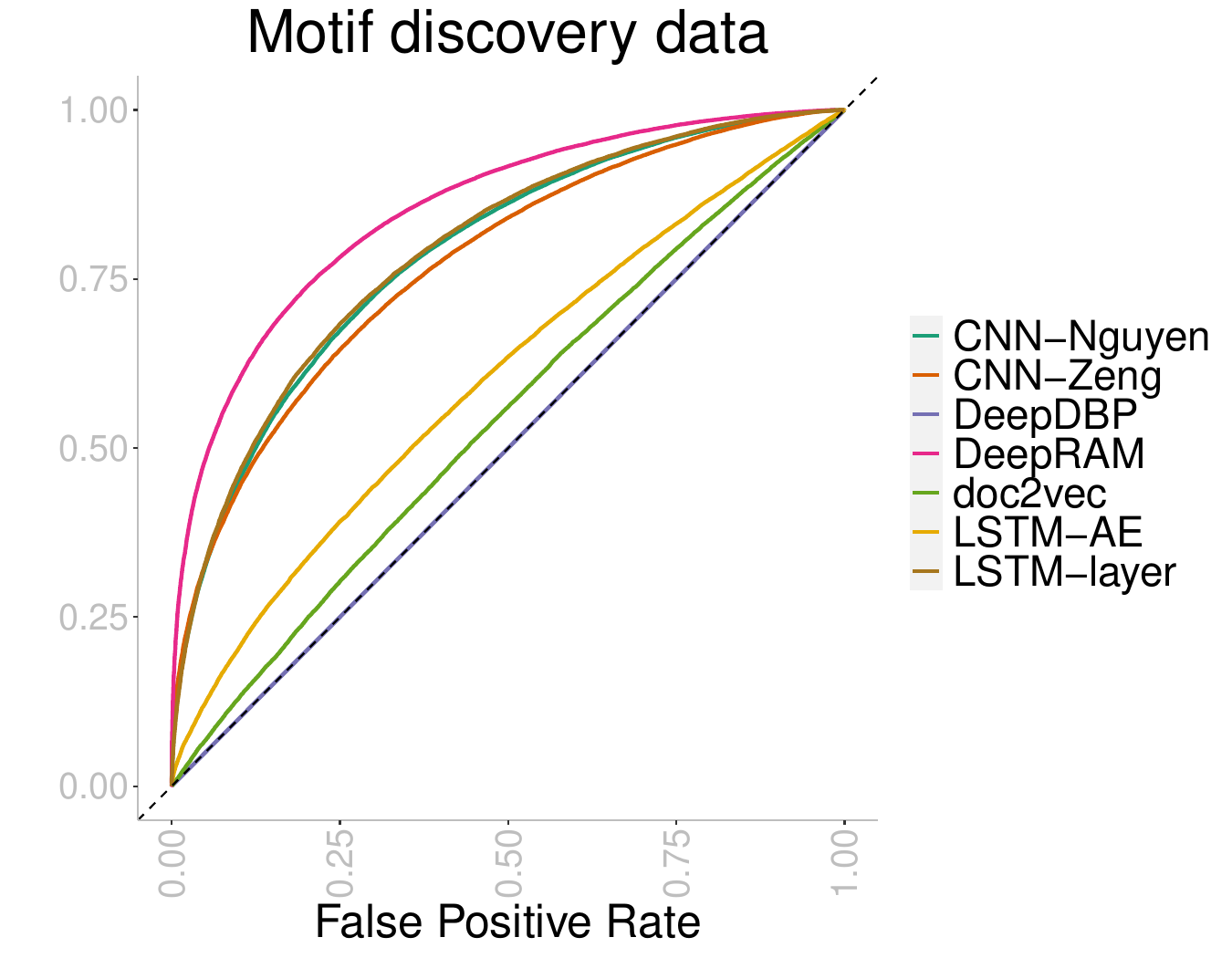}
    \caption{ROC curves of CNN models of increasing number of layers on the three datasets of increasing size. The higher the curve, the better performance with a 45$^{\circ}$ dashed line to represent random prediction. The splice data has three panels since ROC curves assume binary classification and the splice dataset has three classes (0, 1, 2). Each panel corresponds to prediction one class vs the other two combined. DeepRAM outperforms all models in the histone and motif discovery datasets, and behaves well on the splice data. The CNN models (Nguyen and Zeng) outperform all models on the splice data. DeepDBP behaves as a random predictor on the histone and motif discovery data.}
    \label{fig:best-roc}
\end{figure}

Figure \ref{fig:best-pr} shows the precision-recall curves where the same conclusions are confirmed with DeepRAM outperforming all models in the histone and motif discovery datasets. The CNN models (CNN-Nguyen and CNN-Zeng) outperform all models for the splice data. We note that prediction of class 0 in the splice data appears to be harder than prediction of the other two classes as evidenced by lower overall curves. The doc2vec model performs poorly on all datasets. Similar conclusions are drawn with the ROC curves (Figure \ref{fig:best-roc}) with DeepDBP behaving as a random predictor on the histone and motif discovery data.

\section*{Discussion}

Neural network models provide endless opportunities for prediction and classification in biological applications \citep{Peng2020, alber2019integrating}, yet much remains unknown regarding the transferability of the performance across datasets. Robustness across datasets of similar nature is a key ingredient to translate neural network models into medical, agricultural or environmental practice.
Here, we study the performance on genomic data of convolutional neural networks (CNN) and \revision{neural network models assisted by} natural language processing (NLP). \revision{We highlight that the conclusions we found are restricted to the datasets and the models selected, and thus, more work is needed to be able to extend conclusions beyond the current study.}

We find that DeepRAM outperforms all other models especially the recurrent version (RNN) in terms of prediction accuracy, overfitting, and robustness across datasets. Compared to CNN models (CNN-Nguyen and CNN-Zeng) whose prediction accuracy dramatically decreases with larger datasets, DeepRAM models experience a much smaller accuracy decrease. Furthermore, the accuracy levels of DeepRAM that we find here are comparable to those reported in the original DeepRAM paper \citep{Trabelsi2019-yt} and thus, we can conclude that the DeepRAM models are more robust, transferable and generalizable across genomic datasets with varied characteristics.
It is interesting to notice that DeepRAM outperforms CNN-Nguyen and CNN-Zeng even when we are using the original datasets in both CNN papers \citep{Nguyen2016-jj, Zeng2016-nu}.

DeepDBP lacks robustness across datasets at least for the datasets
compared in this work. The original paper of DeepDBP
\citep{Shadab2020-sb} reported prediction accuracy levels of 84.31\%
and while we find a prediction accuracy of around 90\% for the splice
data, for the histone and motif discovery datasets, the DeepDBP
prediction accuracy barely exceeds 50\%. Furthermore, the DeepDBP
paper did not provide usable reproducible scripts that we could
follow\revision{, so the poor performance could be due to
  discrepancies between the model implemented here and the model
  implemented in the original DeepDBP. Given that our goal was to
  approach this study from the perspective of a domain scientist
  (biomedical researcher), we believe that such researcher would read
  a NN article (like the DeepDBP), and then try to fit such model on
  their own data. When NN papers provide clear code (python notebooks,
  for example), they facilitate this task to domain scientists. The authors of DeepDBP, however, did not provide clear code to fit their models, so we test the performance of a model based on the paper description (which is what the biomedical researcher would do). We hope to bring attention to reproducibility practices to help domain scientists fit NN models that appear in literature in their own datasets.} 

In terms of overfitting, the gap between training and testing accuracy increases as the number of layers increases for CNN models. This behavior is more evident for larger datasets (histone and motif discovery) than smaller datasets (splice). \revision{We reiterate that this atypical performance could be due to the choice of data partition into 48.7\% training set for the motif discovery data (far from the standard 70-15-15 data partition). While this choice was made in an attempt to reduce bias caused by heterogeneous input sequences, it is far from perfect. Future work should address implications in prediction due to data partition choices when faced with highly heterogeneous datasets.} 

It is noteworthy that more LSTM layers do not seem to increase overfitting since LSTM-layer (one layer), LSTM-AE (one layer) and DeepRAM-RNN (two layers) have no noticeable overfitting patterns though a more thorough investigation of LSTM layers is still lacking.
The only overfitting case for DeepRAM happens on the motif discovery data in the combined model (CNN-RNN) with embedding data encoding. It seems advisable to utilize one-hot encoding for DeepRAM models to prevent the potential of overfitting.

The doc2vec encoding performed poorly on all scenarios. Given that the prediction model for doc2vec and LSTM-AE \citep{Agarwal2019-pr} is the same (the simple NN in Figure \ref{fig:simple-nn}) and LSTM-AE dramatically outperforms doc2vec, we do not recommend the use of doc2vec for data embedding and recommend the use of LSTM autoencoders instead. This is especially true for the case of shorter sequences.
For the LSTM-AE model, the batch size made no difference in performance and this model seems to be very robust to overfitting, yet we do see smaller accuracy with larger datasets.

We conclude by raising awareness to the importance of reproducibility in science. In many instances, it was impossible to replicate the results of existing publications given the lack of reproducible well-documented scripts and available data. Reproducibility is crucial not just for the sake of open science, but to maximize the applicability of our machine-learning findings into a biological or medical community who might not have a strong programming background.

\subsection*{Practical advice for domain scientists}

\revision{Among the models here compared,} recurrent neural network models (specifically DeepRAM-RNN \citep{Trabelsi2019-yt}) outperform convolutional neural network models in terms of prediction accuracy, overfitting and transferability across datasets. More LSTM layers produce higher prediction accuracy without overfitting, unlike more convolutional layers which tend to produce modestly higher accuracy, but also a larger gap between training and testing accuracy. We recommend accompanying extra convolutional layers with regularization. Convolutional neural networks have a reasonable performance overall, but their accuracy is affected by the size of the data with larger \revision{(more heterogeneous)} datasets having lower prediction accuracy, a behavior not seen with RNN.
For data encoding, the intuitive nature of doc2vec does not translate into good prediction performance and less interpretable encoders like LSTM-AE \citep{Agarwal2019-pr} should be preferred, especially for the case of shorter sequences \revision{as illustrated in the three datasets here used}.
The doc2vec encoder followed by a simple NN performs poorly (accuracy barely exceeding 50\%) in all tested scenarios unlike LSTM-AE \citep{Agarwal2019-pr} followed by the same simple NN which manages moderately good accuracy (comparable to CNN models) across datasets and without too evident overfitting. 
In terms of model characteristics, embedding size and batch size do
not seem to play any important role in our comparisons, while the
optimizer in conjunction with the patience parameter do seem to play a
role in the comparisons (see also \citep{nado2021large}).
We conclude by highlighting that while this work is intended to provide practical advice to domain scientists who are interested in fitting neural network models on their data, it does not intend for domain scientists to work in isolation as nothing can replace the powerful interdisciplinary connections between the domain scientific community and the machine-learning community.

\section*{Appendix}

\begin{figure}[h]
    \centering
    \includegraphics[scale=0.3]{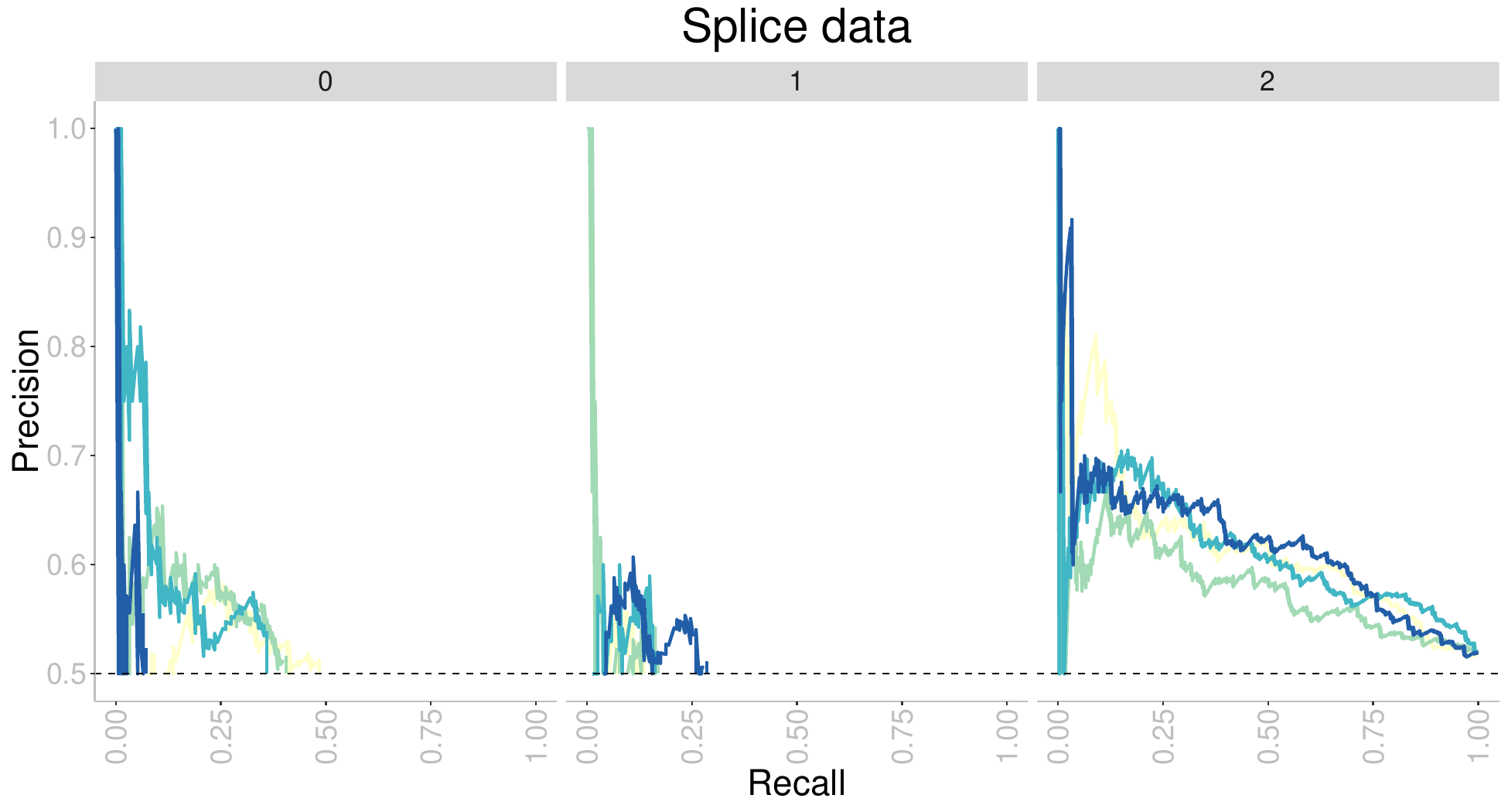}\\[0.25cm]
    \includegraphics[scale=0.25]{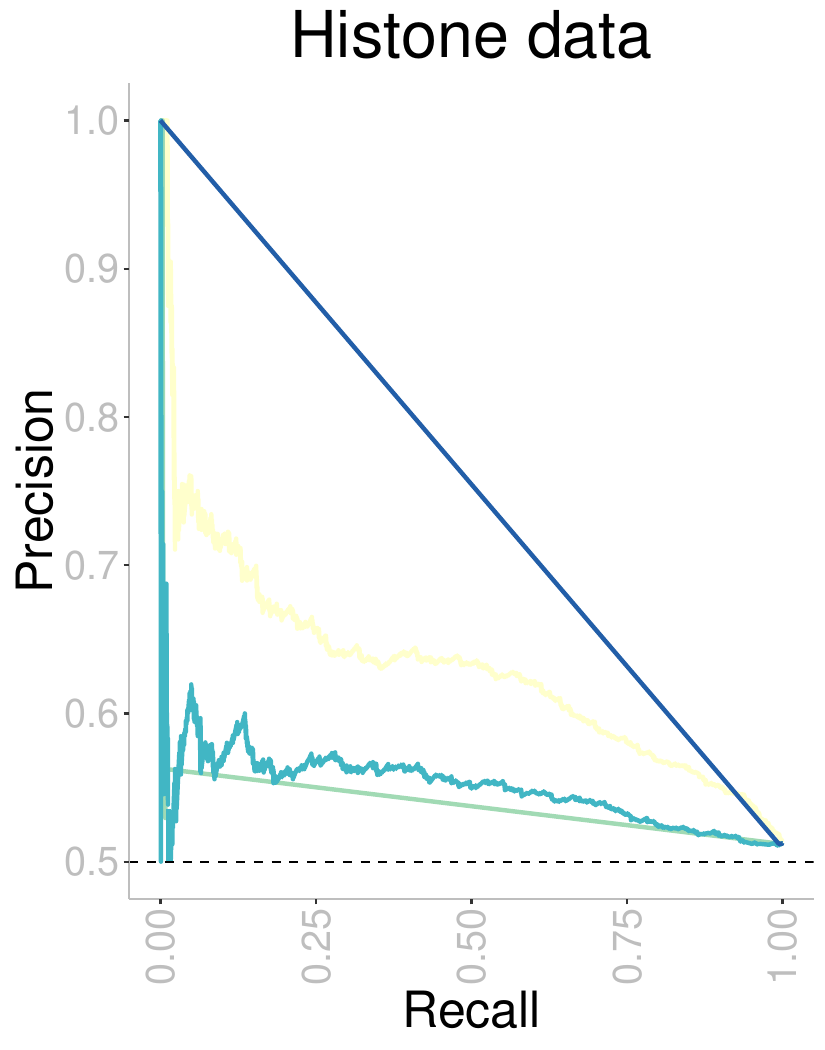}
    \includegraphics[scale=0.25]{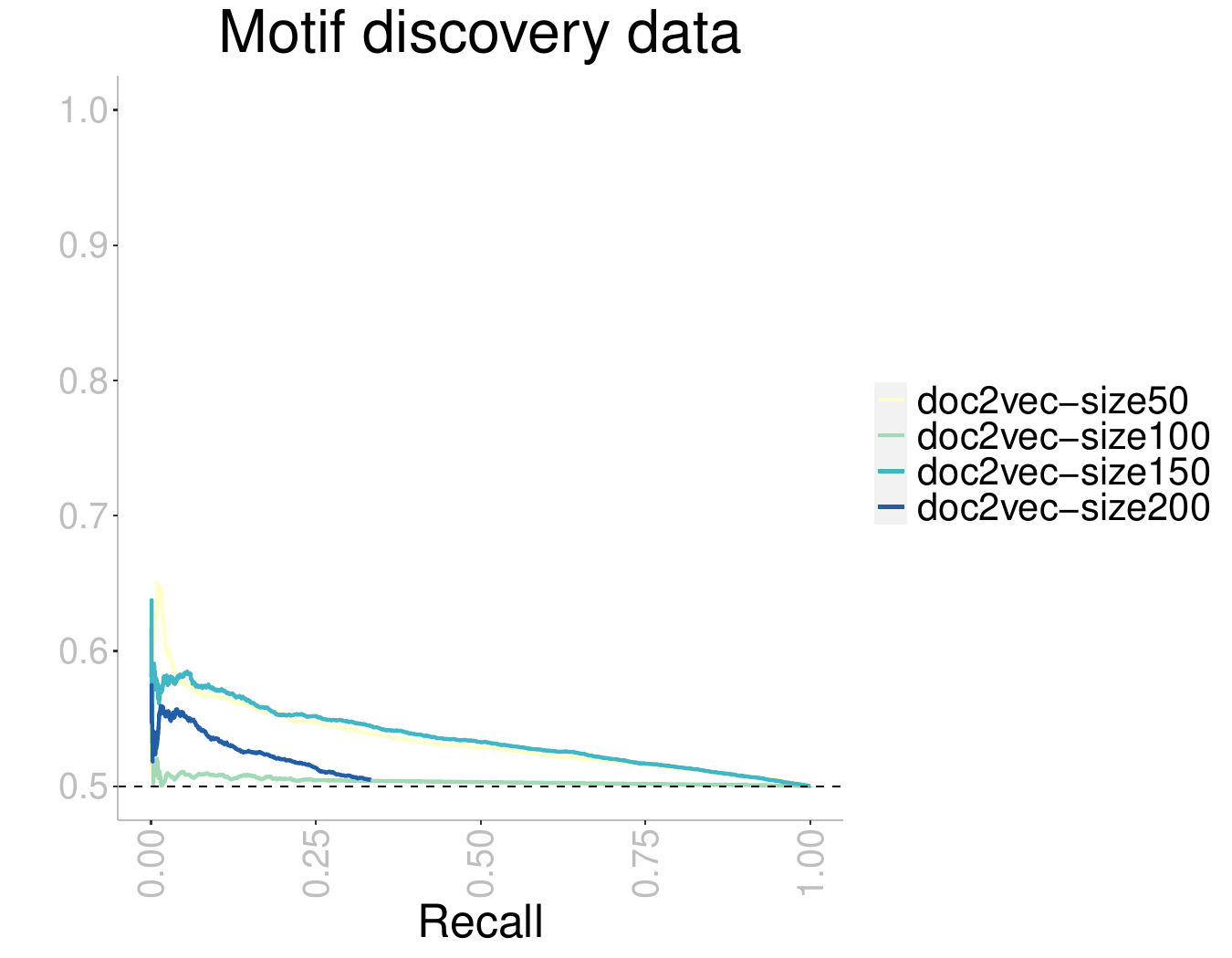}
    \caption{Precision-recall curves of doc2vec+NN model with four different embedding sizes (50, 100, 150, 200) on three datasets of increasing size. The higher the curve, the better performance with a horizontal dashed line to represent random prediction. An ideal precision-recall curve would cross the (1,1) point. The splice data has three panels since precision-recall curves assume binary classification and the splice dataset has three classes (0, 1, 2). Each panel corresponds to prediction one class vs the other two combined. The performance is poor in all cases.}
    \label{fig:embed-size-pr}
\end{figure}

\begin{figure}[h]
    \centering
    \includegraphics[scale=0.3]{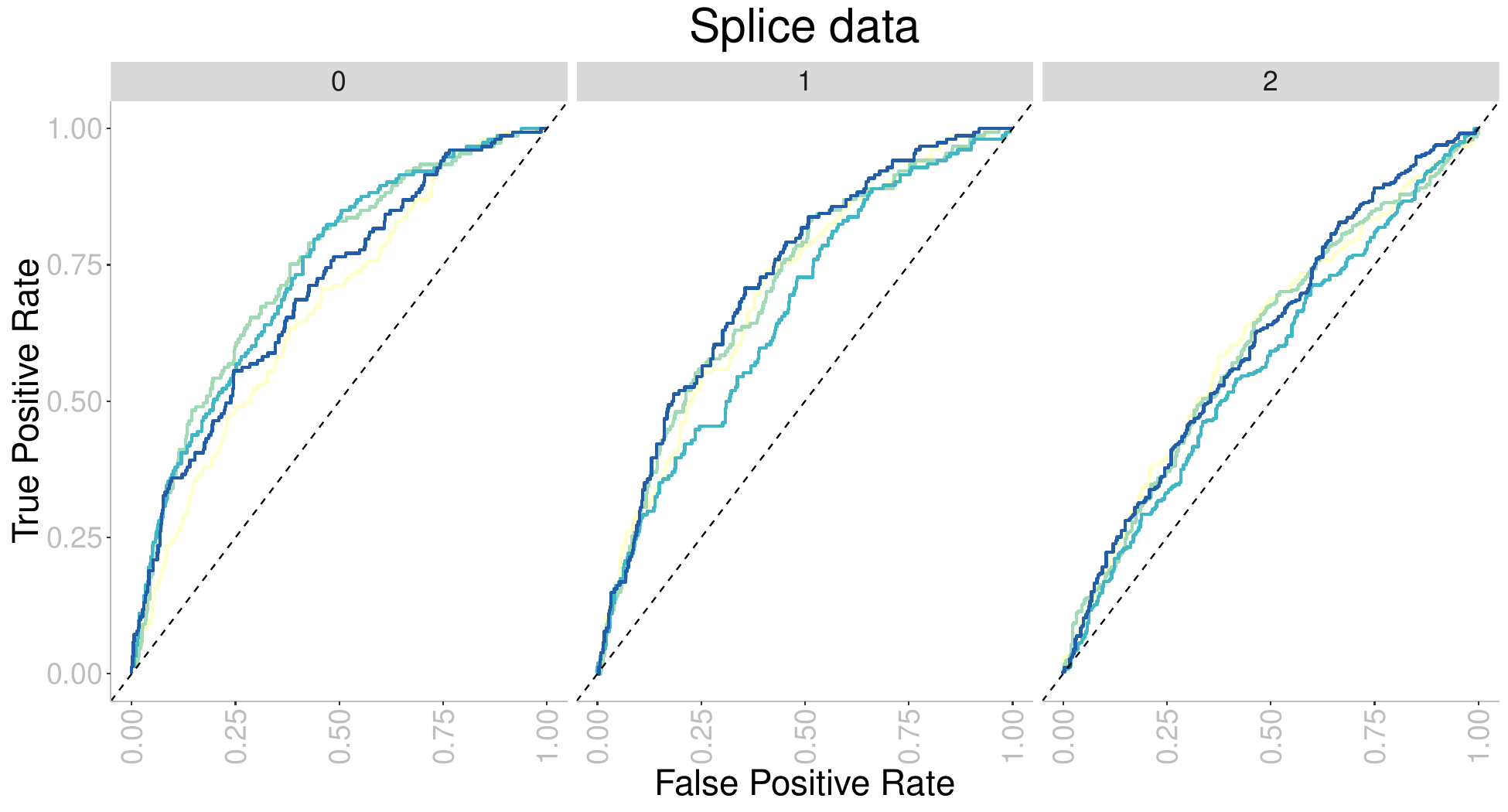}\\[0.25cm]
    \includegraphics[scale=0.25]{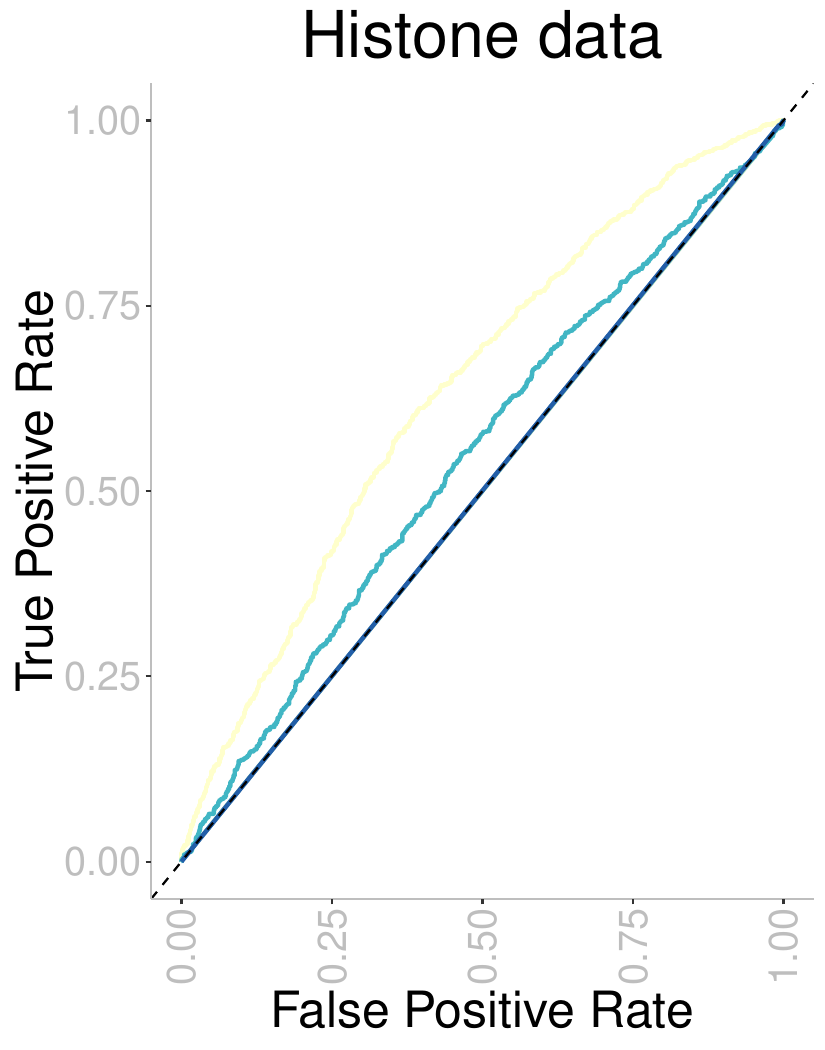}
    \includegraphics[scale=0.25]{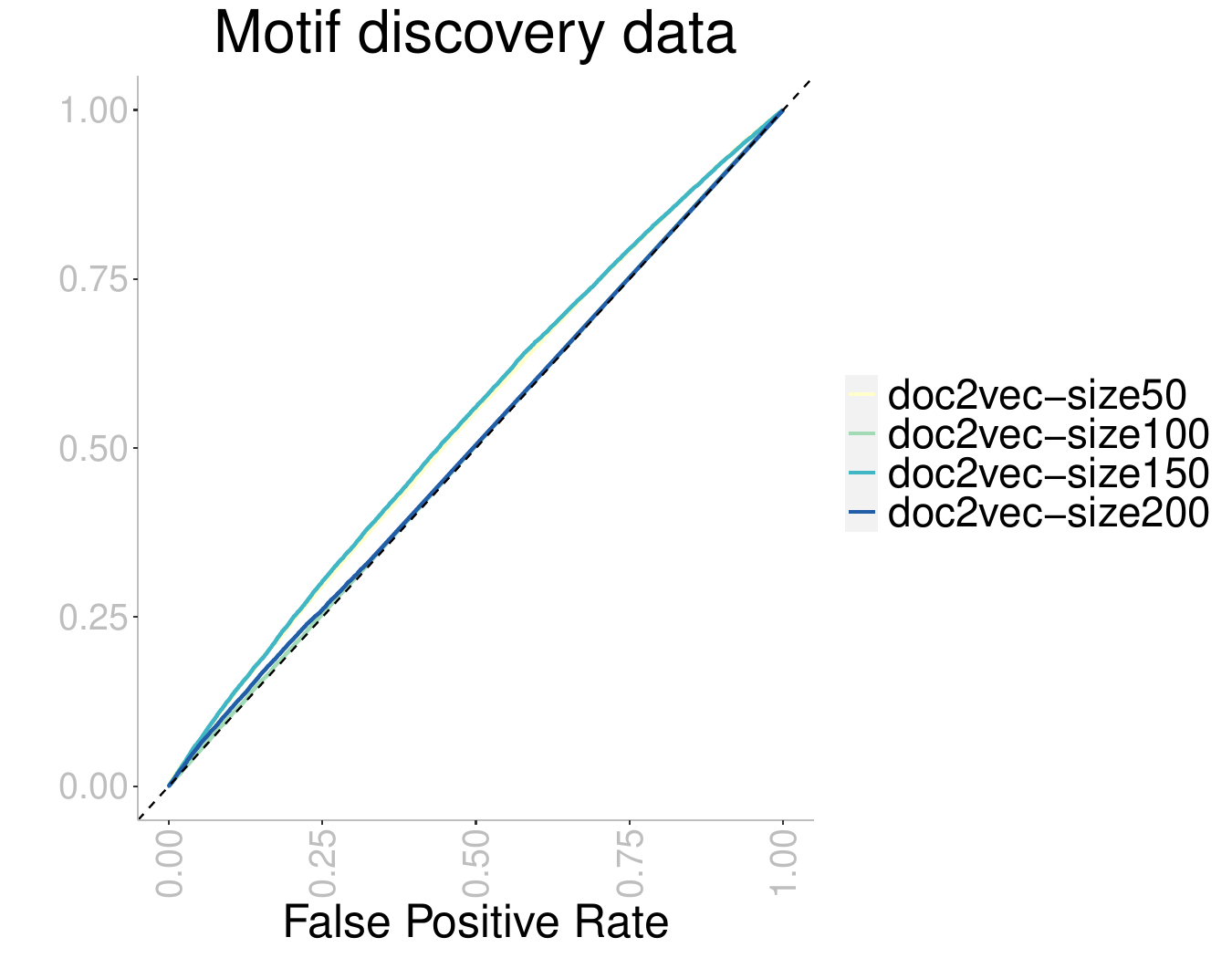}
    \caption{ROC curves of doc2vec+NN model with four different embedding sizes (50, 100, 150, 200) on three datasets of increasing size. The higher the curve, the better performance with a 45$^{\circ}$ dashed line to represent random prediction. The splice data has three panels since ROC curves assume binary classification and the splice dataset has three classes (0, 1, 2). Each panel corresponds to prediction one class vs the other two combined. The performance is poor in all cases.}
    \label{fig:embed-size-roc}
\end{figure}

\begin{figure}[h]
    \centering
    \includegraphics[scale=0.175]{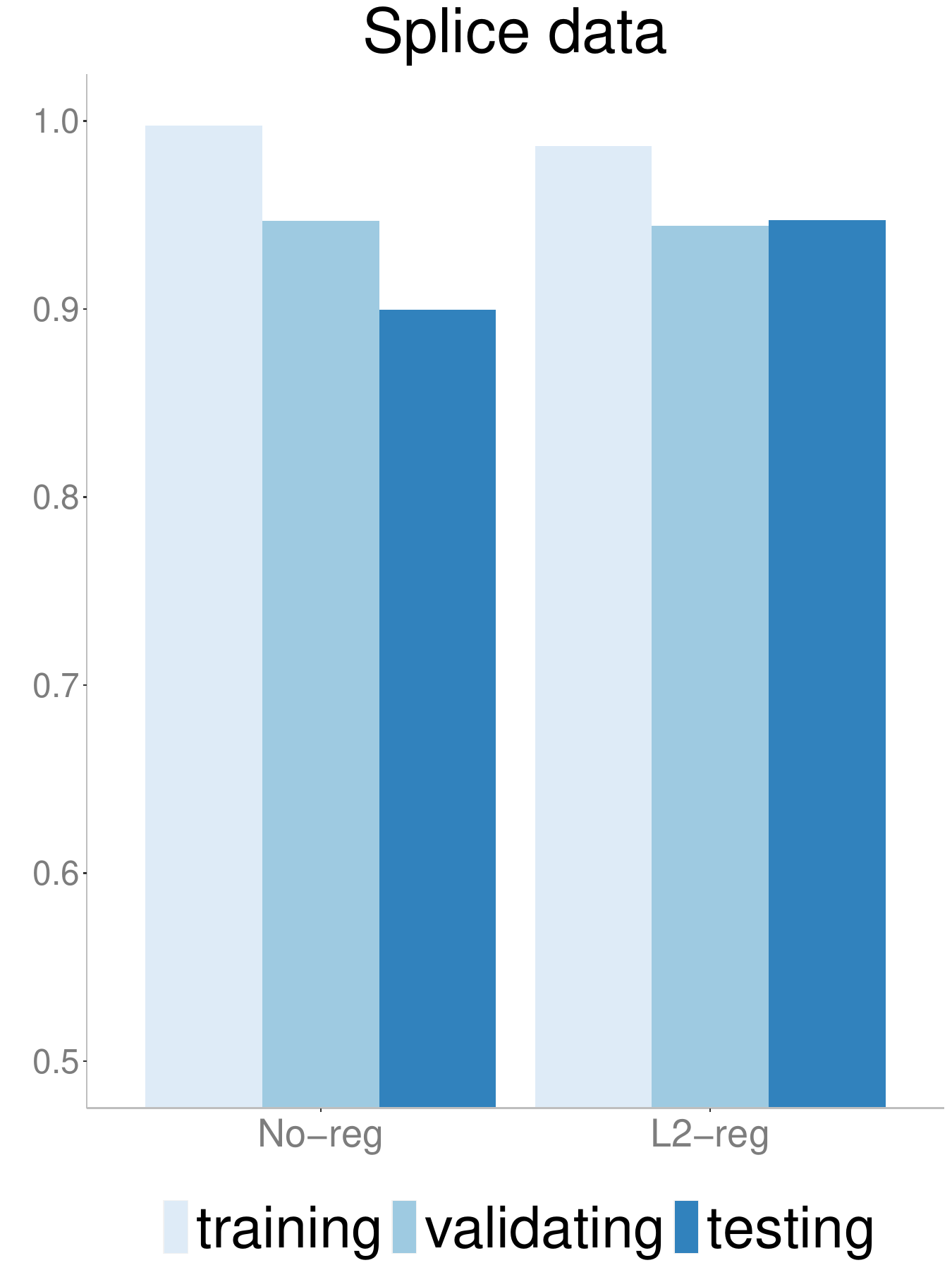}
    \includegraphics[scale=0.175]{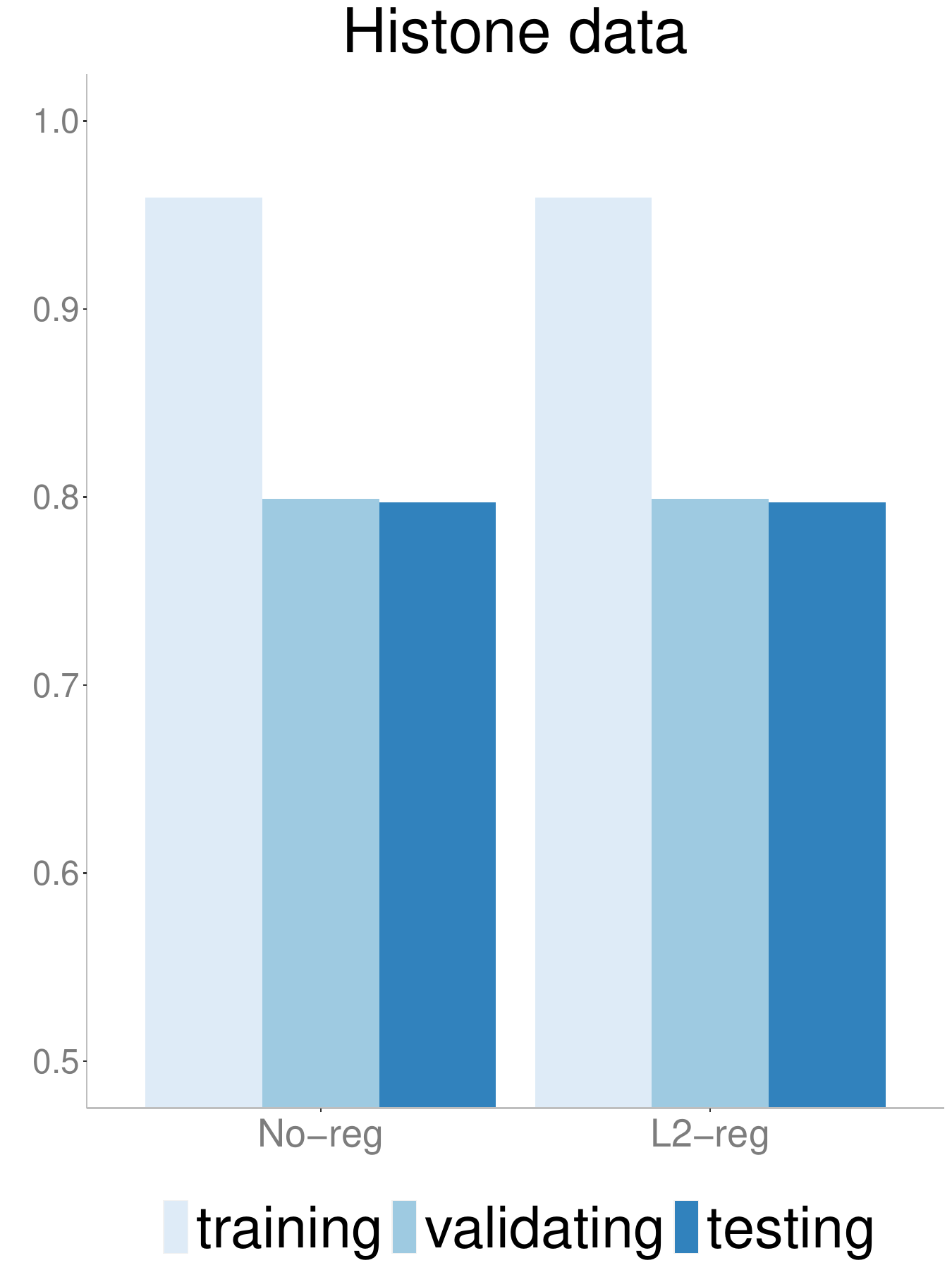}
    \includegraphics[scale=0.175]{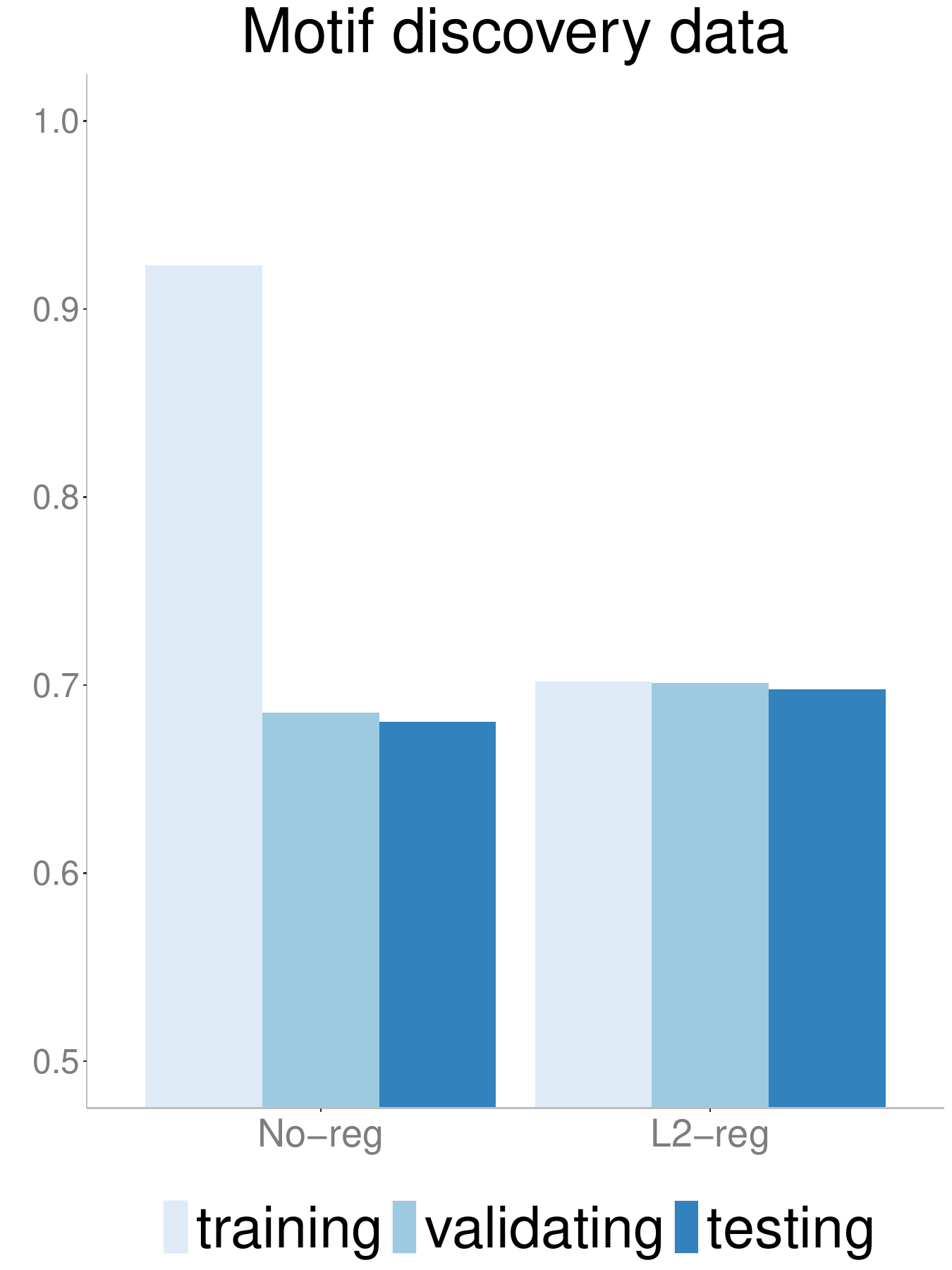}
    \caption{Accuracy of CNN-Zeng models with four 2D layer on three datasets of increasing size with and without regularization. The L2 regularization indeed reduces overfitting for the motif discovery data, but not for the histone data.}
    \label{fig:l2-acc}
\end{figure}

\begin{figure}[h]
    \centering
    \includegraphics[scale=0.3]{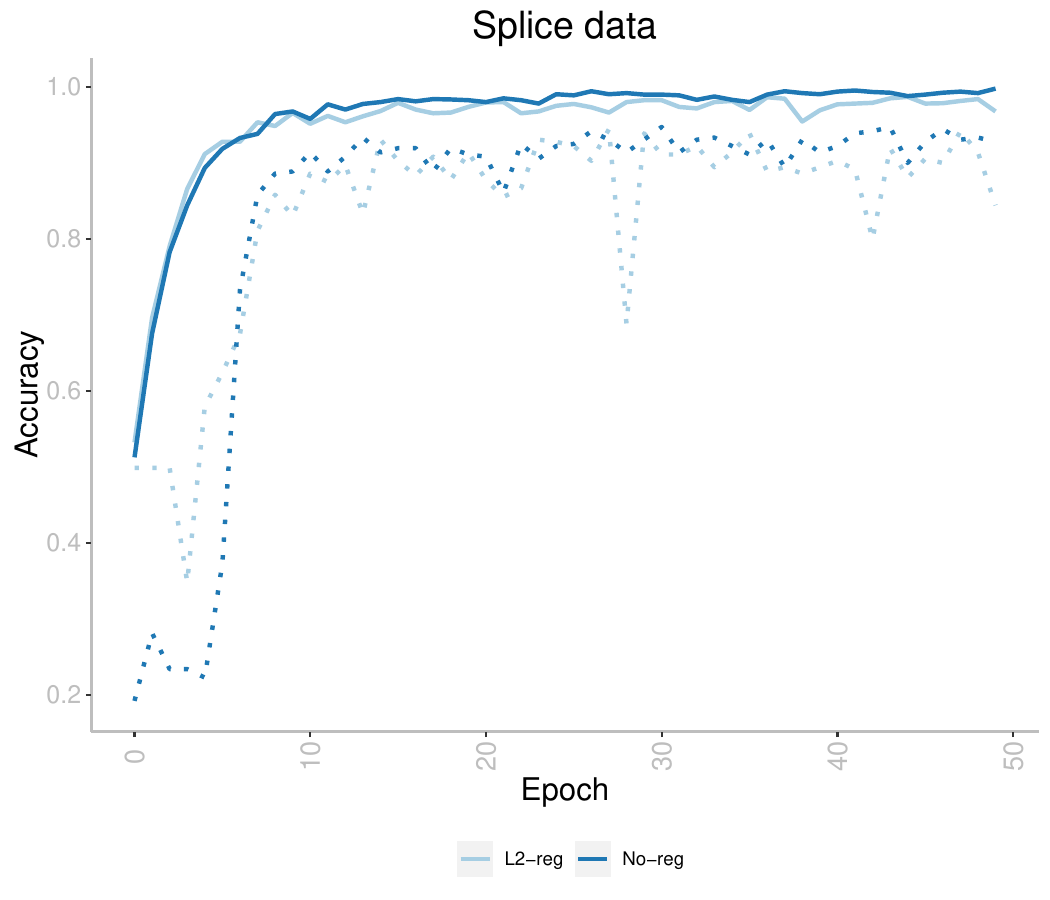}
    \includegraphics[scale=0.3]{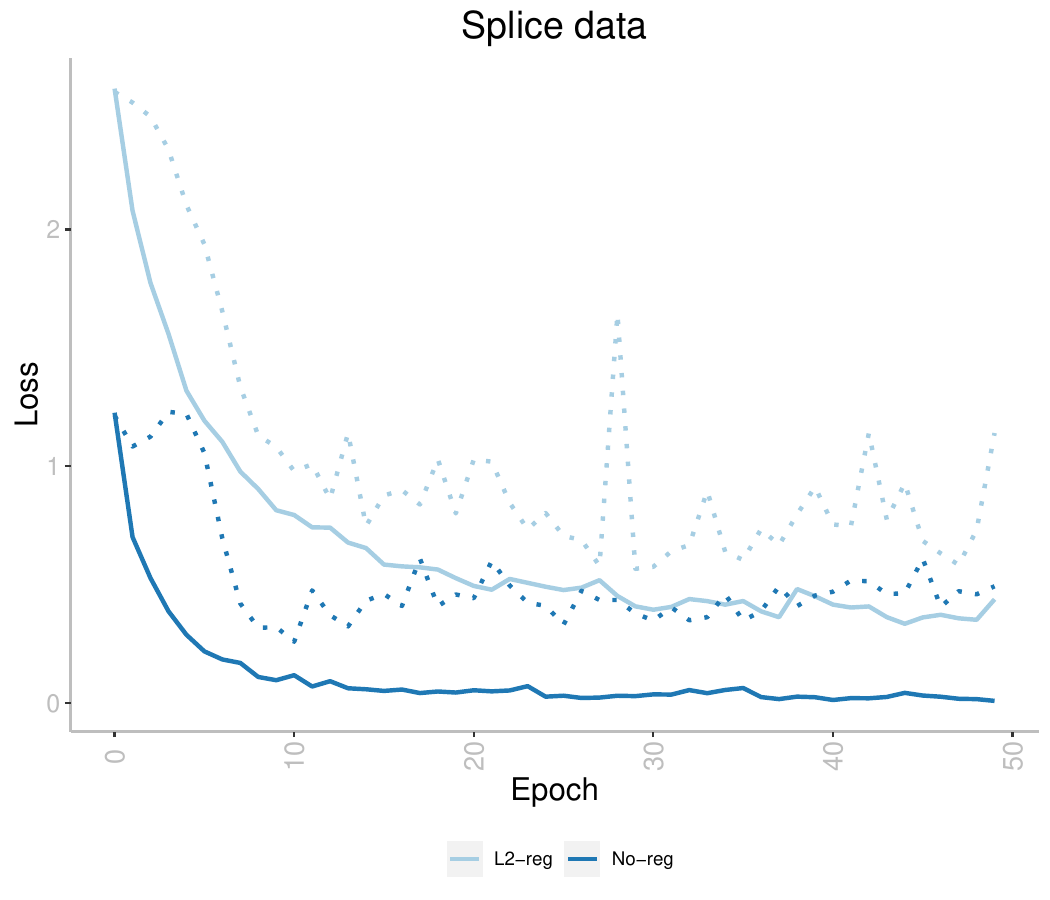}
    \includegraphics[scale=0.3]{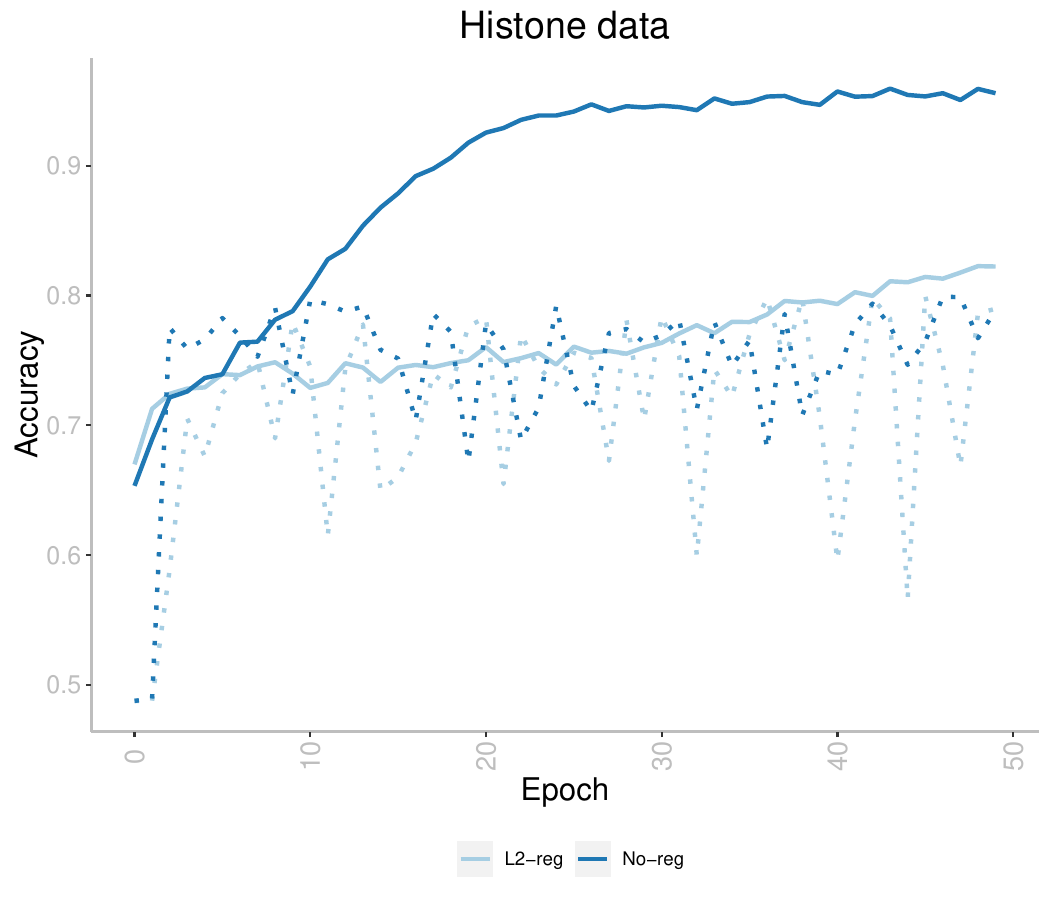}
    \includegraphics[scale=0.3]{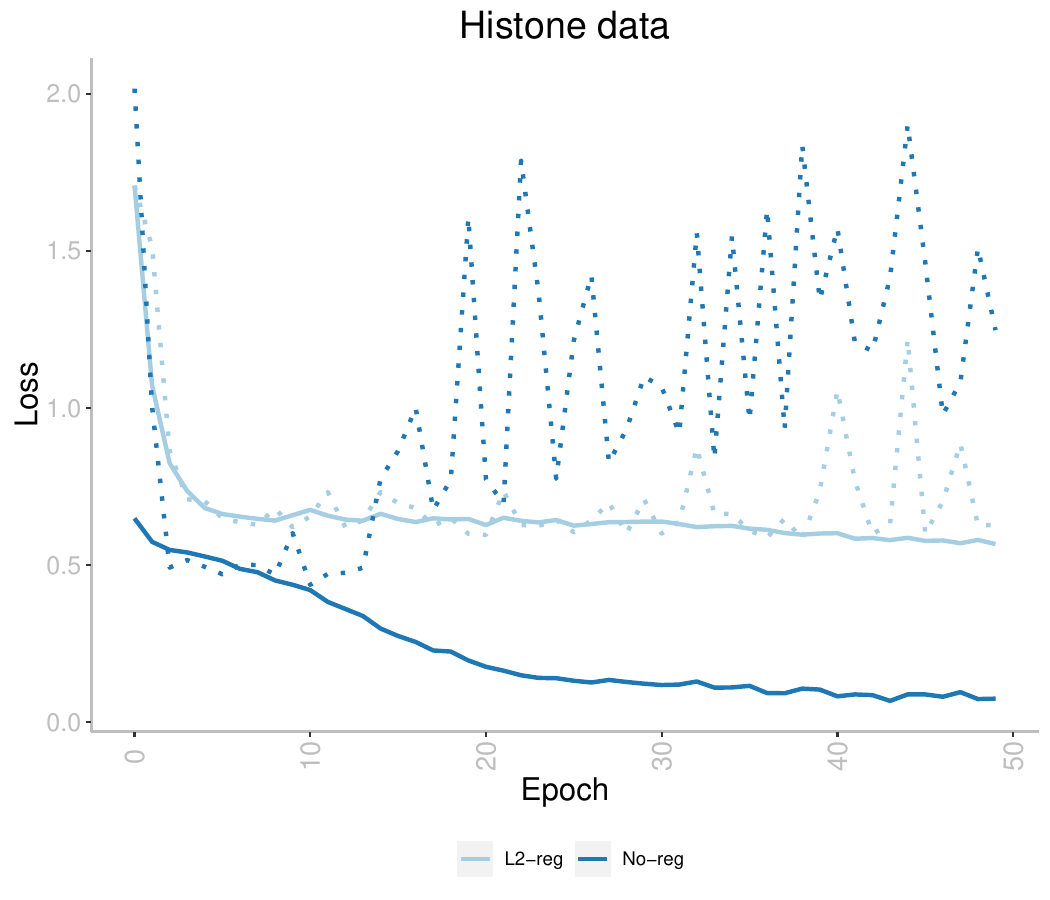}
    \includegraphics[scale=0.3]{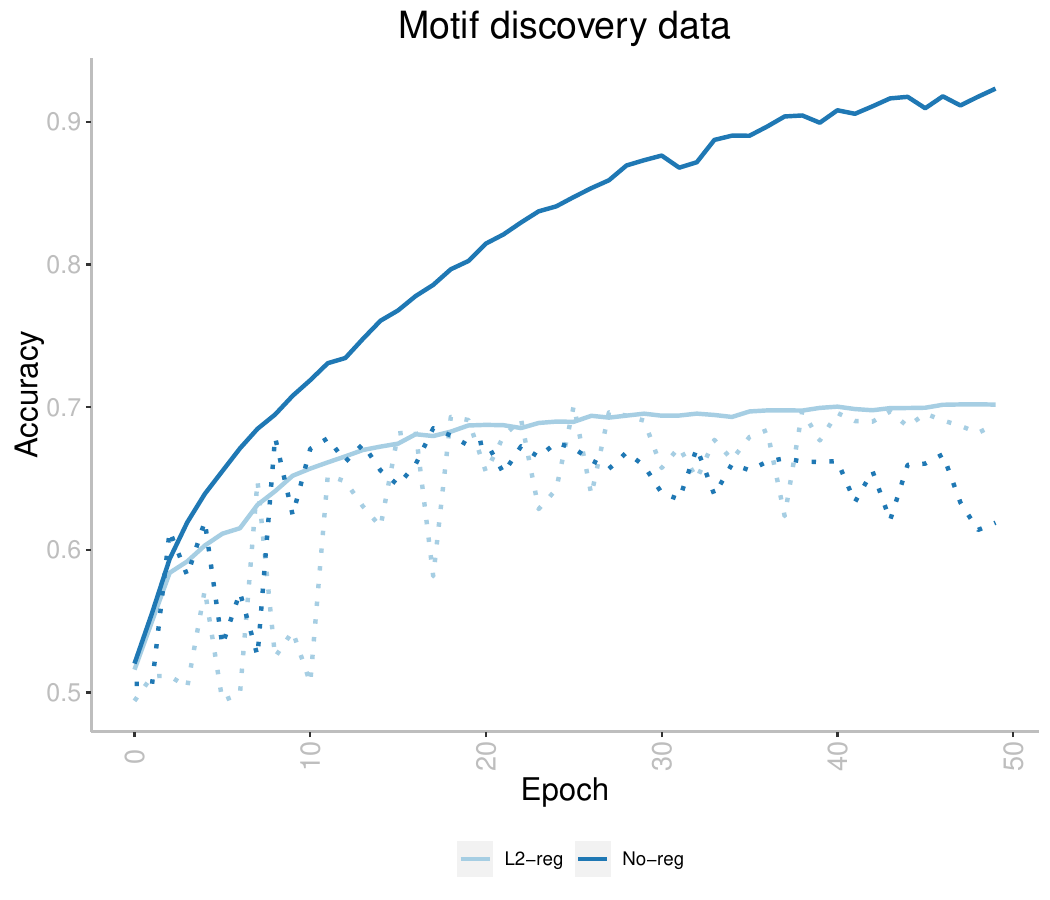}
    \includegraphics[scale=0.3]{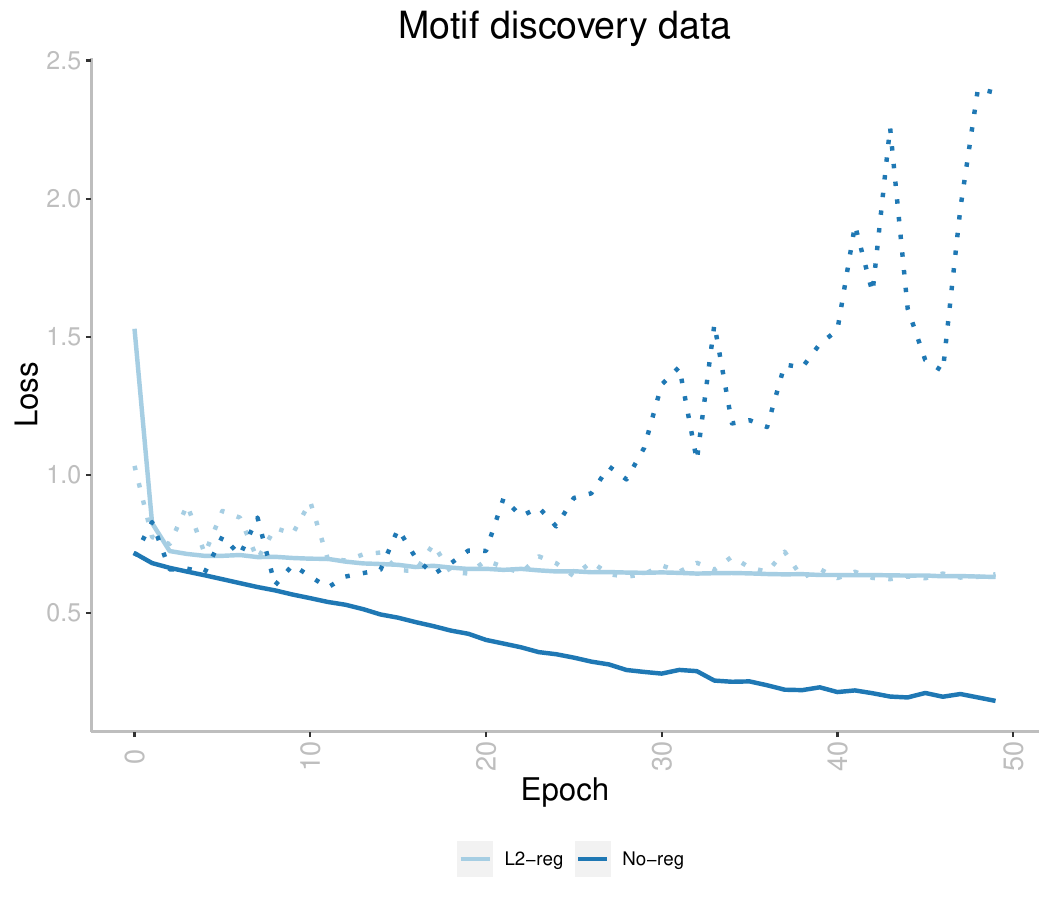}
    \caption{Learning dynamics for CNN-Zeng models with four 2D layer on three datasets of increasing size with and without regularization. Solid lines correspond to training accuracy/loss and dashed lines correspond to validating accuracy/loss. Colors represent no regularization or L2 regularization.}
    \label{fig:l2-dynamics}
\end{figure}

\begin{figure}[h]
    \centering
    \includegraphics[scale=0.3]{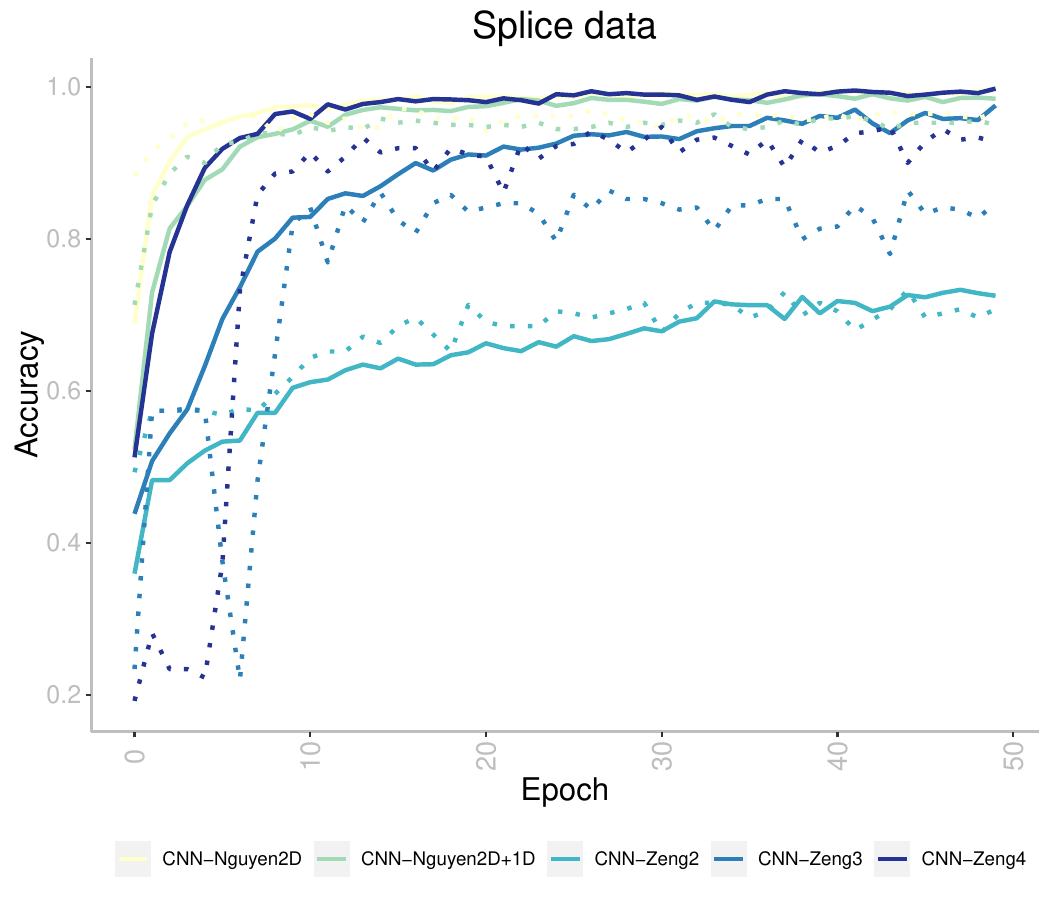}
    \includegraphics[scale=0.3]{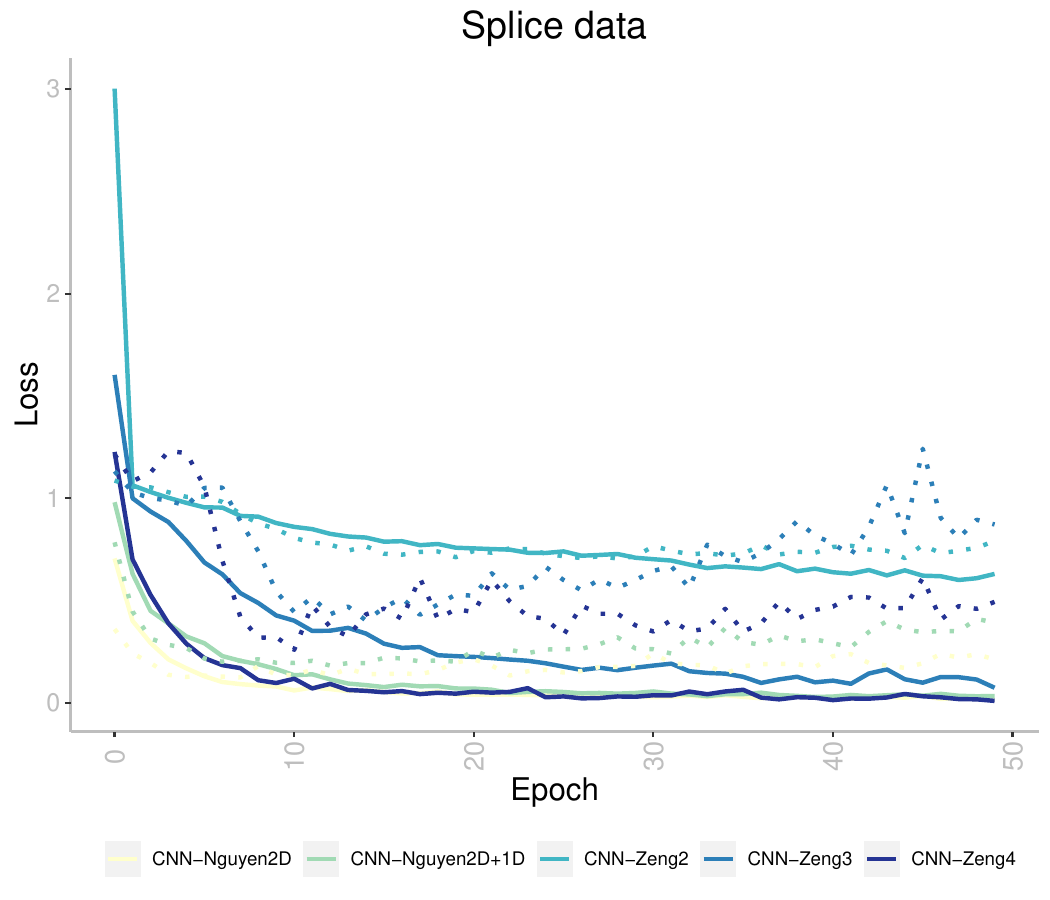}
    \includegraphics[scale=0.3]{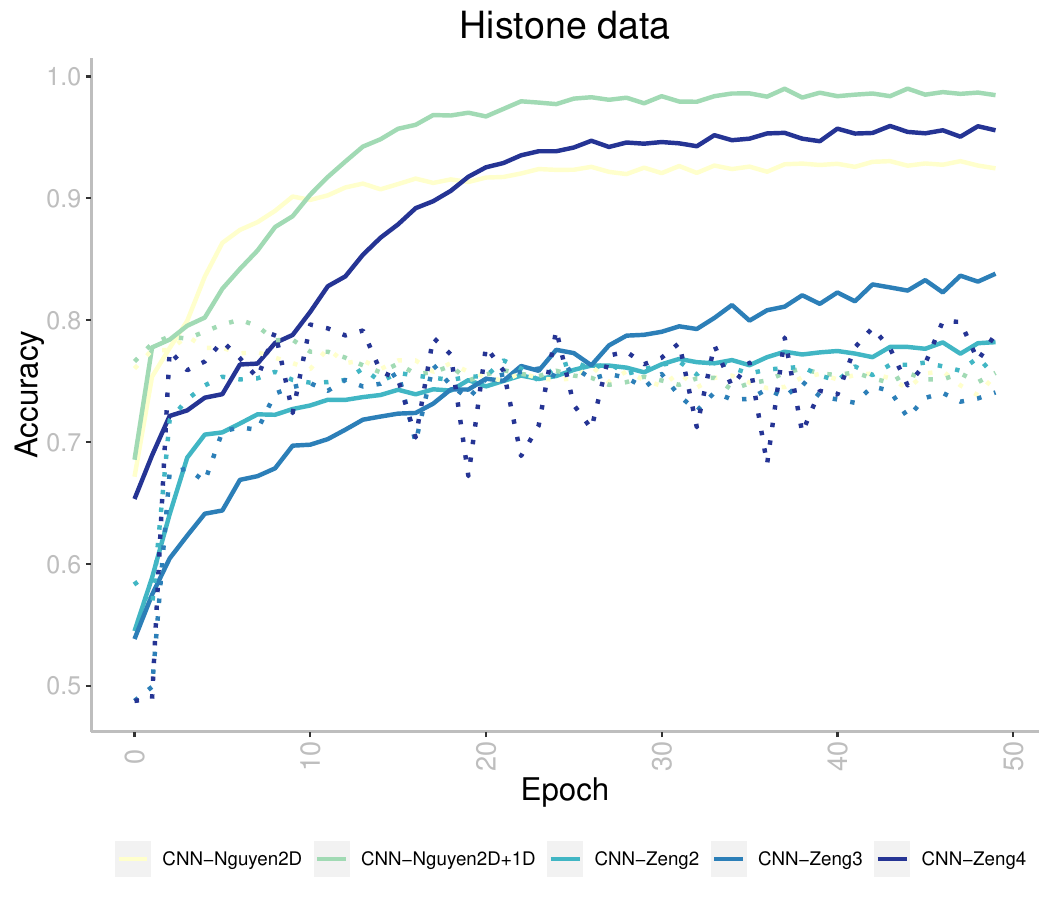}
    \includegraphics[scale=0.3]{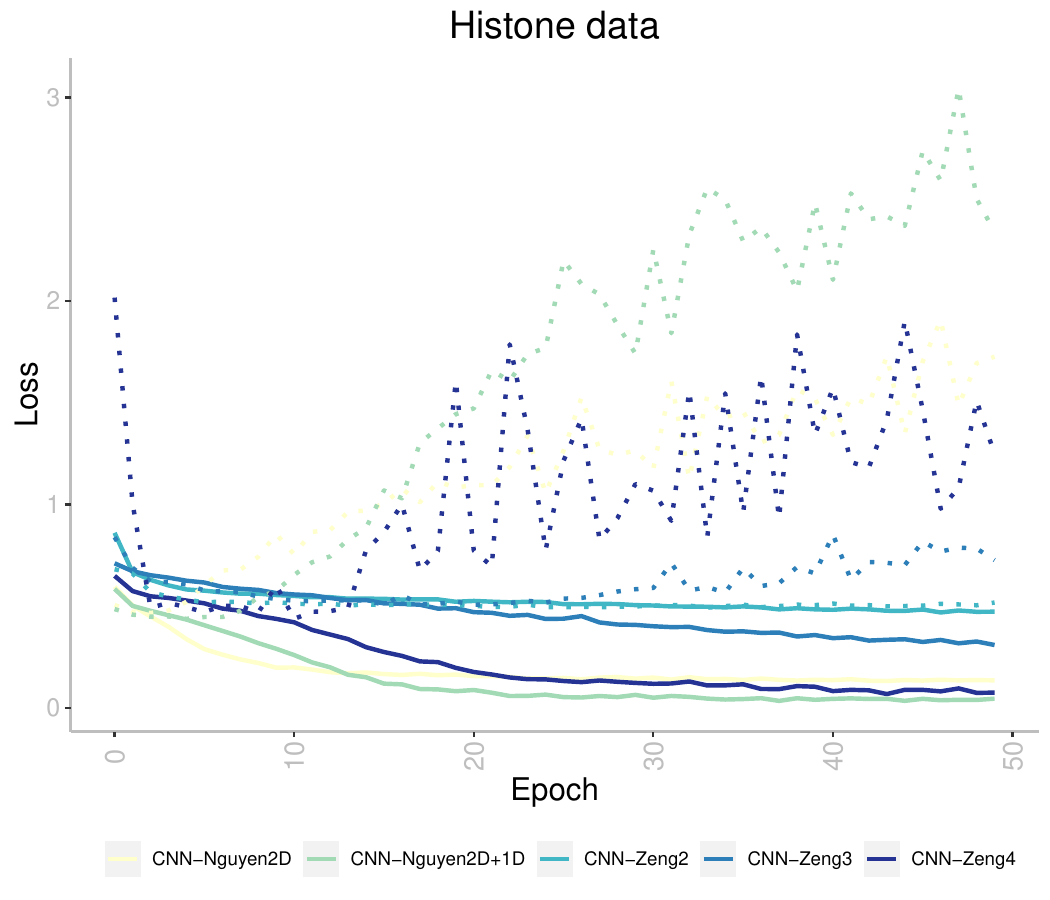}
    \includegraphics[scale=0.3]{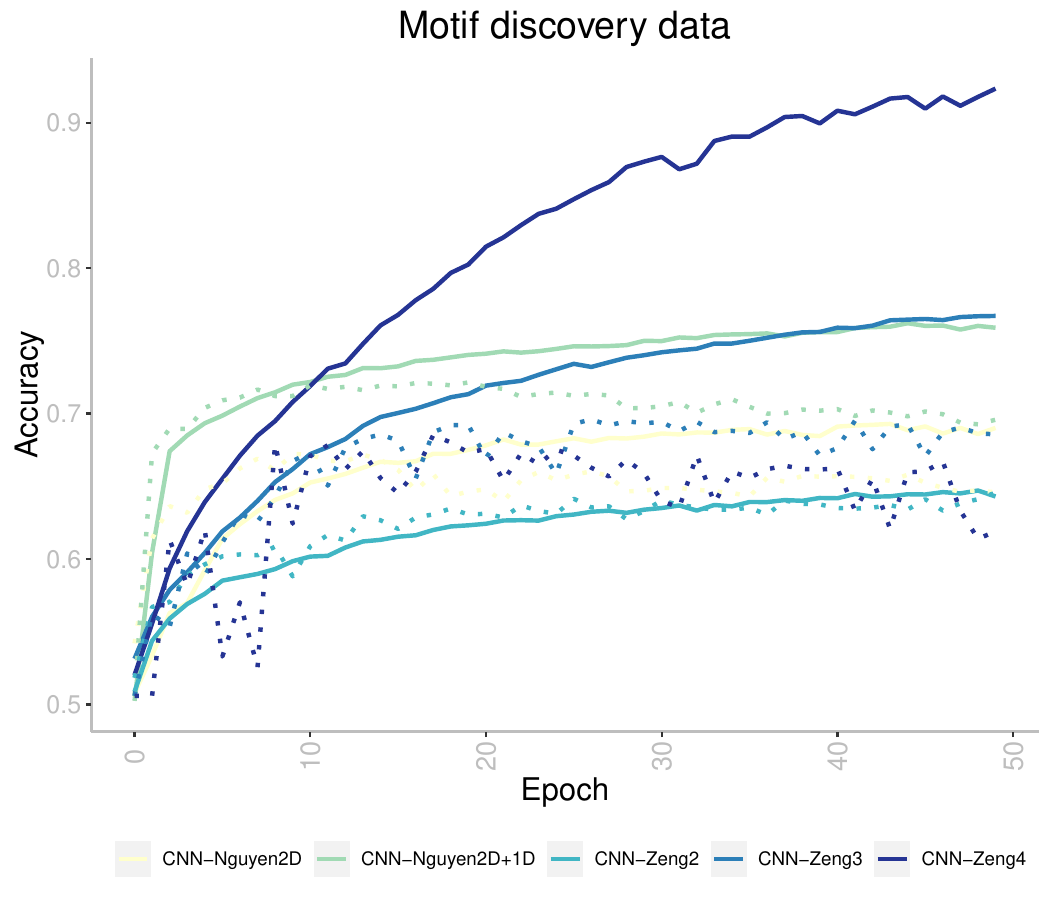}
    \includegraphics[scale=0.3]{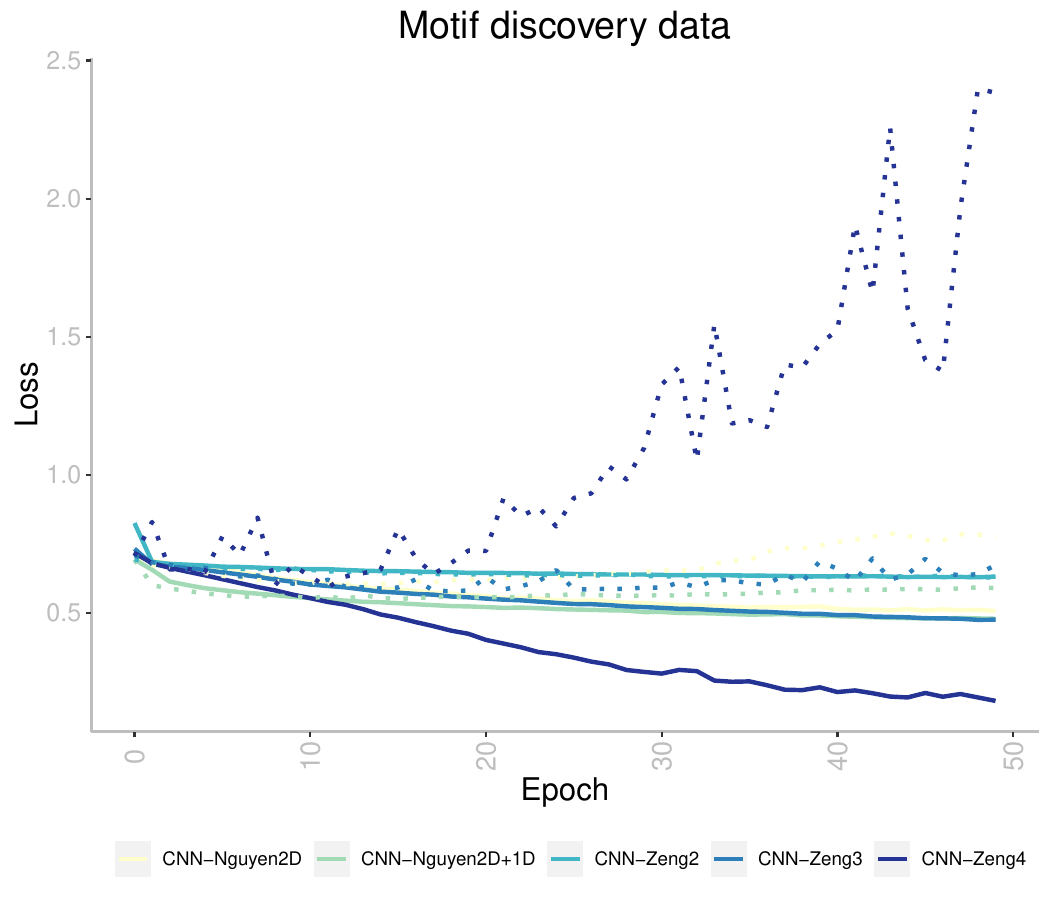}
    \caption{Learning dynamics for CNN models on different datasets. Solid lines correspond to training accuracy/loss and dashed lines correspond to validating accuracy/loss. Colors represent different models.}
    \label{fig:dimension-dynamics}
\end{figure}

\begin{figure}[h]
    \centering
    \includegraphics[scale=0.3]{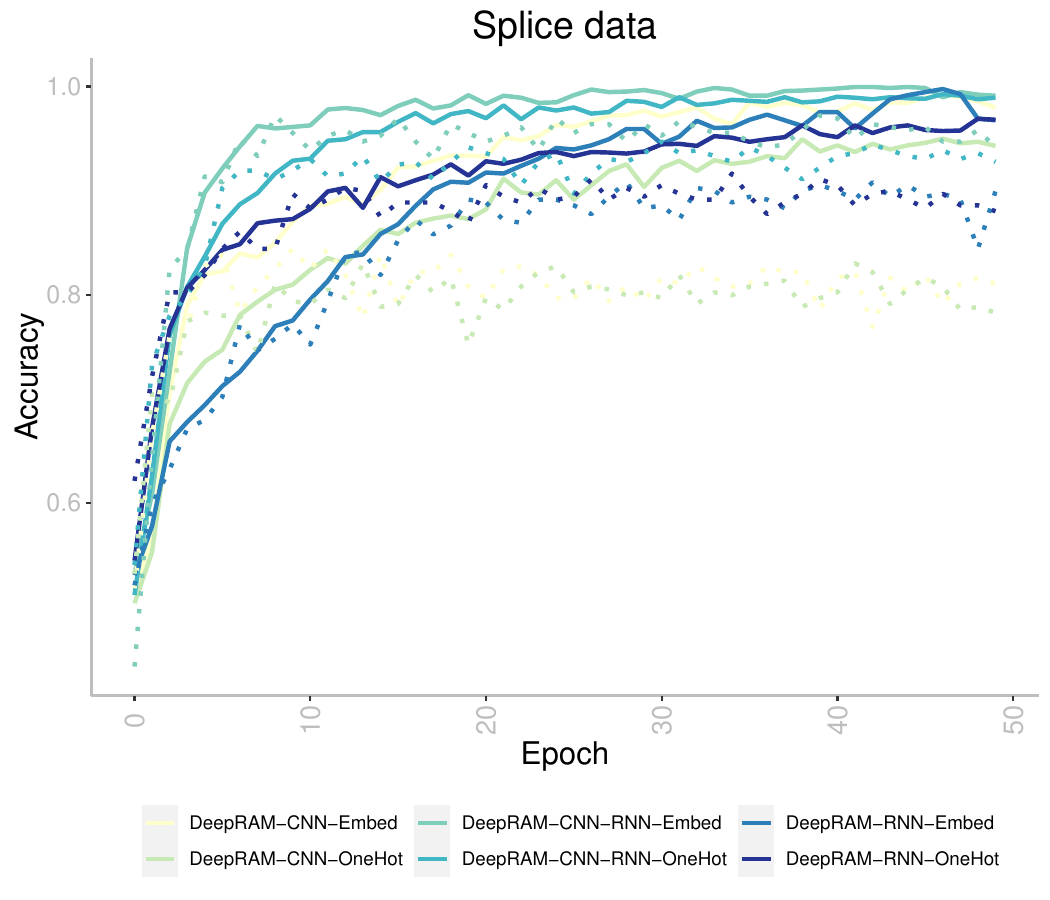}
    \includegraphics[scale=0.3]{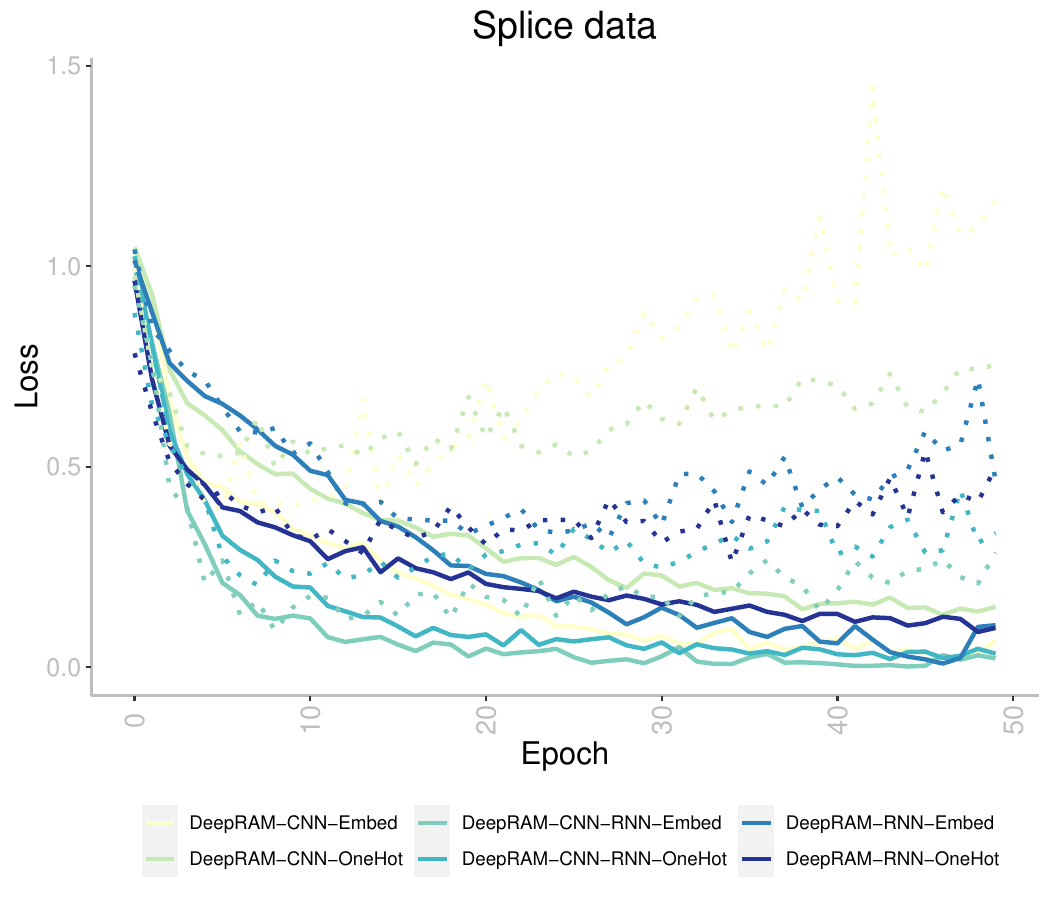}
    \includegraphics[scale=0.3]{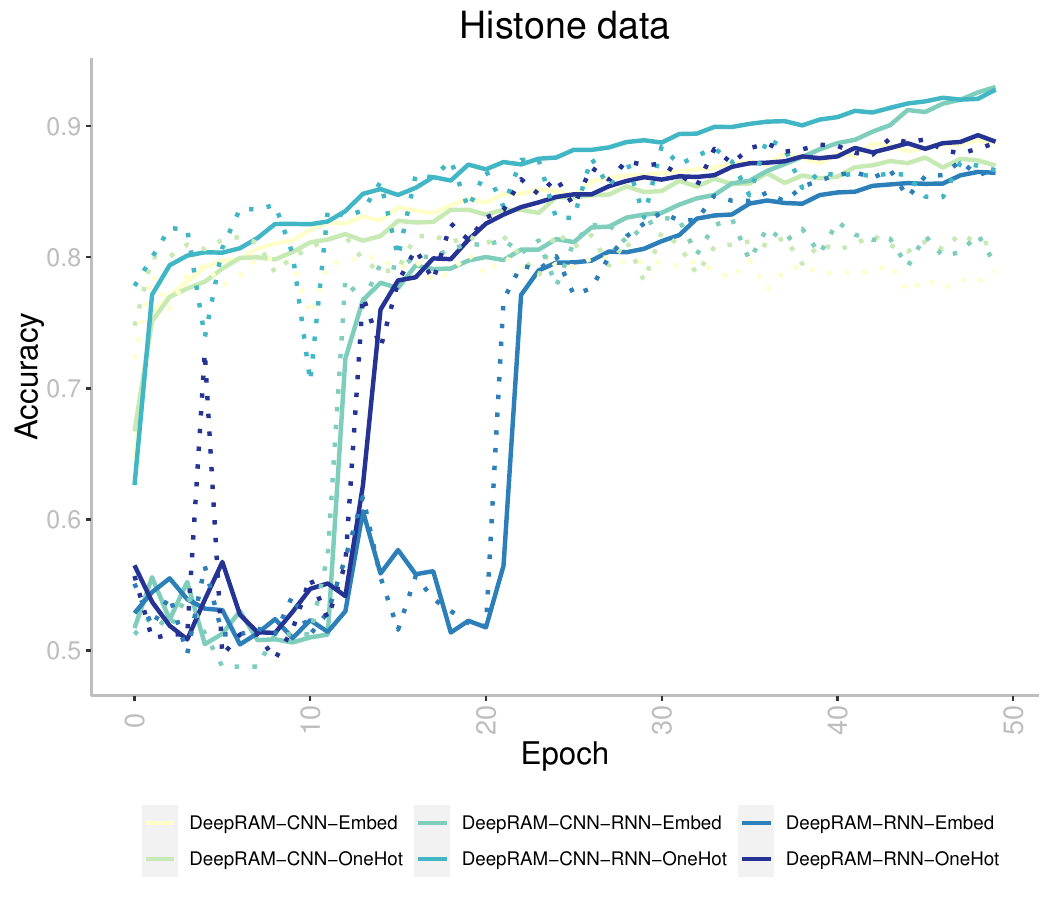}
    \includegraphics[scale=0.3]{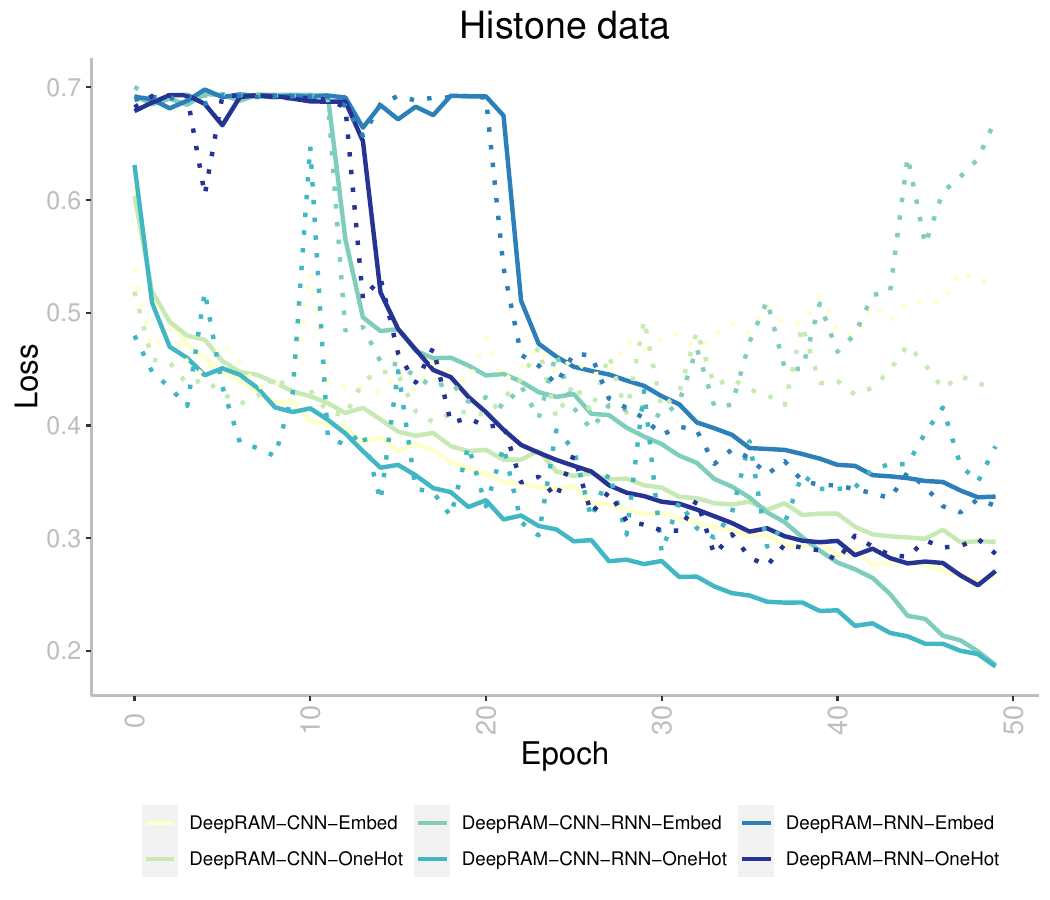}
    \includegraphics[scale=0.3]{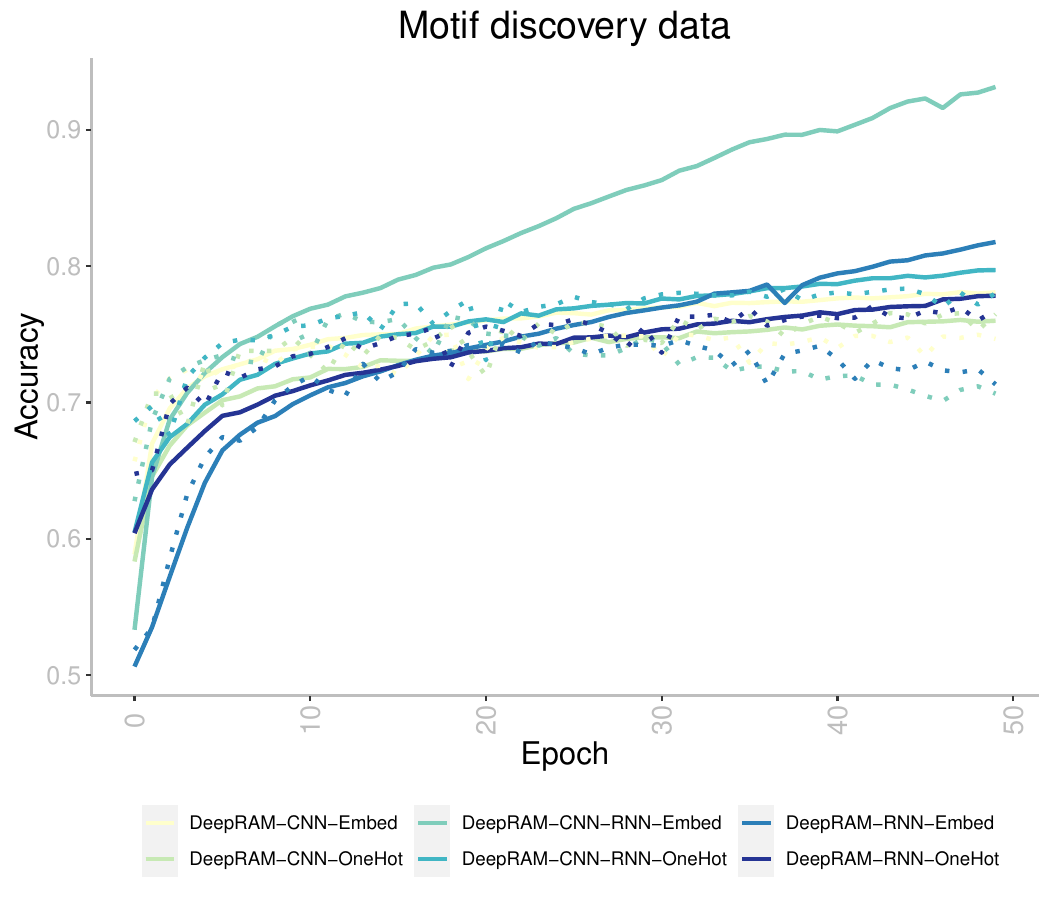}
    \includegraphics[scale=0.3]{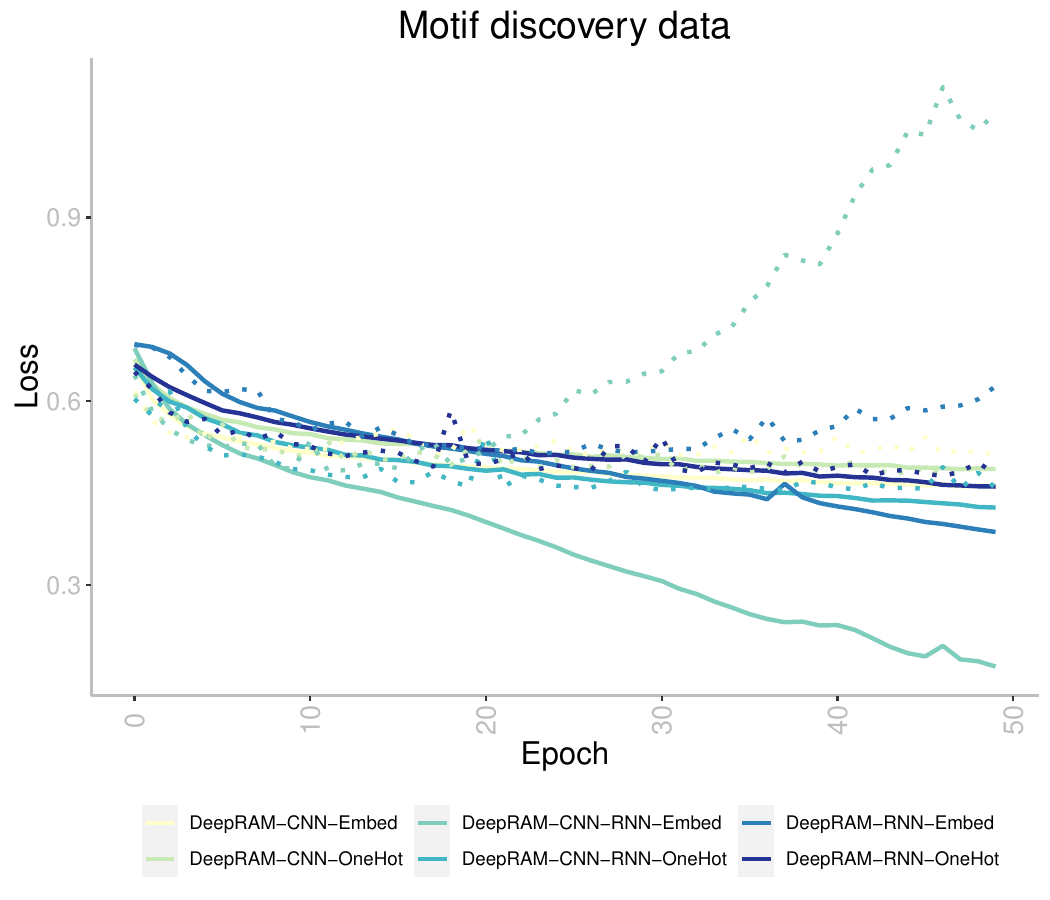}
    \caption{Learning dynamics for DeepRAM models on different datasets and different data embedding. Solid lines correspond to training accuracy/loss and dashed lines correspond to validating accuracy/loss. Colors represent different models and different data embedding.}
    \label{fig:embedding-dynamics}
\end{figure}

\begin{figure}[h]
    \centering
    \includegraphics[scale=0.3]{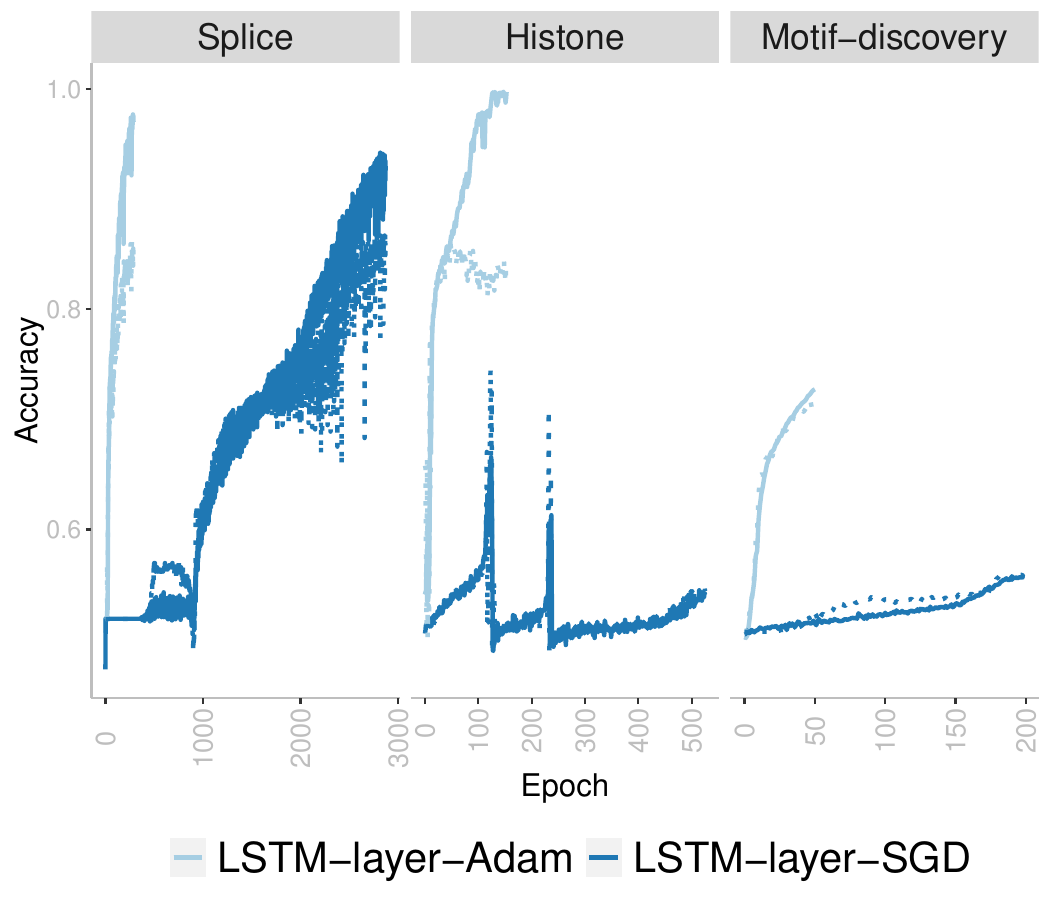}
    \includegraphics[scale=0.3]{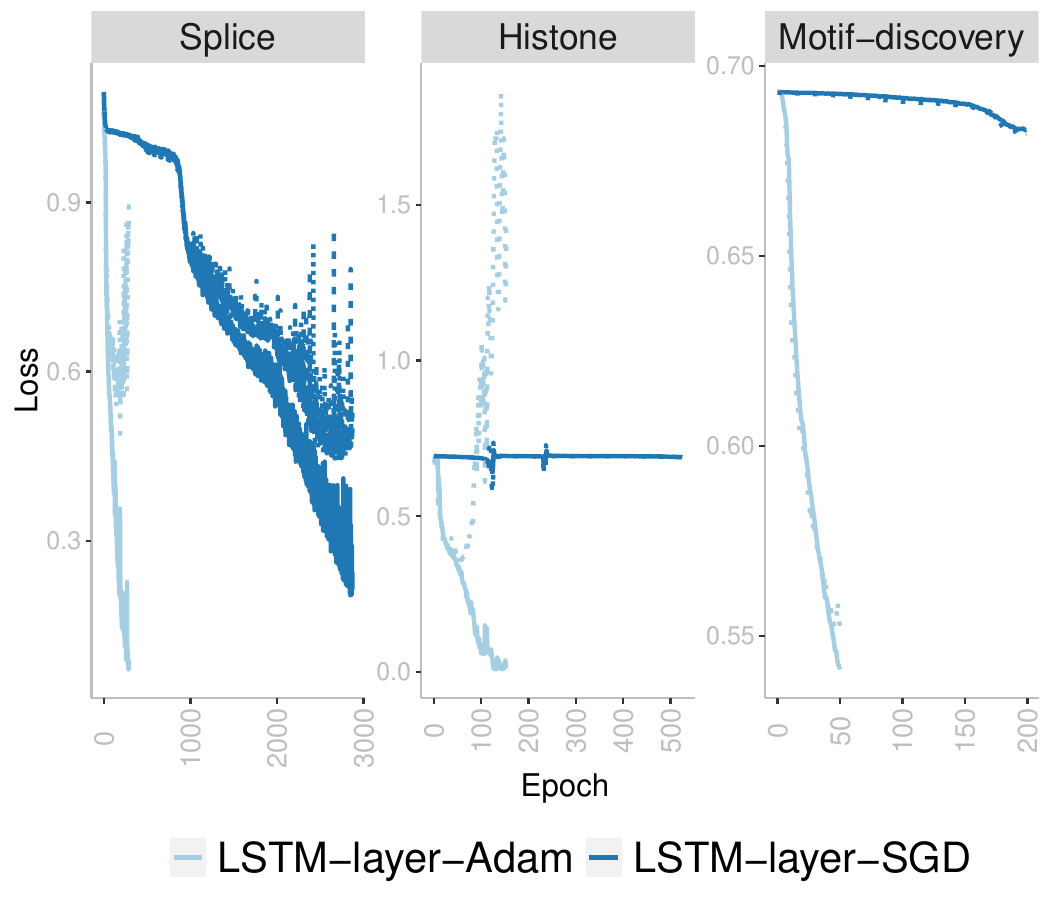}
    \caption{Learning dynamics for LSTM-layer model on different datasets and different optimizer. Solid lines correspond to training accuracy/loss and dashed lines correspond to validating accuracy/loss. Colors represent different optimizer.}
    \label{fig:optimizer-dynamics}
\end{figure}

\begin{figure}[h]
    \centering
    \includegraphics[scale=0.3]{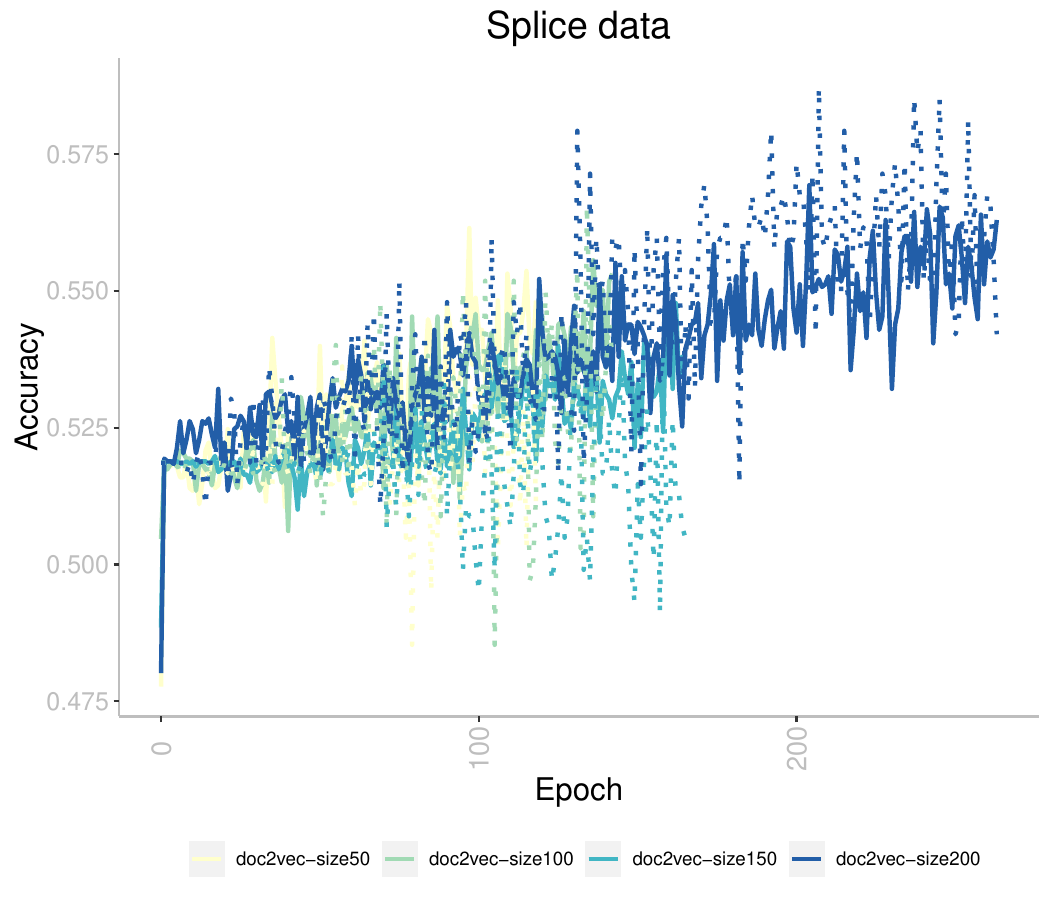}
    \includegraphics[scale=0.3]{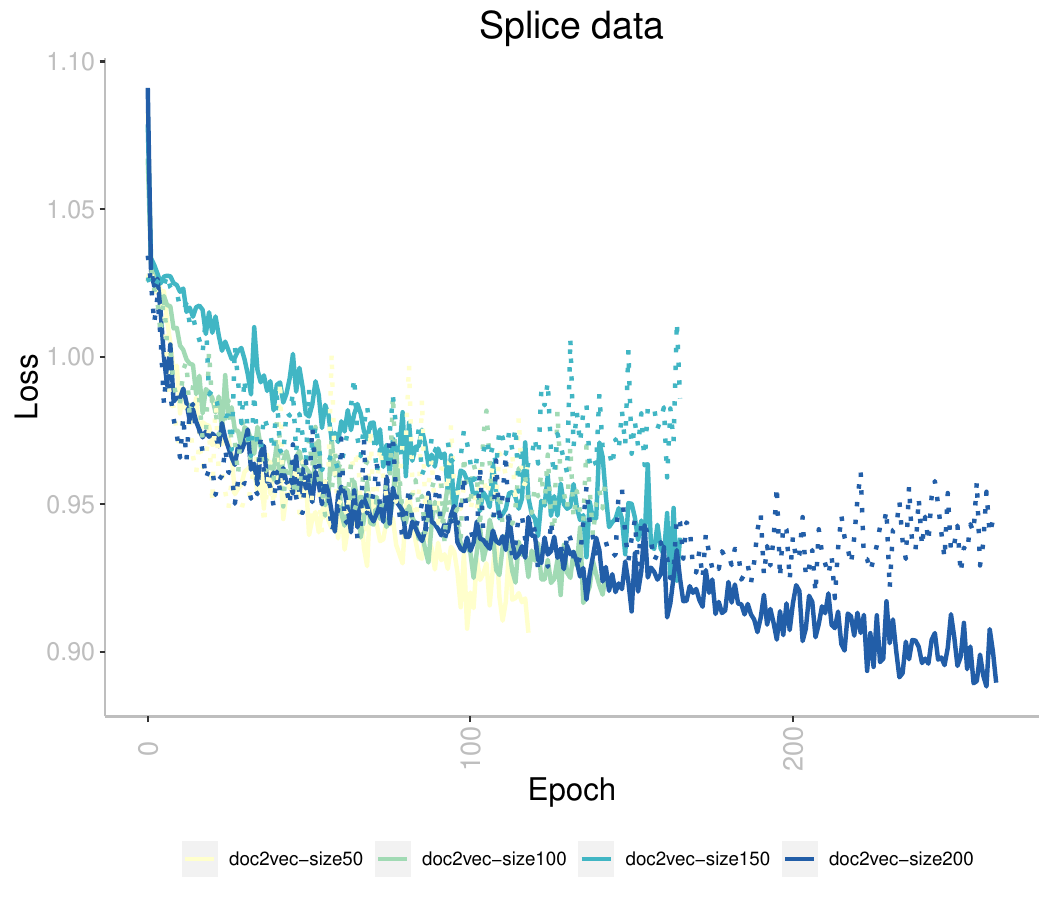}
    \includegraphics[scale=0.3]{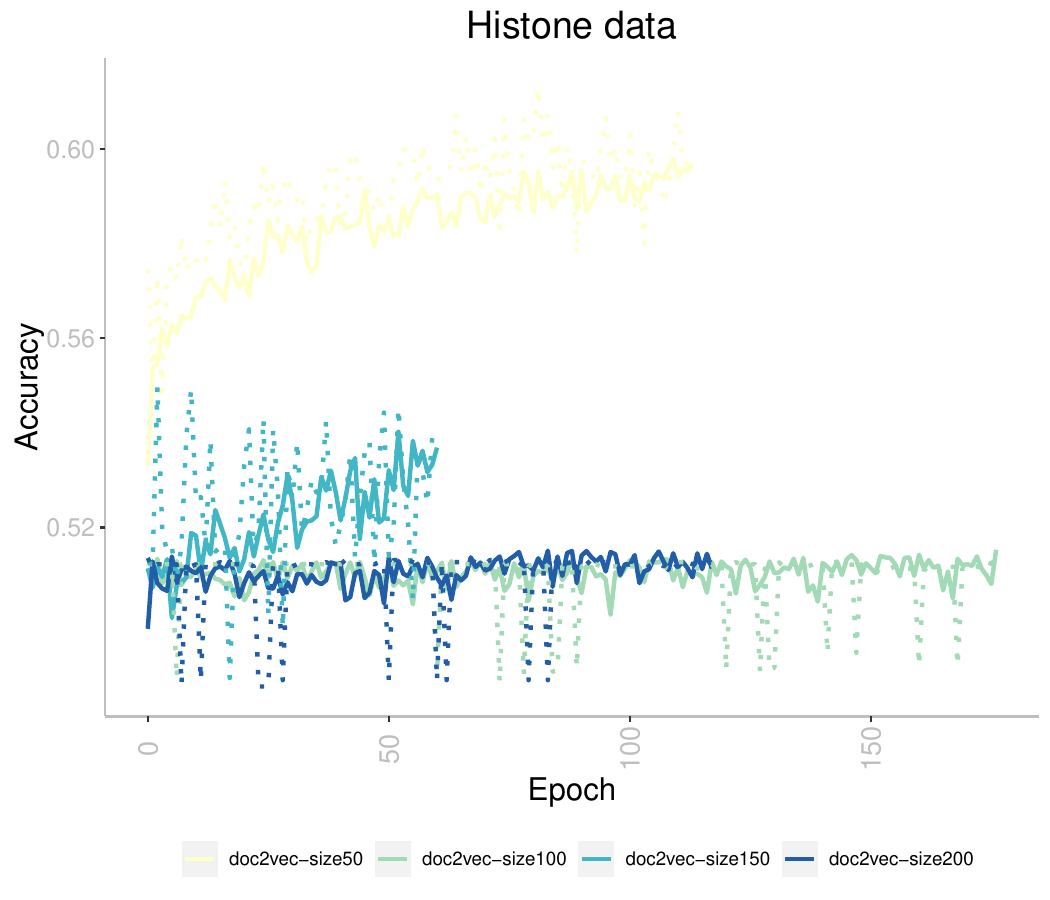}
    \includegraphics[scale=0.3]{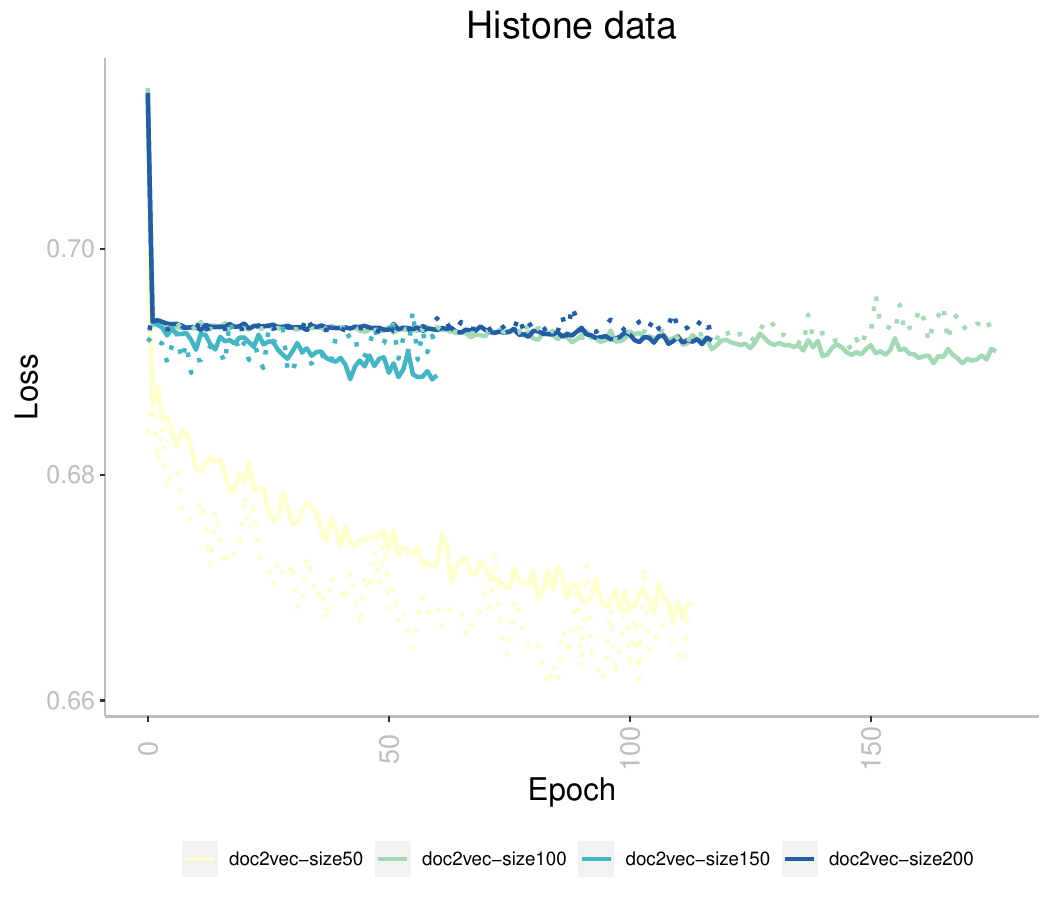}
    \includegraphics[scale=0.3]{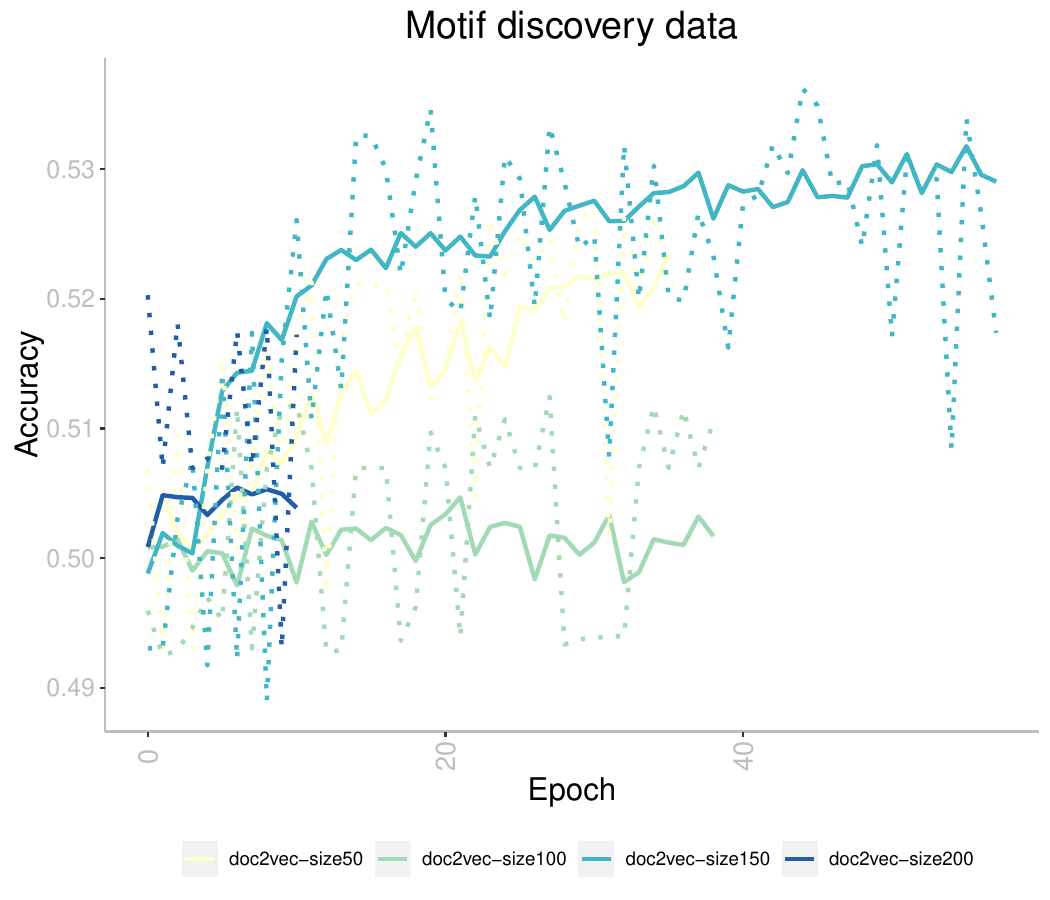}
    \includegraphics[scale=0.3]{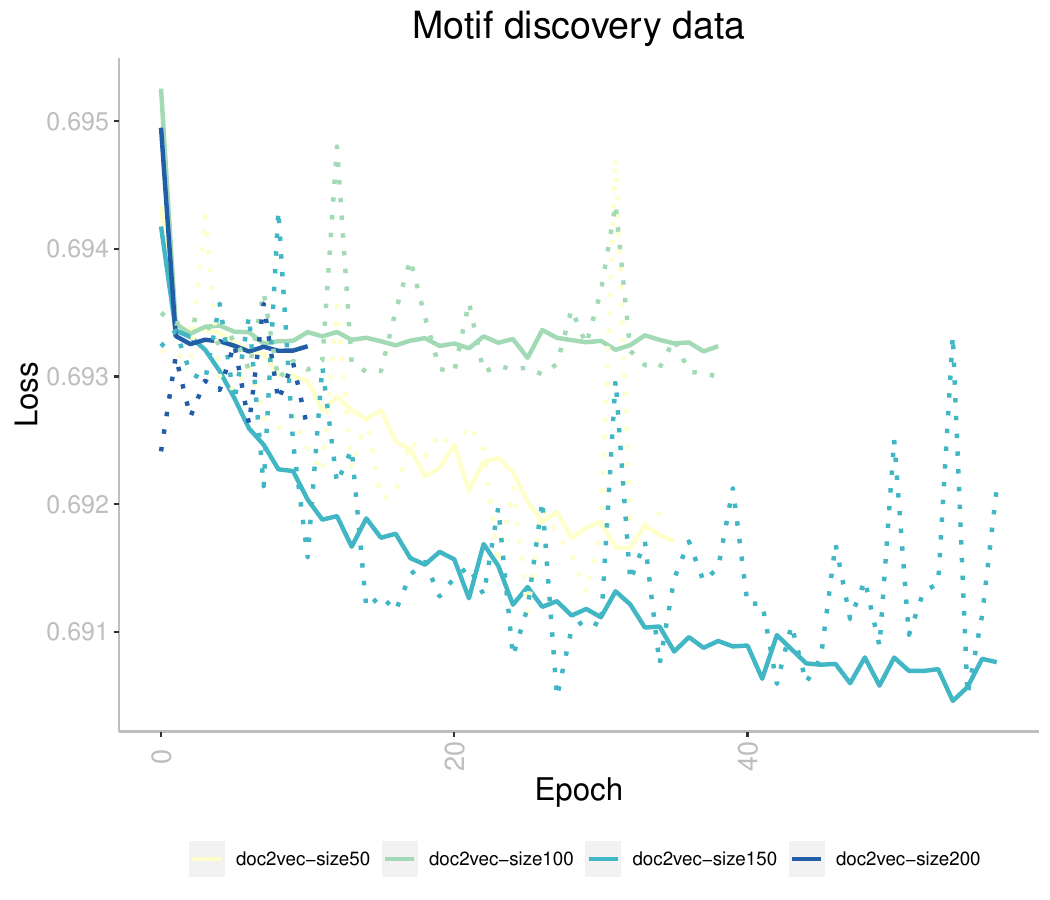}
    \caption{Learning dynamics for doc2vec+NN model on different datasets and different embedding sizes. Solid lines correspond to training accuracy/loss and dashed lines correspond to validating accuracy/loss. Colors represent different embedding sizes.}
    \label{fig:embed-size-dynamics}
\end{figure}

\begin{figure}[h]
    \centering
    \includegraphics[scale=0.3]{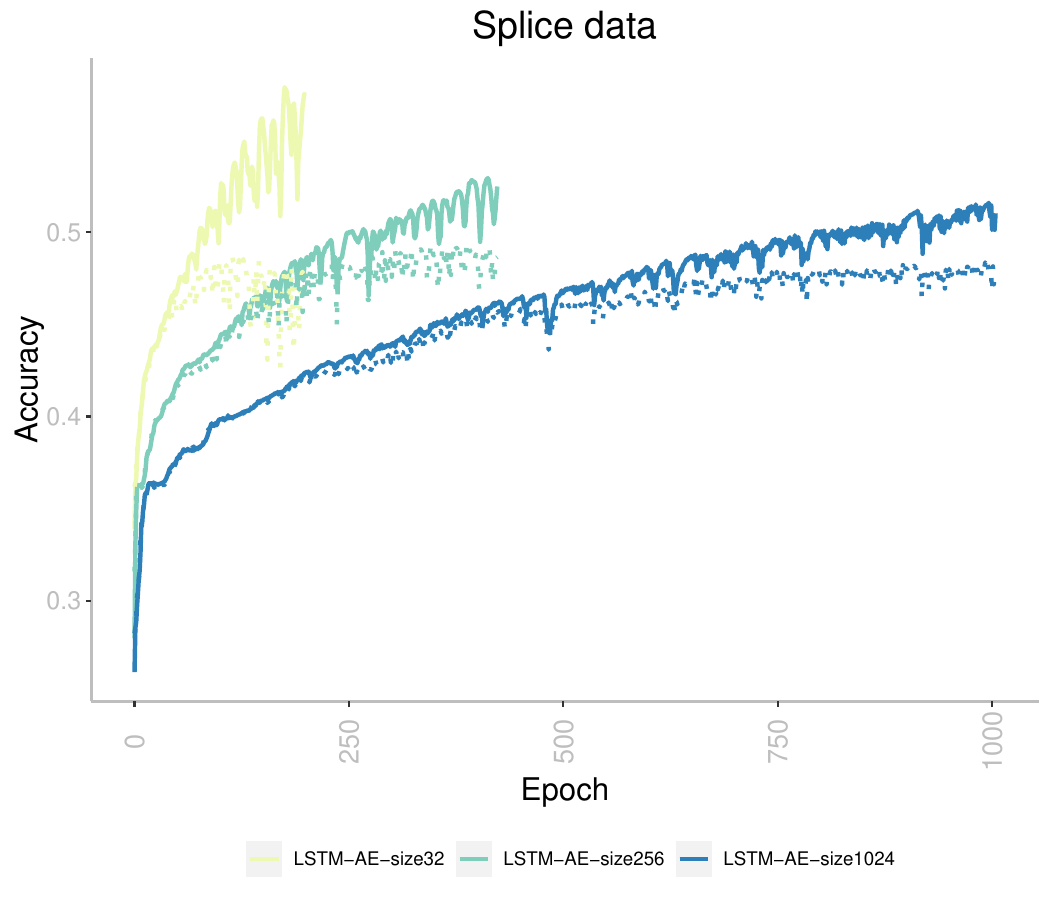}
    \includegraphics[scale=0.3]{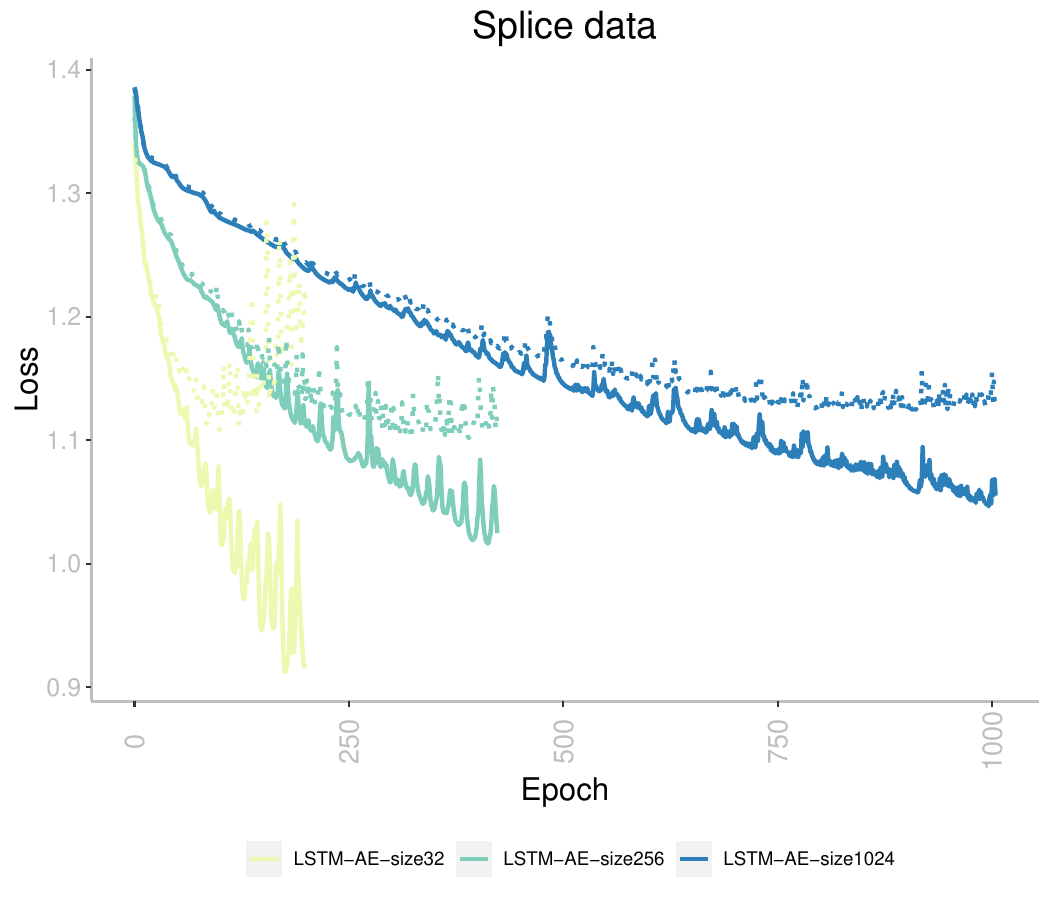}
    \includegraphics[scale=0.3]{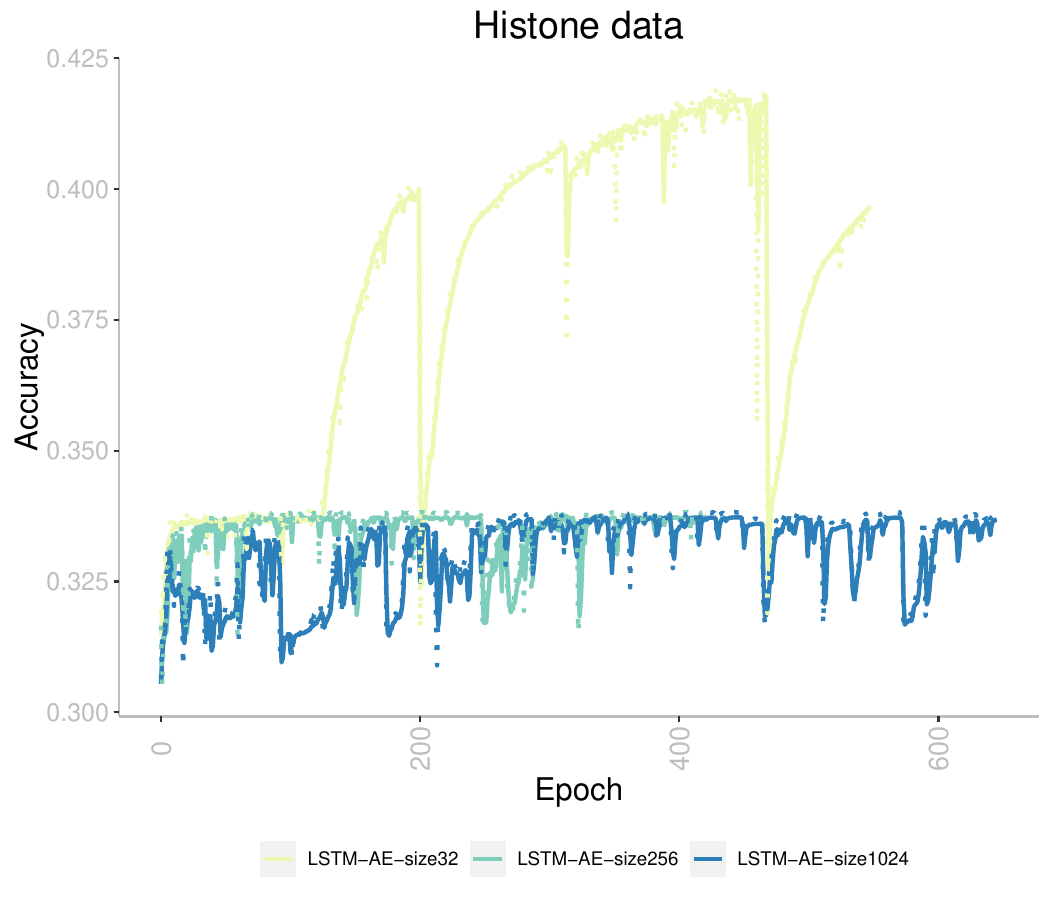}
    \includegraphics[scale=0.3]{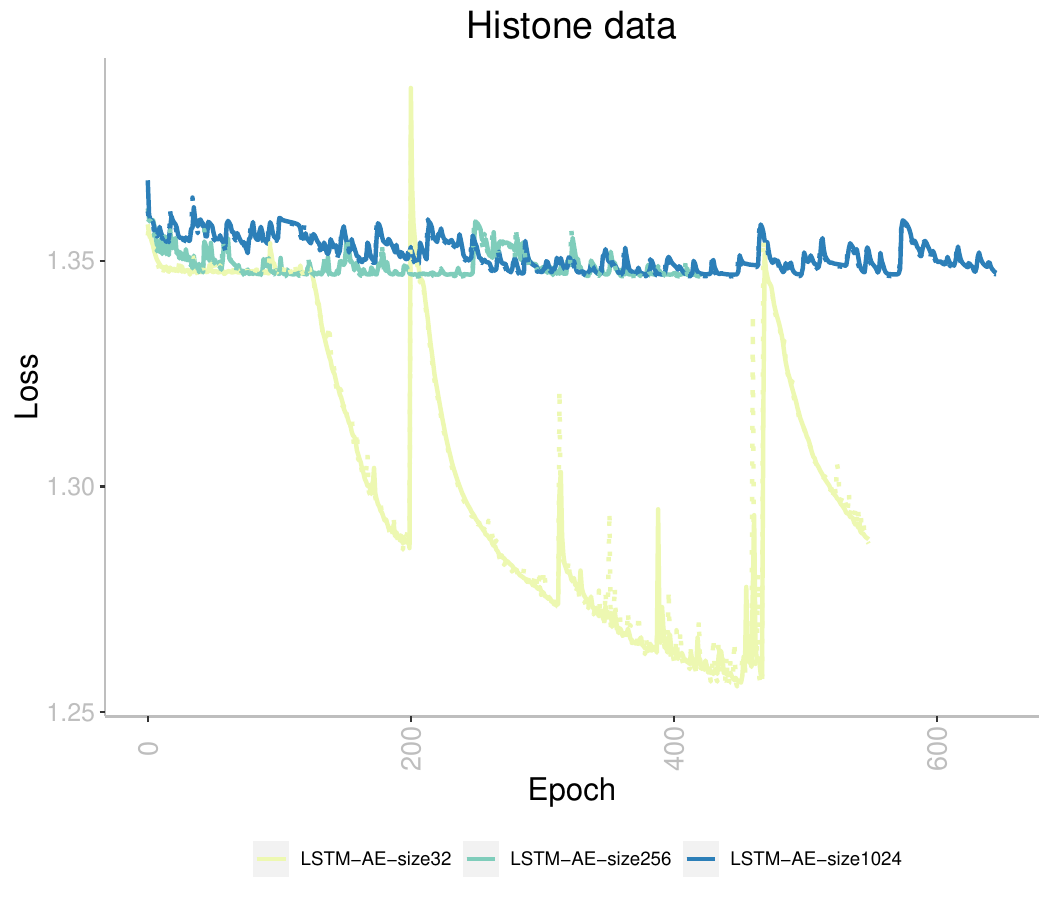}
    \includegraphics[scale=0.3]{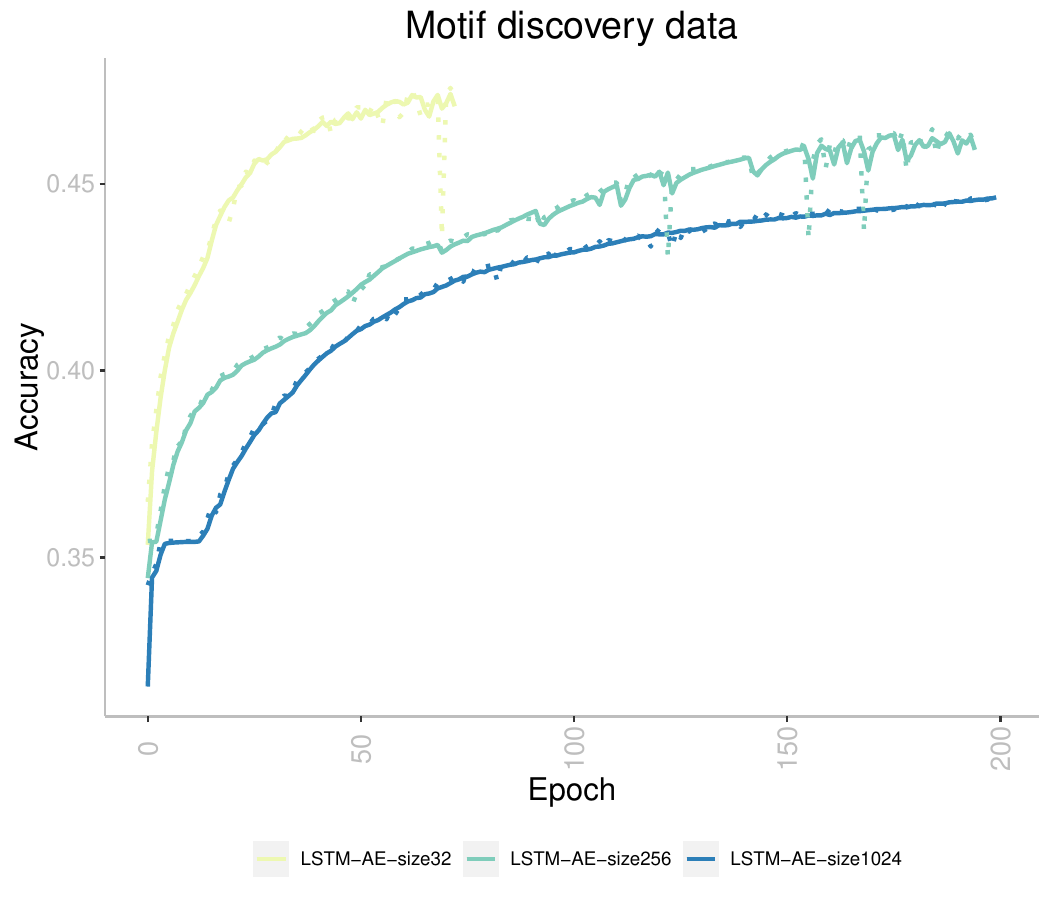}
    \includegraphics[scale=0.3]{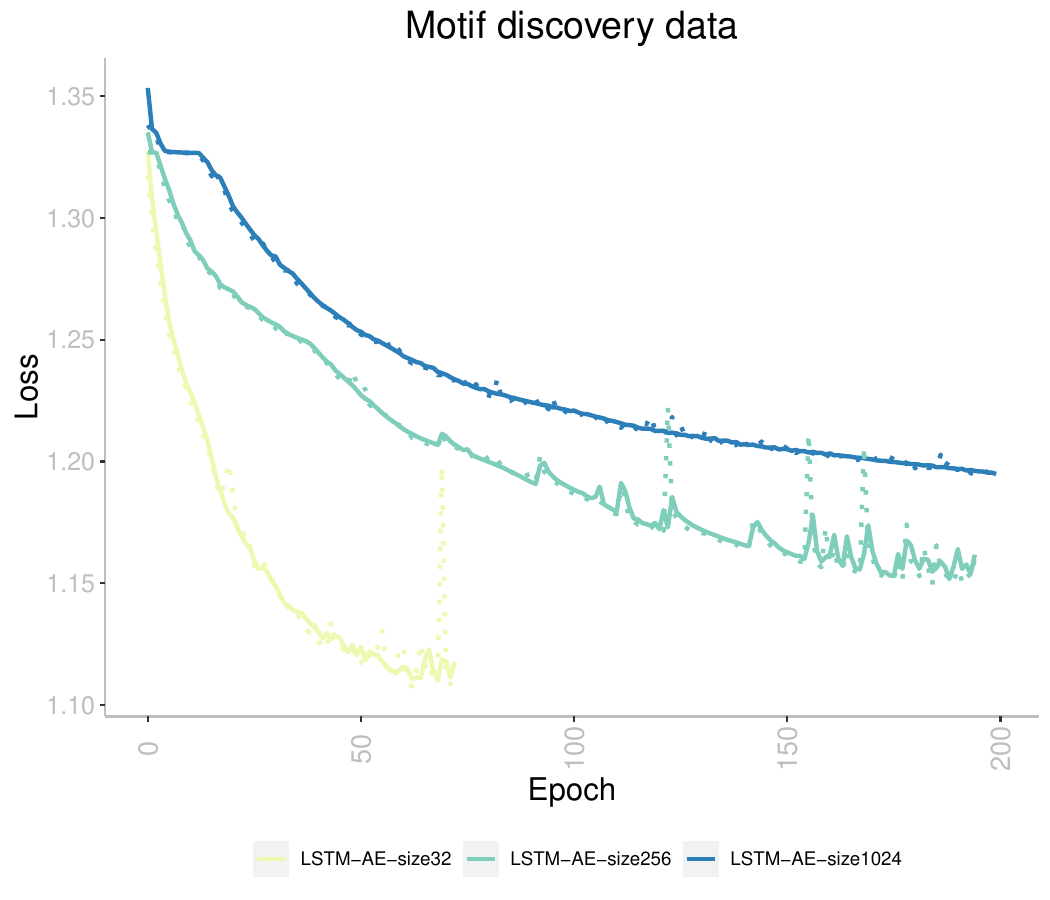}
    \caption{Learning dynamics for LSTM-AE+NN model on different datasets and different batch sizes. Solid lines correspond to training accuracy/loss and dashed lines correspond to validating accuracy/loss. Colors represent different batch sizes.}
    \label{fig:batch-size-dynamics}
\end{figure}

See Figure \ref{fig:embed-size-pr} for the precision-recall plot on the role of the embedding size for the doc2vec model, and see Figure \ref{fig:embed-size-roc} for the ROC curve plot.
See Figures \ref{fig:l2-acc} and \ref{fig:l2-dynamics} for the effect of regularization on the CNN-Zeng model.
Finally, Figures \ref{fig:dimension-dynamics} to \ref{fig:batch-size-dynamics} show the learning dynamics of loss and accuracy for all the models presented in this work.

\subsection*{Reproducible Julia script for random order of first authors}

\begin{verbatim}
using Random
s = 313627913
Random.seed!(s)
people = ["songyang","zhaoyi"]
people[randperm(length(people))]
\end{verbatim}

\section*{Acknowledgements}
We thank Dr. Aurelie Rakotondrafara and Helena Jaramillo Mesa for the motivation to compare neural network models on plant viral data. We acknowledge the work in \citep{Hotaling2020} which helped us improve the scientific writing of this manuscript. \revision{We thank the associate editor and the three anonymous reviewers for the insightful comments and suggestions which greatly improved the manuscript.}

\section*{Funding}
This work is supported by the Department of Energy [DE-SC0021016 to CSL]. 


\section*{Availability of data and materials}
The data was made publicly available by the original manuscripts. All the scripts developed in this work are publicly available in the GitHub repository \url{https://github.com/solislemuslab/dna-nn-theory}.


\section*{Competing interests}
The authors declare that they have no competing interests.


\section*{Authors' contributions}
ZZ and SC ran all the analyses, researched the literature to find the candidate models to compare, programmed all the open-source code (\url{https://github.com/solislemuslab/dna-nn-theory}) and wrote an initial draft of the manuscript. CSL developed the idea, created all the plots in the manuscript and wrote the final version of the manuscript.


\vskip 0.2in
\typeout{}
\bibliography{references}

\begin{thebibliography}{45}
\providecommand{\natexlab}[1]{#1}
\providecommand{\url}[1]{\texttt{#1}}
\expandafter\ifx\csname urlstyle\endcsname\relax
  \providecommand{\doi}[1]{doi: #1}\else
  \providecommand{\doi}{doi: \begingroup \urlstyle{rm}\Url}\fi

\bibitem[Agarwal et~al.(2019)Agarwal, Jayanth Kumar~Reddy, and
  Anand]{Agarwal2019-pr}
Vishal Agarwal, N~Jayanth Kumar~Reddy, and Ashish Anand.
\newblock Unsupervised representation learning of {DNA} sequences.
\newblock \emph{arXiv}, \penalty0 (1906.03087), June 2019.

\bibitem[Alber et~al.(2019)Alber, Tepole, Cannon, De, Dura-Bernal, Garikipati,
  Karniadakis, Lytton, Perdikaris, Petzold, et~al.]{alber2019integrating}
Mark Alber, Adrian~Buganza Tepole, William~R Cannon, Suvranu De, Salvador
  Dura-Bernal, Krishna Garikipati, George Karniadakis, William~W Lytton, Paris
  Perdikaris, Linda Petzold, et~al.
\newblock Integrating machine learning and multiscale modeling—perspectives,
  challenges, and opportunities in the biological, biomedical, and behavioral
  sciences.
\newblock \emph{NPJ digital medicine}, 2\penalty0 (1):\penalty0 1--11, 2019.

\bibitem[Andrew and Delong(2015)]{andrew2015weirauch}
Babak~Alipanahi Andrew and Matthew~T Delong.
\newblock Weirauch frey brendan j: Predicting the sequence specificities of
  dna-and rna-binding proteins by deep learning.
\newblock \emph{Nat Biotechnol}, 10, 2015.

\bibitem[Ashley(2015)]{ashley2015precision}
Euan~A Ashley.
\newblock The precision medicine initiative: a new national effort.
\newblock \emph{Jama}, 313\penalty0 (21):\penalty0 2119--2120, 2015.

\bibitem[Bokulich et~al.(2018)Bokulich, Dillon, Bolyen, Kaehler, Huttley, and
  Caporaso]{bokulich2018q2}
Nicholas~A Bokulich, Matthew~R Dillon, Evan Bolyen, Benjamin~D Kaehler, Gavin~A
  Huttley, and J~Gregory Caporaso.
\newblock q2-sample-classifier: machine-learning tools for microbiome
  classification and regression.
\newblock \emph{Journal of open research software}, 3\penalty0 (30), 2018.

\bibitem[Carrieri et~al.(2019)Carrieri, Rowe, Winn, and
  Pyzer-Knapp]{carrieri2019fast}
Anna~Paola Carrieri, Will~PM Rowe, Martyn Winn, and Edward~O Pyzer-Knapp.
\newblock A fast machine learning workflow for rapid phenotype prediction from
  whole shotgun metagenomes.
\newblock In \emph{Proceedings of the AAAI Conference on Artificial
  Intelligence}, volume~33, pages 9434--9439, 2019.

\bibitem[Chang et~al.(2017)Chang, Haudenshield, Bowen, and
  Hartman]{chang2017metagenome}
Hao-Xun Chang, James~S Haudenshield, Charles~R Bowen, and Glen~L Hartman.
\newblock Metagenome-wide association study and machine learning prediction of
  bulk soil microbiome and crop productivity.
\newblock \emph{Frontiers in Microbiology}, 8:\penalty0 519, 2017.

\bibitem[Chen et~al.(2019)Chen, Liu, and Peng]{chen2018}
Po-Hsuan~Cameron Chen, Yun Liu, and Lily Peng.
\newblock How to develop machine learning models for healthcare.
\newblock \emph{Nature Materials}, 18\penalty0 (5):\penalty0 410--414, 2019.

\bibitem[Coelho et~al.(2013)Coelho, Carri{\c{c}}o, Knight, Mart{\'\i}nez,
  Morrissey, Oggioni, and Freitas]{coelho2013use}
Joana~Rosado Coelho, Jo{\~a}o~Andr{\'e} Carri{\c{c}}o, Daniel Knight, Jose-Luis
  Mart{\'\i}nez, Ian Morrissey, Marco~Rinaldo Oggioni, and Ana~Teresa Freitas.
\newblock The use of machine learning methodologies to analyse antibiotic and
  biocide susceptibility in staphylococcus aureus.
\newblock \emph{PLoS One}, 8\penalty0 (2):\penalty0 e55582, 2013.

\bibitem[Dacrema et~al.(2019)Dacrema, Cremonesi, and Jannach]{dacrema2019}
Maurizio~Ferrari Dacrema, Paolo Cremonesi, and Dietmar Jannach.
\newblock Are we really making much progress? a worrying analysis of recent
  neural recommendation approaches.
\newblock In \emph{Proceedings of the 13th ACM Conference on Recommender
  Systems}, RecSys '19, page 101–109, New York, NY, USA, 2019. Association
  for Computing Machinery.
\newblock ISBN 9781450362436.
\newblock \doi{10.1145/3298689.3347058}.
\newblock URL \url{https://doi.org/10.1145/3298689.3347058}.

\bibitem[Dua and Graff(2017)]{Dua:2019}
Dheeru Dua and Casey Graff.
\newblock {UCI} machine learning repository, 2017.
\newblock URL \url{http://archive.ics.uci.edu/ml}.

\bibitem[Duarte-Carvajalino et~al.(2018)Duarte-Carvajalino, Alzate, Ramirez,
  Santa-Sepulveda, Fajardo-Rojas, and Soto-Su{\'a}rez]{duarte2018evaluating}
Julio~M Duarte-Carvajalino, Diego~F Alzate, Andr{\'e}s~A Ramirez, Juan~D
  Santa-Sepulveda, Alexandra~E Fajardo-Rojas, and Mauricio Soto-Su{\'a}rez.
\newblock Evaluating late blight severity in potato crops using unmanned aerial
  vehicles and machine learning algorithms.
\newblock \emph{Remote Sensing}, 10\penalty0 (10):\penalty0 1513, 2018.

\bibitem[Dutta et~al.(2018)Dutta, Singh, and Anand]{Dutta2018}
T.~Dutta, A.and~Dubey, K.~K. Singh, and A.~Anand.
\newblock Splicevec: Distributed feature representations for splice junction
  prediction.
\newblock \emph{Computational biology and chemistry}, 74:\penalty0 434--441,
  2018.

\bibitem[Egmont-Petersen et~al.(2002)Egmont-Petersen, de~Ridder, and
  Handels]{egmont2002image}
Michael Egmont-Petersen, Dick de~Ridder, and Heinz Handels.
\newblock Image processing with neural networks: a review.
\newblock \emph{Pattern recognition}, 35\penalty0 (10):\penalty0 2279--2301,
  2002.

\bibitem[Ekins et~al.(2019)Ekins, Puhl, Zorn, Lane, Russo, Klein, Hickey, and
  Clark]{ekins2019}
Sean Ekins, Ana~C. Puhl, Kimberley~M. Zorn, Thomas~R. Lane, Daniel~P. Russo,
  Jennifer~J. Klein, Anthony~J. Hickey, and Alex~M. Clark.
\newblock Exploiting machine learning for end-to-end drug discovery and
  development.
\newblock \emph{Nature Materials}, 18\penalty0 (5):\penalty0 435--441, 2019.

\bibitem[Fjell et~al.(2009)Fjell, Jenssen, Hilpert, Cheung, Pante, Hancock, and
  Cherkasov]{fjell2009identification}
Christopher~D Fjell, H{\aa}vard Jenssen, Kai Hilpert, Warren~A Cheung, Nelly
  Pante, Robert~EW Hancock, and Artem Cherkasov.
\newblock Identification of novel antibacterial peptides by chemoinformatics
  and machine learning.
\newblock \emph{Journal of medicinal chemistry}, 52\penalty0 (7):\penalty0
  2006--2015, 2009.

\bibitem[Hjelm{\aa}s and Low(2001)]{hjelmaas2001face}
Erik Hjelm{\aa}s and Boon~Kee Low.
\newblock Face detection: A survey.
\newblock \emph{Computer vision and image understanding}, 83\penalty0
  (3):\penalty0 236--274, 2001.

\bibitem[Ho et~al.(2019)Ho, Schierding, Wake, Saffery, and
  O’Sullivan]{ho2019machine}
Daniel Sik~Wai Ho, William Schierding, Melissa Wake, Richard Saffery, and
  Justin O’Sullivan.
\newblock Machine learning snp based prediction for precision medicine.
\newblock \emph{Frontiers in Genetics}, 10:\penalty0 267, 2019.

\bibitem[Hochreiter and Schmidhuber(1997)]{Hochreiter1997}
Sepp Hochreiter and J\"{u}rgen Schmidhuber.
\newblock Long short-term memory.
\newblock \emph{Neural Comput.}, 9\penalty0 (8):\penalty0 1735–1780, November
  1997.
\newblock ISSN 0899-7667.
\newblock \doi{10.1162/neco.1997.9.8.1735}.
\newblock URL \url{https://doi.org/10.1162/neco.1997.9.8.1735}.

\bibitem[Hotaling(2020)]{Hotaling2020}
Scott Hotaling.
\newblock Simple rules for concise scientific writing.
\newblock \emph{Limnology and Oceanography Letters}, 5\penalty0 (6):\penalty0
  379--383, 2020.
\newblock \doi{https://doi.org/10.1002/lol2.10165}.
\newblock URL
  \url{https://aslopubs.onlinelibrary.wiley.com/doi/abs/10.1002/lol2.10165}.

\bibitem[Ip et~al.(2018)Ip, Ang, Seng, Broster, and Pratley]{ip2018big}
Ryan~HL Ip, Li-Minn Ang, Kah~Phooi Seng, JC~Broster, and JE~Pratley.
\newblock Big data and machine learning for crop protection.
\newblock \emph{Computers and Electronics in Agriculture}, 151:\penalty0
  376--383, 2018.

\bibitem[Jonas et~al.(2018)Jonas, Bobra, Shankar, Hoeksema, and
  Recht]{jonas2018flare}
Eric Jonas, Monica Bobra, Vaishaal Shankar, J~Todd Hoeksema, and Benjamin
  Recht.
\newblock Flare prediction using photospheric and coronal image data.
\newblock \emph{Solar Physics}, 293\penalty0 (3):\penalty0 48, 2018.

\bibitem[Katuwal and Chen(2016)]{katuwal2016machine}
Gajendra~Jung Katuwal and Robert Chen.
\newblock Machine learning model interpretability for precision medicine.
\newblock \emph{arXiv preprint arXiv:1610.09045}, 2016.

\bibitem[Kavvas et~al.(2018)Kavvas, Catoiu, Mih, Yurkovich, Seif, Dillon,
  Heckmann, Anand, Yang, Nizet, et~al.]{kavvas2018machine}
Erol~S Kavvas, Edward Catoiu, Nathan Mih, James~T Yurkovich, Yara Seif,
  Nicholas Dillon, David Heckmann, Amitesh Anand, Laurence Yang, Victor Nizet,
  et~al.
\newblock Machine learning and structural analysis of mycobacterium
  tuberculosis pan-genome identifies genetic signatures of antibiotic
  resistance.
\newblock \emph{Nature communications}, 9\penalty0 (1):\penalty0 1--9, 2018.

\bibitem[Kelley et~al.(2016)Kelley, Snoek, and Rinn]{kelley2016basset}
David~R Kelley, Jasper Snoek, and John~L Rinn.
\newblock Basset: learning the regulatory code of the accessible genome with
  deep convolutional neural networks.
\newblock \emph{Genome research}, 26\penalty0 (7):\penalty0 990--999, 2016.

\bibitem[Kimothi et~al.(2016)Kimothi, Soni, Biyani, and
  Hogan]{kimothi2016distributed}
Dhananjay Kimothi, Akshay Soni, Pravesh Biyani, and James~M. Hogan.
\newblock Distributed representations for biological sequence analysis.
\newblock \emph{arXiv}, \penalty0 (1608.05949), 2016.

\bibitem[Krittanawong et~al.(2017)Krittanawong, Zhang, Wang, Aydar, and
  Kitai]{krittanawong2017artificial}
Chayakrit Krittanawong, HongJu Zhang, Zhen Wang, Mehmet Aydar, and Takeshi
  Kitai.
\newblock Artificial intelligence in precision cardiovascular medicine.
\newblock \emph{Journal of the American College of Cardiology}, 69\penalty0
  (21):\penalty0 2657--2664, 2017.

\bibitem[Kucuk et~al.(2017)Kucuk, Banda, and Angryk]{kucuk2017large}
Ahmet Kucuk, Juan~M Banda, and Rafal~A Angryk.
\newblock A large-scale solar dynamics observatory image dataset for computer
  vision applications.
\newblock \emph{Scientific data}, 4:\penalty0 170096, 2017.

\bibitem[Le and Mikolov(2014)]{le2014distributed}
Quoc Le and Tomas Mikolov.
\newblock Distributed representations of sentences and documents.
\newblock In Eric~P. Xing and Tony Jebara, editors, \emph{Proceedings of the
  31st International Conference on Machine Learning}, volume~32 of
  \emph{Proceedings of Machine Learning Research}, pages 1188--1196, Bejing,
  China, 22--24 Jun 2014. PMLR.
\newblock URL \url{http://proceedings.mlr.press/v32/le14.html}.

\bibitem[Lee et~al.(2018)Lee, Celik, Logsdon, Lundberg, Martins, Oehler, Estey,
  Miller, Chien, Dai, et~al.]{lee2018machine}
Su-In Lee, Safiye Celik, Benjamin~A Logsdon, Scott~M Lundberg, Timothy~J
  Martins, Vivian~G Oehler, Elihu~H Estey, Chris~P Miller, Sylvia Chien, Jin
  Dai, et~al.
\newblock A machine learning approach to integrate big data for precision
  medicine in acute myeloid leukemia.
\newblock \emph{Nature communications}, 9\penalty0 (1):\penalty0 1--13, 2018.

\bibitem[Li et~al.(2018)Li, Yin, and Zhang]{li2018tracking}
Li-Guan Li, Xiaole Yin, and Tong Zhang.
\newblock Tracking antibiotic resistance gene pollution from different sources
  using machine-learning classification.
\newblock \emph{Microbiome}, 6\penalty0 (1):\penalty0 1--12, 2018.

\bibitem[Maino et~al.(2018)Maino, Umina, and Hoffmann]{maino2018climate}
James~L Maino, Paul~A Umina, and Ary~A Hoffmann.
\newblock Climate contributes to the evolution of pesticide resistance.
\newblock \emph{Global Ecology and Biogeography}, 27\penalty0 (2):\penalty0
  223--232, 2018.

\bibitem[Nado et~al.(2021)Nado, Gilmer, Shallue, Anil, and Dahl]{nado2021large}
Zachary Nado, Justin~M Gilmer, Christopher~J Shallue, Rohan Anil, and George~E
  Dahl.
\newblock A large batch optimizer reality check: Traditional, generic
  optimizers suffice across batch sizes.
\newblock \emph{arXiv preprint arXiv:2102.06356}, 2021.

\bibitem[Nguyen et~al.(2016)Nguyen, Tran, Ngo, Phan, Lumbanraja, Faisal,
  Abapihi, Kubo, and Satou]{Nguyen2016-jj}
Ngoc~Giang Nguyen, Vu~Anh Tran, Duc~Luu Ngo, Dau Phan, Favorisen~Rosyking
  Lumbanraja, Mohammad~Reza Faisal, Bahriddin Abapihi, Mamoru Kubo, and Kenji
  Satou.
\newblock {DNA} sequence classification by convolutional neural network.
\newblock \emph{JBiSE}, 09\penalty0 (05):\penalty0 280--286, 2016.

\bibitem[Peng et~al.(2020)Peng, Alber, Buganza~Tepole, Cannon, De, Dura-Bernal,
  Garikipati, Karniadakis, Lytton, Perdikaris, Petzold, and Kuhl]{Peng2020}
Grace C.~Y. Peng, Mark Alber, Adrian Buganza~Tepole, William~R. Cannon, Suvranu
  De, Savador Dura-Bernal, Krishna Garikipati, George Karniadakis, William~W.
  Lytton, Paris Perdikaris, Linda Petzold, and Ellen Kuhl.
\newblock Multiscale modeling meets machine learning: What can we learn?
\newblock \emph{Archives of Computational Methods in Engineering}, 2020.

\bibitem[Pesesky et~al.(2016)Pesesky, Hussain, Wallace, Patel, Andleeb,
  Burnham, and Dantas]{pesesky2016evaluation}
Mitchell~W Pesesky, Tahir Hussain, Meghan Wallace, Sanket Patel, Saadia
  Andleeb, Carey-Ann~D Burnham, and Gautam Dantas.
\newblock Evaluation of machine learning and rules-based approaches for
  predicting antimicrobial resistance profiles in gram-negative bacilli from
  whole genome sequence data.
\newblock \emph{Frontiers in microbiology}, 7:\penalty0 1887, 2016.

\bibitem[Reddi et~al.(2019)Reddi, Kale, and Kumar]{reddi2019convergence}
Sashank~J. Reddi, Satyen Kale, and Sanjiv Kumar.
\newblock On the convergence of adam and beyond.
\newblock \emph{arXiv}, \penalty0 (1904.09237), 2019.

\bibitem[Rost et~al.(2016)Rost, Radivojac, and Bromberg]{rost2016protein}
Burkhard Rost, Predrag Radivojac, and Yana Bromberg.
\newblock Protein function in precision medicine: deep understanding with
  machine learning.
\newblock \emph{FEBS letters}, 590\penalty0 (15):\penalty0 2327--2341, 2016.

\bibitem[Shadab et~al.(2020)Shadab, Khan, Neezi, Adilina, and
  {others}]{Shadab2020-sb}
S~Shadab, M~T~A Khan, N~A Neezi, S~Adilina, and {others}.
\newblock {DeepDBP}: Deep neural networks for identification of {DNA-binding}
  proteins.
\newblock \emph{Informatics in Medicine}, 2020.

\bibitem[Teschendorff(2019)]{teschendorff2019avoiding}
Andrew~E Teschendorff.
\newblock Avoiding common pitfalls in machine learning omic data science.
\newblock \emph{Nature Materials}, 18\penalty0 (5):\penalty0 422--427, 2019.

\bibitem[Trabelsi et~al.(2019)Trabelsi, Chaabane, and Ben-Hur]{Trabelsi2019-yt}
Ameni Trabelsi, Mohamed Chaabane, and Asa Ben-Hur.
\newblock Comprehensive evaluation of deep learning architectures for
  prediction of {DNA/RNA} sequence binding specificities.
\newblock \emph{Bioinformatics}, 35\penalty0 (14):\penalty0 i269--i277, July
  2019.

\bibitem[Wolpert and Macready(1997)]{Wolpert1997}
D.H. Wolpert and W.G. Macready.
\newblock No free lunch theorems for optimization.
\newblock \emph{IEEE Transactions on Evolutionary Computation}, 1\penalty0
  (1):\penalty0 67--82, 1997.
\newblock \doi{10.1109/4235.585893}.

\bibitem[Yang and Guo(2017)]{yang2017machine}
Xin Yang and Tingwei Guo.
\newblock Machine learning in plant disease research.
\newblock \emph{European Journal of BioMedical Research}, 3\penalty0
  (1):\penalty0 6--9, 2017.

\bibitem[Zeng et~al.(2016)Zeng, Edwards, Liu, and Gifford]{Zeng2016-nu}
Haoyang Zeng, Matthew~D Edwards, Ge~Liu, and David~K Gifford.
\newblock Convolutional neural network architectures for predicting
  {DNA-protein} binding.
\newblock \emph{Bioinformatics}, 32\penalty0 (12):\penalty0 i121--i127, June
  2016.

\bibitem[Zhou and Troyanskaya(2015)]{zhou2015predicting}
Jian Zhou and Olga~G Troyanskaya.
\newblock Predicting effects of noncoding variants with deep learning--based
  sequence model.
\newblock \emph{Nature methods}, 12\penalty0 (10):\penalty0 931--934, 2015.

\end{thebibliography}

\section*{Figures}

\section*{Tables}
\begin{table}[h!]
    \centering
    \caption{Datasets used to test the neural network models.}
    \begin{tabular}{cccc}
        \hline
        Dataset & Sample size & Sequence Length (bp) & Reference \\
        \hline
        Splice & 3190 & 60 & \citep{Nguyen2016-jj} \\
        Histone &  14965 & 500 & \citep{Nguyen2016-jj} \\
        Motif discovery & 269100 & 101 & \citep{Zeng2016-nu} \\
        \hline
    \end{tabular}
    \label{tab:datasets}
\end{table}

\begin{table}[ht]
    \centering
    \caption{CNN models along with the datasets used and the performance tests}
    \begin{tabular}{p{0.19\textwidth} p{0.15\textwidth} p{0.15\textwidth} p{0.17\textwidth} p{0.19\textwidth}}
        \hline
         & Model & Own datasets & Outside datasets & Performance tests\\
         \hline
        CNN-Nguyen \citep{Nguyen2016-jj} & Figure \ref{fig:cnn-nguyen} & splice, histone & motif discovery & number of layers, dimension \\
        CNN-Zeng \citep{Zeng2016-nu} & Figure \ref{fig:cnn-zeng} & motif discovery & splice, histone & number of layers \\
        DeepDBP \citep{Shadab2020-sb} & Figure \ref{fig:deepdbp} & & splice, histone, motif discovery &   \\
        DeepRAM \citep{Trabelsi2019-yt} & Figures \ref{fig:deepram-cnn}, \ref{fig:deepram-rnn}, \ref{fig:deepram-cnnrnn} & & splice, histone, motif discovery & data encoding \\
        \hline
    \end{tabular}
    \label{tab:analyses}
\end{table}

\begin{table}[h!]
    \centering
    \caption{NLP models along with the datasets used and the performance tests}
    \begin{tabular}{p{0.19\textwidth} p{0.15\textwidth} p{0.15\textwidth} p{0.17\textwidth} p{0.19\textwidth}}
        \hline
         & Model & Own datasets & Outside datasets & Performance tests\\
         \hline
        LSTM-layer & Figure \ref{fig:lstm-layer} &  & splice, histone, motif discovery & optimizer \\
        doc2vec+NN \citep{le2014distributed, kimothi2016distributed} & Figure \ref{fig:simple-nn} & & splice, histone, motif discovery & embedding size \\
        LSTM-AE+NN \citep{Agarwal2019-pr} & Figures \ref{fig:lstm-ae-encoder}, \ref{fig:simple-nn} & & splice, histone, motif discovery & batch size  \\
        \hline
    \end{tabular}
    \label{tab:analyses2}
\end{table}

\begin{table}[h!]
    \centering
    \caption{\revision{Training details on NLP models. "Max. It." means maximum number of iterations allowed.}}
    \begin{tabular}{p{0.25\textwidth} p{0.1\textwidth} p{0.1\textwidth} p{0.12\textwidth} p{0.1\textwidth} p{0.22\textwidth}}
        \hline
         Model & Data & Max. It. & Patience & Early Stop & Optimizer\\
         \hline
        LSTM-AE (32) & Splice & 2000 & 100 & 1474 & Adam (LR=0.001) \\
        LSTM-AE+NN (32) & Splice & 1000 & 100 & 123 & SGD (LR=0.01) \\
        LSTM-AE (256) & Splice & 2000 & 100 & 424 & Adam (LR=0.001) \\
        LSTM-AE+NN (256) & Splice & 1000 & 100 & 114 & SGD (LR=0.01) \\
        LSTM-AE (1024) & Splice & 2000 & 100 & 1005 & Adam (LR=0.001) \\
        LSTM-AE+NN (1024) & Splice & 1000 & 100 & 200 & SGD (LR=0.01) \\
        LSTM-layer & Splice & 4000 & 100 & 289 & Adam (LR=0.001) \\
        LSTM-layer & Splice & 4000 & 100 & 2872 & SGD (LR=0.01) \\
        doc2vec+NN (50) & Splice & 1000 & 50 & 119 & SGD (LR=0.01) \\
        doc2vec+NN (100) & Splice & 1000 & 50 & 143 & SGD (LR=0.01) \\
        doc2vec+NN (150) & Splice & 1000 & 50 & 166 & SGD (LR=0.01) \\
        doc2vec+NN (200) & Splice & 1000 & 50 & 264 & SGD (LR=0.01) \\
        \hline
        LSTM-AE (32) & Histone & 4000 & 100 & 549 & Adam (LR=0.001) \\
        LSTM-AE+NN (32) & Histone & 1000 & 100 & 212 & SGD (LR=0.001) \\
        LSTM-AE (256) & Histone & 4000 & 200 & 422 & Adam (LR=0.001) \\
        LSTM-AE+NN (256) & Histone & 1500 & 200 & 805 & SGD (LR=0.001) \\
        LSTM-AE (1024) & Histone & 4000 & 200 & 646 & Adam (LR=0.001) \\
        LSTM-AE+NN (1024) & Histone & 1500 & 200 & 999 & SGD (LR=0.001) \\
        LSTM-layer & Histone & 3000 & 100 & 154 & Adam (LR=0.001) \\
        LSTM-layer & Histone & 4000 & 400 & 526 & SGD (LR=0.01) \\
        doc2vec+NN (50) & Histone & 1000 & 30 & 114 & SGD (LR=0.01) \\
        doc2vec+NN (100) & Histone & 1000 & 30 & 177 & SGD (LR=0.01) \\
        doc2vec+NN (150) & Histone & 1000 & 30 & 61 & SGD (LR=0.01) \\
        doc2vec+NN (200) & Histone & 1000 & 30 & 118 & SGD (LR=0.01) \\
        \hline
        LSTM-AE (32) & Motif & 200 & 10 & 78 & Adam (LR=0.001) \\
        LSTM-AE+NN (32) & Motif & 500 & 10 & 185 & SGD (LR=0.01) \\
        LSTM-AE (256) & Motif & 200 & 10 & 195 & Adam (LR=0.001) \\
        LSTM-AE+NN (256) & Motif & 500 & 10 & 146 & SGD (LR=0.01) \\
        LSTM-AE (1024) & Motif & 200 & 10 & 200 & Adam (LR=0.001) \\
        LSTM-AE+NN (1024) & Motif & 500 & 10 & 62 & SGD (LR=0.01) \\
        LSTM-layer & Motif & 200 & 5 & 51 & Adam (LR=0.001) \\
        LSTM-layer & Motif & 200 & 5 & 200 & SGD (LR=0.01) \\
        doc2vec+NN (50) & Motif & 400 & 10 & 36 & SGD (LR=0.01) \\
        doc2vec+NN (100) & Motif & 400 & 10 & 39 & SGD (LR=0.01) \\
        doc2vec+NN (150) & Motif & 400 & 10 & 58 & SGD (LR=0.01) \\
        doc2vec+NN (200) & Motif & 400 & 10 & 11 & SGD (LR=0.01) \\
        \hline
    \end{tabular}
    \label{tab:training}
  \end{table}

\begin{table}[h!]
    \centering
    \caption{Models with the highest testing accuracy for each dataset.}
    \begin{tabular}{cccc}
        \hline
        Model & Splice & Histone & Motif discovery \\
        \hline
        CNN-Nguyen & original (two 2D layer) & extra 1D layer & extra 1D layer \\
        CNN-Zeng & four layers L2-reg & four layers L2-reg & four layers L2-reg \\
        DeepRAM & RNN-Embed & RNN-OneHot & RNN-OneHot \\
        LSTM-layer & SGD & ADAM & ADAM \\
        LSTM-AE &  batch size 1024 & batch size 1024 & batch size 1024 \\
        doc2vec & embed size 150 & embed size 50 & embed size 150 \\
        \hline
    \end{tabular}
    \label{tab:best}
\end{table}



\end{document}